\documentclass{article}

\usepackage[utf8]{inputenc}
\usepackage[english]{babel}
\usepackage[subpreambles=false]{standalone}
\usepackage{amsthm, amssymb, amsmath}
\usepackage{import}
\usepackage{csquotes}
\usepackage[table]{xcolor}
\usepackage{booktabs, multirow, slashbox}
\usepackage[inline]{enumitem}
\usepackage{todonotes}
\usepackage{upgreek}
\usepackage{adjustbox}
\usepackage{graphicx}
\usepackage{subcaption}
\usepackage{bookmark}
\usepackage{breakcites} 

\usepackage{xfrac}
\usepackage{pgfplots}
\usepackage{import}
\pgfplotsset{compat=newest}
\pgfplotsset{plot coordinates/math parser=false}
\newlength\figureheight
\newlength\figurewidth

\usepackage{pifont}
\usepackage[toc,page]{appendix}
\usepackage[section]{placeins}
\usepackage{fmtcount}
\usetikzlibrary{calc, patterns, intersections}
\usepackage[sort,square,numbers]{natbib}
\usepackage{hyperref}
\usepackage[capitalize,nameinlink]{cleveref}

\newcommand{\cmark}{{\color{green}{\ding{51}}}}%
\newcommand{\xmark}{{\color{red}{\ding{55}}}}%

\newcommand\MyHead[2]{%
  \multicolumn{1}{l}{\parbox{#1}{\centering #2}}
}
\newcommand*{\tabhead}[1]{\textbf{#1}}%

\newcommand\InsertEmptyLine{~\\\indent}

\DeclareRobustCommand{\insertimage}[1]{%
	\tikz{%
		\draw[#1] (0, 0) rectangle (1, 1);%
	}%
}

\newtheorem{theorem}{Theorem}[section]
\newtheorem{lemma}[theorem]{Lemma}
\newtheorem{claim}[theorem]{Claim}

\theoremstyle{definition}
\newtheorem{definition}[theorem]{Definition}
\newtheorem{observation}[theorem]{Observation}

\makeatletter
\newtheorem*{rep@theorem}{\rep@title}
\newcommand{\newreptheorem}[2]{%
\newenvironment{rep#1}[1]{%
 \def\rep@title{#2 \ref{##1}}%
 \begin{rep@theorem}}%
 {\end{rep@theorem}}}
\makeatother
\newreptheorem{theorem}{Theorem}

\newcommand{\MultiAgentSortingMainDirectory}{arXiv.masmd}

\newcommand{\StateMachineTablesDirectory}{arXiv.tables}
\newcommand{\ProcessedDataRelativeToSortingMainDirectory}{data}
\newcommand{\ProcessedDataRelativeToSortingTwoLaneDirectory}{data}

\newcommand{\RootDirectory}{./}

\graphicspath{{\RootDirectory\MultiAgentSortingMainDirectory/clipart/}}

\title{Emerging cooperation on the road by myopic local interactions}
\author{Dmitry Rabinovich \\ 
\normalsize dmitry.ra@cs.technion.ac.il \\
\normalsize Technion - IIT, Israel
\and Alfred M. Bruckstein \\ 
\normalsize freddy@cs.technion.ac.il \\
\normalsize Technion - IIT, Israel}

\newcommand\mydoclevel{0}

\begin{document}

\ifnum\mydoclevel=0
	\maketitle
	\begin{abstract}
		We study a combinatorial problem inspired by the following scenario:
	fully autonomous vehicles drive on a multi-lane ($m \geq 2$) road. Each
	vehicle heads to its own destination and is allowed to exit the road only 
	through a single designated off-ramp lane. However, an individual vehicle has a severely 
	limited memory and sensing capabilities, and, moreover, does not 
	communicate with its peers. In this work we present a distributed
	algorithm that, nonetheless, allows vehicles to get to the desired lane
	without collisions and in timely manner.

\end{abstract}

\fi

	\section{Introduction}
		In recent years the research in the field of autonomous vehicles has gained considerable momentum,
		and the idea of relieving the burden of driving from humans starts to lose its futuristic science
		fiction aura. Some people believe that autonomous traffic is \enquote{our last hope} of relief from the 
		frequent road-jams, we now witness in even mid-size urban areas. 
		We envision roads of the future with fully autonomous vehicles, that not only track
		the lane, keep safe distance and assist the driver, but essentially liberate humans from
		driving related activities altogether.
		
		Contemporary vehicles already posses a relatively high-degree autonomy.
		Today, low-level procedures for collision-free driving are routinely executed by the consumer vehicles.
		Furthermore, the advancement on the strategic level is also impressive.
		We can plan optimal routes from point A to point B. Moreover, we can accomplish such planning, accounting
		for numerous rules, e.g. considering current road conditions, intermediate point visiting 
		requirements and meeting stringent time constraints. These are high-level tasks that technology
		provided solutions for us even before the DARPA proposed the Urban Challenge back in $2007$ (\cite{buehler2009darpa}).
		
		However, before revolutionary changes will greatly improve our quality of transportation, 
		a different line of challenges should also be addressed. Currently, intermediate level responsibilities 
		are still handed back to the driver. The list of such tasks includes merging
		into the traffic on the entry ramps and exiting it via the off-ramps. Some initial efforts to address the associated 
		complexity were already made by the automotive community. Various approaches attempted include
		modeling the problem, developing communication protocols and numerous centralized controller-based solutions. 
		On the other hand, we still lack truly scalable distributed solutions addressing these challenges, solutions 
		to mitigate	single points of failure of centralized controllers.
		
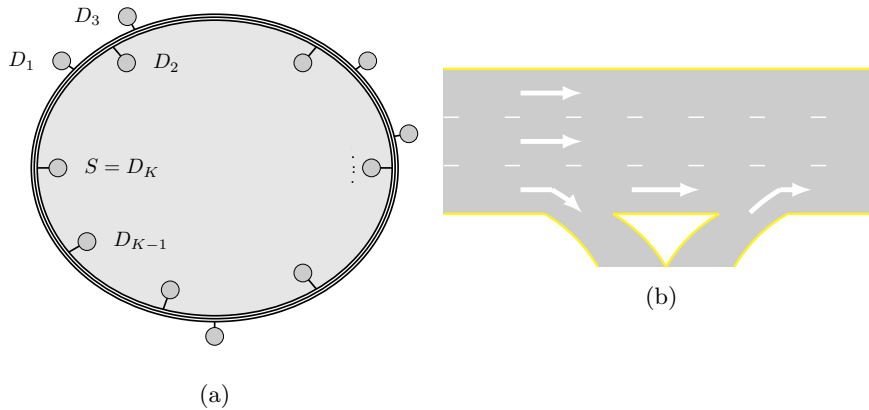
\begin{figure}
			\centering
			\begin{subfigure}{.48\textwidth}
				\resizebox{\textwidth}{!}{	\begin{tikzpicture}

		\pgfmathsetmacro{\bigradius}{3}
		\pgfmathsetmacro{\smallradius}{2.5}
		\pgfmathsetmacro{\inneroffset}{0.35}
		\pgfmathsetmacro{\cityradius}{0.15}
		
		\clip (-\bigradius - 2 * \inneroffset, -\bigradius - \inneroffset) rectangle (\bigradius + 2 * \inneroffset, \bigradius + \inneroffset);
		
		\draw[thick, name path = road, fill=black!10] (0, 0) ellipse (\bigradius cm and \smallradius cm);
		\draw[thick] 
			let 
				\n1={\bigradius+0.05},
				\n2={\smallradius+0.05}
			in
			(0, 0) ellipse (\n1 cm and \n2 cm);
		\draw[thick, name path = outroad] 
			let 
				\n1={\bigradius+0.1},
				\n2={\smallradius+0.1}
			in
			(0, 0) ellipse (\n1 cm and \n2 cm);
		\path[name path = cities] 
			let 
				\n1={\bigradius - \inneroffset},
				\n2={\smallradius - \inneroffset}
			in
			(0, 0) ellipse (\n1 cm and \n2 cm);
		\foreach \angle in {-50,0,50,130,180,210,250}
		{
			\path [name path = tocity\angle] (0, 0) -- (\angle:{2*\bigradius});
			\path [name intersections={of=cities and tocity\angle, by=city\angle}];
			\path [name intersections={of=road and tocity\angle, by=exit\angle}];
			\draw[fill = black!20] (city\angle) circle (\cityradius cm);
			\draw[thick] (exit\angle) -- ($(city\angle)+(\angle:\cityradius)$);
		}
			
		\path[name path = ocities] 
			let 
				\n1={\bigradius + \inneroffset},
				\n2={\smallradius + \inneroffset}
			in
			(0, 0) ellipse (\n1 cm and \n2 cm);
		\foreach \angle in {10,35,120,145,270}
		{
			\path [name path = tocity\angle] (0, 0) -- (\angle:{2*\bigradius});
			\path [name intersections={of=ocities and tocity\angle, by=city\angle}];
			\path [name intersections={of=outroad and tocity\angle, by=exit\angle}];
			\draw[fill = black!20] (city\angle) circle (\cityradius cm);
			\draw[thick] (exit\angle) -- ($(city\angle)-(\angle:\cityradius)$);
		}
		
		\pgfmathsetmacro{\labeldistance}{2}
		\node[label={[label distance=\labeldistance mm]180:$D_1$}] at (city145) {};
		\node[label={[label distance=\labeldistance mm]0:$D_2$}] at (city130) {};
		\node[label={[label distance=\labeldistance mm]180:$D_3$}] at (city120) {};
		\node[label={[label distance=\labeldistance mm]0:$S = D_K$}] at (city180) {};
		\node[label={[label distance=\labeldistance mm]0:$D_{K-1}$}] at (city210) {};
		
		\begin{scope}
			\pgfmathsetmacro{\cropangle}{8}
			\clip (0, 0) -- (\cropangle:{\bigradius + \smallradius}) arc (\cropangle:-\cropangle:{\bigradius + \smallradius})
				-- cycle;
			\draw[thick, loosely dotted] 
				let 
					\n1={\bigradius - \inneroffset - 2 * \cityradius},
					\n2={\smallradius - \inneroffset - 2 * \cityradius}
				in
				(0, 0) ellipse (\n1 cm and \n2 cm);
		\end{scope}
	\end{tikzpicture}}
				\caption{}
				\label{sub:circular.road}
			\end{subfigure}
			\begin{subfigure}{.48\textwidth}
				\resizebox{\textwidth}{!}{	\begin{tikzpicture}
		\pgfmathsetmacro{\m}{3}
		\pgfmathsetmacro{\length}{9}
		\fill[black!20] (0, 0) rectangle (\length, \m);
		\foreach \y in {\m}
		{
			\draw[line width = 0.5mm, yellow!90] (0,\y) -- (\length, \y);
		}
		\foreach \y in {2, 3, ..., \m}
		{
			\draw[line width = 0.3mm, dash pattern=on 9pt off 27pt, white] (0,\y - 1) -- (\length, \y - 1);
		}
		
		\pgfmathsetmacro{\exitradius}{3}
		\pgfmathsetmacro{\exitwidth}{1.41}
		\begin{scope}[xshift=3.5 cm]
			\path (0,0) coordinate (E1) arc (60:30:\exitradius) coordinate (O1);
			\path ($(E1) - (\exitwidth, 0)$) coordinate (E2) arc (60:30:\exitradius) coordinate (O2);
			\path (O1) arc (150:120:\exitradius) coordinate (E3);
			\path ($(O1)+(\exitwidth,0)$) coordinate (O4) arc (150:120:\exitradius) coordinate (E4);
		\end{scope}
		
		\fill[black!20] (E2) 
							-- (E1) 
							arc (60:30:\exitradius) 
							arc (150:120:\exitradius) 
							-- (E4) 
							arc (120:150:\exitradius) 
							-- (O2)
							arc (30:60:\exitradius);
							
		\draw[line width = 0.5mm, yellow!90] (E1) arc (60:30:\exitradius);
		\draw[line width = 0.5mm, yellow!90] (E2) arc (60:30:\exitradius);
		\draw[line width = 0.5mm, yellow!90] (O1) arc (150:120:\exitradius);
		\draw[line width = 0.5mm, yellow!90] (O4) arc (150:120:\exitradius);

		\draw[line width = 0.5mm, yellow!90] (0, 0) -- (E2);					
		\draw[line width = 0.5mm, yellow!90] (E1) -- (E3);					
		\draw[line width = 0.5mm, yellow!90] (E4) -- (\length, 0);

		\begin{scope}[shift={(E2)}]
			\draw[line width = 1 mm, white] (-0.5, 0.5) -- + (0.65,0) coordinate (A1);
			\draw[line width = 1 mm, white, -latex] (A1) arc (60:45:\exitradius) coordinate (A3);
			
			\foreach \lane in {2, ..., \m}
			{
				\draw[line width = 1 mm, white, -latex] 
					let \p1=(A3)
					in
				(-0.5, \lane - 0.5) -- (\x1, \lane - 0.5);
			}
			
			\draw[line width = 1 mm, white, -latex] (1.8, 0.5) -- (3.2, 0.5);

		\end{scope}
		\begin{scope}[shift={(E4)}]
			\draw[line width = 1 mm, white, -latex] (-0.15, 0.5) coordinate (A4) -- + (0.65,0);
			\draw[line width = 1 mm, white] (A4) arc (120:135:\exitradius);
		\end{scope}
	\end{tikzpicture}}
				\caption{}
				\label{sub:intersection}
			\end{subfigure}
			\caption{(\subref{sub:circular.road}) A circular road around few urban destinations 
			($D_1$, $D_2$, $\ldots$, $D_K$). (\subref{sub:intersection}) An exit/entry section of the road. }
		\end{figure}

		In this work we envision autonomous vehicle traffic on a multi-lane freeway. A road is assumed to 
		connect a number of locations $D_1$, $D_2$, $\ldots$, $D_K$ along the road. 
		We can consider a specific location, say $D_K = S$, as the source and model, w.l.o.g., a circular road
		to the other locations as destinations (\cref{sub:circular.road}). 
		A similar model of integrated system of corridors and the urban network was previously mentioned in
		\cite{amirgholy2020traffic}. However, our interest will mainly be focused on a single sliding section 
		of the road after an on-ramp entrance and before the next off-ramp road exit.
		
		We consider the following scenario: autonomous cars enter the freeway through an on-ramp entry at 
		some location. The specific car entrance point is of no	importance. Each car heads on its own, forward along the road, 
		and it is expected to depart through a desired off-ramp exit. We shall assume further that a vehicle is required
		to move on the right-most lane in order to exit at the destination off-ramp (see \cref{sub:intersection} for a 
		depiction of an departure-merger section).
		Hence, we assume Right-hand Traffic (RHT), however, clearly, the very same principles could be applied
		to the Left-hand Traffic roads with appropriate changes (if in England, for example!)
		
		Throughout the work, we assume vehicles possess full autonomy. They execute all the necessary low level
		maneuvers, which in our model boil down to \enquote{follow the lane at constant speed}. Additionally, we require
		a vehicle to be able to execute some high-level complex commands. The list of supported commands is as follows: 
		do nothing(\enquote{continue to move in the same lane at constant speed $v$}), change lane to the left or 
		right, move forward/backward, where forward and backward is
		relative to vehicles that continue to move at constant speed $v$ on neighboring lanes.
		
		In this paper we develop a series of completely distributed, polynomial time, communication-less local
		algorithms to re-arrange vehicles on the road. The work culminates in a geometric sorting algorithm, that allows
		autonomous vehicles to exit the freeway using myopic local interactions only, without explicitly
		communicating with others.
		
	\section{Related work}
		\subsection{Traffic management}
		Highway congestion poses a serious burden on the road infrastructure and its users 
		all over the globe. Until recent technological advances we were helpless while car fleets were 
		surging at unprecedented rates.		
		
		In a prototype sketch \cite{varaiya1991sketch} and later in a seminal paper \cite{varaiya1993smart}, Varaiya proposed 
		a concept of an Intelligent Vehicle Highway System (IHVS). The proposed $4$-layer IHVS makes a heavy use of
		platooning, where groups of vehicles move in tightly-spaced coordinated fashion towards 
		possibly different destinations, while sharing the same road. The platooning concept came to solve 
		a problem of road infrastructure waste incurred by road safety gap requirements.

		However, an actual protocol for this centralized multi-layer controller was estimated to be a huge $500$ 
		thousand states big state machine (\cite{hsu1993protocol}). Furthermore, Kurzhanskiy et al. 
		have shown, in \cite{kurzhanskiy2015traffic}, that \textbf{bad management} of the existing highways 
		\textit{is} the main cause of congestion,
		while excessive demand is only the next culprit. Thus, designing a simple and, at the same time, efficient
		local traffic management system becomes a crucial challenge.
		
		\subsection{Vehicular control}
		Automated longitudinal control of vehicles gained track since at least $60$s. 
		Currently, with the design 
		of (semi-) autonomous vehicles, the topic of lane-changing algorithms and protocols raised to prominence.
		Lane-changing could be divided into three main scenarios.
		First, \textit{merging} - the entrance of the vehicle to the freeway from an on-ramp. Second, -
		\textit{diverging} - the exit of the vehicle from the freeway and the subsequent integration into the 
		urban mesh traffic. Integration into, and departure from a freeway traffic are even more complicated at
		\enquote{weaving} sections, where both scenarios are executed simultaneously by a number of vehicles,
		and lane-changing has to be carried out in-between. 
		Currently, the lane-changing process, while mostly a discretionary activity
		causes a significant portion of road accidents and is a leading cause of the freeway congestion. 
		In fully autonomous settings the requirement to change lanes in a safe way will clearly remain. 
		Although, in the future, we should expect a higher traffic efficiency and stability due to cooperation, 
		better sensing, faster reaction, and mostly due to an automated deterministic decision-making.
		
		A centralized optimal freeway merging strategy of fully automated vehicles is presented in 
		\cite{letter2017efficient}. A dedicated roadside short-range communication controller
		assigns merging priority through adjusting acceleration
		profiles (optimizing vehicle trajectories). The proposed algorithms are shown to maximize average travel speed under
		safety gap requirements. A gap metering approach is suggested by Jin et al., in \cite{jin2017gap}, and later
		improved by Jiang et al., in \cite{jiang2020dynamic}. The gap metering system, installed near the merging
		section on the road, advises drivers to provide a gap in favor of incoming merging traffic.
		A significant on-ramp delay decrease was shown in simulations even under low 
		driver compliance rate.
	
		In the diverging scenario a significant effort was invested to develop lane-changing strategies that minimize
		congestion and increase safety. Huang et al. conducted a series of simulations \cite{huang2019determining} to
		investigate the cause of heavy congestions in two Beijing off-ramp areas. It was found that 
		lane-spacings and the lane-changing spacings are the main factors that define traffic throughput,
		safety and efficacy. Moreover, as average traffic speeds rise both should increase accordingly in order
		to allow the same off-ramp throughput flow. Zheng et al., \cite{zheng2019cooperative}, proposed a cooperative
		strategy between the exiting vehicles and the through traffic vehicles near the off-ramp. The strategy calls for 
		cooperative	vehicles deceleration to create safe lane gaps that allow smoother lane changing for 
		off-ramp vehicles.
		
		A research spotlight in the third lane-changing scenario focuses on the safety of the maneuver
		and the opportunities it creates to minimize travel time. Ji et al., in \cite{ji2020review}, found
		that individuals tend to maximize their \enquote{payoffs} in terms of safety and travel times,
		given others are doing the same. Hence, lane-changing is seen as an opportunity to increase
		\enquote{gains}. While different drivers weigh factors differently and adopt diverse behavior
		strategies, the whole traffic system eventually reaches an equilibrium. Those findings
		were reinforced in \cite{wang2019cooperative}, as it was found that mixing self and overall
		reward boosts an overall traffic efficiency. 
		
		Mainstream works in the field suggest lane changing communication protocols as a
		solution to cooperative lane changing. A vehicle
		broadcasts to the surroundings, or some limited neighborhood, its position, speed and 
		acceleration, and sends its intentions to change the lane to the desired one.
		Lombard et al. suggest, in \cite{lombard2017cooperative}, one such \enquote{automatic courtesy} protocol.
		A lane changing vehicle initiates the protocol and becomes a virtual local \enquote{intersection}
		server that broadcasts intersection coordinates to neighbors. Surrounding vehicles could opt to become
		followers and allow safety gap creations by voluntary deceleration, thus providing the space for
		the initiator's desired maneuver. A promising decentralized cooperative lane changing
		decision model was proposed by Nie et al., in \cite{nie2016decentralized}. Decision making in the model is based
		on three module framework installed on each Connected Autonomous Vehicle (CAV).
		Here the authors show that their approach improves traffic stability, homogeneity and efficiency.
		
		\subsection{Tile-sliding puzzles}
		The research into tile-sliding puzzles traces back to the nineteenth century.
		The $15$-puzzle is a classical tile-sliding puzzle game where
		a player slides arbitrarily placed tiles with numbers from $1$ to $15$ in a $4 \times 4$ matrix. It
		was shown using permutation parity arguments (\cite{johnson1879notes}), that only half of the initial 
		configurations are solvable in the sense that tiles could be arranged in a specific predefined
		order. However, finding a solution that requires a minimum number of tile slides in a
		generalized $n$-puzzle version was shown to be an NP-hard problem (\cite{ratner1986finding}). 
		Despite the hardness, a number of efficient sub-optimal solutions exist, and, in fact, 
		are pretty-straight-forward, see e.g., \cite{ratner1990n2}, \cite{wang2017dsolving}.
		
		An interesting application of tile-sliding puzzles was proposed in a two-lane congested road settings
		by Toy et al., in \cite{toy2002emergency}. An emergency vehicle (EV) moves through the traffic on the IHVS 
		(see reference \cite{varaiya1993smart} above). Two scenarios are addressed:
		\begin{enumerate*}[label=\alph*)]
			\item a rapid transit of EV through traffic,
			and
			\item an EV transition through stopped traffic.
		\end{enumerate*}
		A vertex maneuver, i.e. traffic that circulates \textit{around} the EV, is discussed and is shown
		in simulations to provide a solution for the first scenario. For the second scenario two solutions
		are suggested: the part-and-go maneuver, that allows platoon creation in both lanes, thereby ensuring
		enough space for the EV; and the zig-zag maneuver in case the part-and-go maneuver is
		inapplicable due to lack of initial free space. Moreover, the latter maneuver could be
		performed even in the absence of the free space. Some necessary stationary, platoon-forward-join
		maneuvers can be initiated under all circumstances.
		
		A beautiful problem was suggested by Petig et al., in \cite{petig2018changing}. Suppose, we observe a two-lane road
		in the autonomous cars scenario. Under mild restrictions, cars are allowed to change lanes before the 
		upcoming intersection, according to their intentions to leave the road. Define a \textit{makespan},
		determined by
		the total number of vehicle slots occupied in the final configuration and a \textit{total cost} - the 
		number of steering maneuvers executed to reach that configuration. The paper provides a
		polynomial-time algorithm that finds a sequence of maneuvers that \textbf{minimizes the makespan} and
		at the same time is a \textbf{1.5-approximation of the minimal total cost}. Additionally, it is
		shown that a natural multi-lane extension of the problem is NP-hard. However, the algorithm based on
		the view point of an omniscient being, i.e. it is a centralized solution.
		
		A surprisingly simple solution to a tile-sliding sorting problem was recently proposed in 
		\cite{DBLP:journals/corr/abs-2111-06284}
		. A set of stones is laid on the lower row of a two-row grid. The
		goal is to split stones according to their color (black or white) by moving them from one grid cell to
		another. Black stones should be moved to the left side, while white ones to the right. Stones are moved
		at discrete time ticks, where every moving stone can change its current cell to one of its unoccupied
		neighbors, in a way that no two stones share the same cell at any point in time. The paper provides
		a lower time bound for a solution and presents an amazingly simple instance-optimal approach for the
		stated problem, i.e. no other algorithm can sort stones faster for any initial stone configuration
		on the grid. Furthermore, authors propose a distributed variant of the algorithm, that could require
		a single additional time tick to complete, compared to the centralized version.
		
	\section{The exit the road (ER) problem}\label{section:ER Problem}
		\subsection{The freeway road model}
		We here assume that in the future no human-controlled vehicles will be present on the road. In other words, 
		all vehicles on the road will be fully autonomous and running the same standardized distributed traffic 
		control algorithm. At present, due to the driver reaction time safety concerns, 
		road infrastructure is deeply under-utilized, or it is completely jammed, hence has an extremely low throughput rate. 
		However, autonomous vehicle \textit{reaction} time could be made orders of magnitude lower than that of humans. 
		Therefore, safety distances could be substantially decreased. Moreover, it can be shown, that the safety distance 
		$d$, of a vehicle moving at the speed $v$, behind another vehicle, is given by
		\begin{equation*}
			d = v \cdot \Delta_t,
		\end{equation*}
		where $\Delta_t$ - is the reaction time of the driver (or the system controlling the vehicle).
		
		Human drivers react somewhat slowly to the abrupt changes on the road. For example, an emergency breaking
		of the vehicle in front initiates a reactive breaking delayed by $1-2$ seconds. Therefore, a driver is taught
		to hold a safety distance of about $2$ seconds, as a rule of thumb. (According to the California's Performance 
		Measurement System (PeMS) data \cite{chen2001causes} the average vehicle headway is $1.63$ seconds).		
		At the speed of $35$ m$\cdot$sec$^{-1}$, the safety distance becomes $35-70$ meters. Assuming, that 
		an average vehicle length is around $5$ meters and drivers adhere to the \enquote{$2$ seconds} rule of thumb, 
		we conclude that road 
		occupancy is bounded from above	by a mere $12.5\%$ (see \cite{shladover2009cooperative} for similar 
		back-of-the-envelope calculations).
		
		In the autonomous settings the occupancy could be boosted significantly. A reaction time of $100$ms translates to
		a $3.5$m safety distance and to an occupancy upper bound of nearly $60\%$. Define a \textit{safety buffer} to be
		this $3.5$m long empty space of	one lane. Then, a vehicle itself, a safety buffer 
		in front of it, and an empty space around the vehicle in the same lane may be defined as a 
		\textit{vehicle slot} or as a \textit{slot} for brevity. We shall allow slots to be \textit{empty} 
		or \enquote{vehicle-less}, and require all slots to be of the same physical dimensions.
		
		Thus, we model, superimposed on the road, a possibly infinite, collection of vehicle slots. Slots are 
		considered to move at some 
		predefined constant speed $v$ along the road. We look to maximize the utilization of the road,
		therefore we may set $v$ to equal the speed limit of the road.
		
		A vehicle occupies a slot, and moves within the slot at speed $v$.
		A slot will always move along a lane, however a vehicle is allowed to change lanes. Therefore,
		from time to time a vehicle is allowed to need two slots at the same time, while executing
		a lane-change maneuver, i.e., moving to another neighboring slot in the same lane or to a slot
		in a neighboring lane.
		
		Complex maneuvers resulting in a change of a currently occupied slot to a slot behind and left
		will be executed in two steps, hence only one move will be allowed in any given time interval. 
		First, a vehicle could proceed to the left, temporarily moving from the
		current slot and to its direct left neighbor. Then, by slowing down (and then accelerating to speed $v$)
		a vehicle will move one slot back (see \cite{hedrick1994control} for an optimal trapezoidal acceleration
		profile). Note, that we shall allow a vehicle to temporally exceed the speed limit $v$ to accomplish side 
		and forward-move maneuvers.		
		
		\begin{definition}
			A \textit{frame} is a sliding rectangular grid of slots of size $n \times m$ on an $m$ lane wide road.
		\end{definition}
		
\begin{figure}
			\centering
			\begin{subfigure}{.48\textwidth}
				\resizebox{\textwidth}{!}{\subimport{./images}{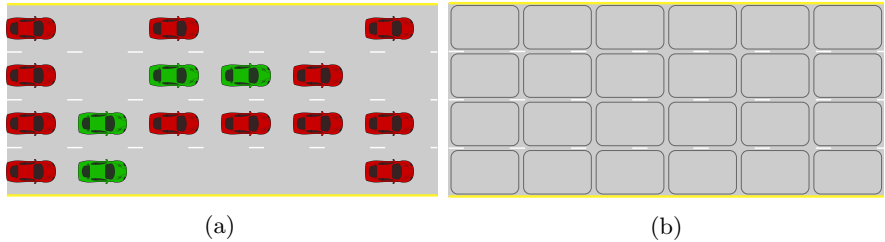}}
				\caption{}
				\label{sub:roadframe.cars}
			\end{subfigure}
			\begin{subfigure}{.48\textwidth}
				\resizebox{\textwidth}{!}{	\begin{tikzpicture}
		\pgfmathsetmacro{\m}{4}
		\pgfmathsetmacro{\length}{9}
		\fill[black!20] (0, 0) rectangle (\length, \m);
		\foreach \y in {0, \m}
		{
			\draw[line width = 0.5mm, yellow!90] (0,\y) -- (\length, \y);
		}
		\foreach \y in {2, 3, ..., \m}
		{
			\draw[line width = 0.3mm, dash pattern=on 9pt off 27pt, white] (0,\y - 1) -- (\length, \y - 1);
		}
		\pgfmathsetmacro{\slotlength}{1.5}
		\pgfmathsetmacro{\smalloffset}{0.05}
		\foreach \x in {0, ..., 5}
			\foreach \y in {0, ..., 3}
			{				
				\draw[black!60, rounded corners] (\x * \slotlength + \smalloffset, \y + \smalloffset) rectangle ++(\slotlength - 2 * \smalloffset , 1 - 2 * \smalloffset);
			}
	\end{tikzpicture}}
				\caption{}
				\label{sub:roadframe.frame}
			\end{subfigure}
			\caption{(\subref{sub:roadframe.cars}) Autonomous vehicles on the road. Safety buffers are small. 
			Green (light colored) vehicles 
			(the block of $4$ vehicles in the second left row and the second left lane) intend to exit at the next destination.
			(\subref{sub:roadframe.frame}) Underlying slots are depicted. Slots could be either empty or occupied,
			by at most a single vehicle. }
		\end{figure}

		A frame, proposed also by Chouchan et al., in \cite{chouhan2020cooperative}, is depicted in 
		\cref{sub:roadframe.frame}. In our assumed road model frames move at 
		constant speed $v$, hence slots are static with respect to the moving frame. Vehicles move from one slot to
		another, however, we assume that they never cross frame boundaries. Thus, vehicle-frame association
		remains constant over time.
		We index the slots in the frame by	row-column index pairs, where rows are counted from the 
		front side of the frame to the
		back, and columns from the left side of the road to the right side. For example, in	\cref{sub:roadframe.cars}, 
		light colored (green in the pdf version) cars are located at positions: $(2, 6), (3, 6), (4, 3)$ in
		the frame.
		
		\subsection{Vehicles as Agents on the Grid}
		We model a vehicle as an agent moving on a grid of slots (cells). As discussed above, we assume that
		agents posses a number of capabilities that allow them to solve various challenges, by moving from
		one grid cell to another. An important agent ability is sensing in the neighborhood. At discrete time ticks,
		agents sense their surroundings, then decide on actions, and, finally, execute them
		(the so called LOOK-COMPUTE-MOVE paradigm).
		
\begin{figure}
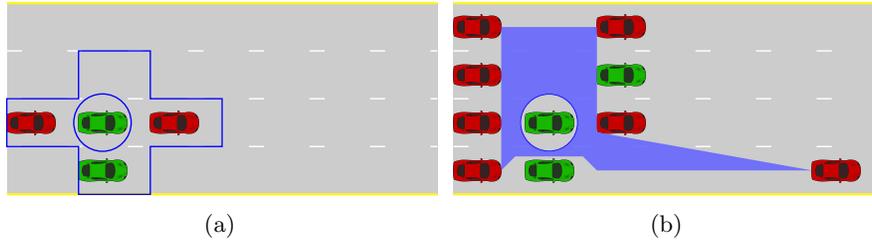

			\centering
			\begin{subfigure}{.48\textwidth}
				\resizebox{\textwidth}{!}{\subimport{./images}{StraightRoad.Neighborhood.L1}}
				\caption{}
				\label{sub:roadframe.L1}
			\end{subfigure}
			\begin{subfigure}{.48\textwidth}
				\resizebox{\textwidth}{!}{\subimport{./images}{StraightRoad.Neighborhood.Linf}}
				\caption{}
				\label{sub:roadframe.Linf}
			\end{subfigure}
			\caption{Different visibility ranges' visualization. Depiction of a visibility range
			around a circled vehicle,
			(\subref{sub:roadframe.L1}) an $L_1$ visibility range of $1$ cell, 				
			(\subref{sub:roadframe.Linf}) an infinite visibility range.}
		\end{figure}

		We define \textit{sensing} as a process of mapping the visible surroundings to a neighborhood representation. 
		The visibility range can be defined in several possible ways: from no sensing at all to everything 
		that is not obscured can be sensed (see \cref{sub:roadframe.Linf}).
		We use an $L_1$-metric with
		a given visibility range and accounting for occlusions, hence obstacles (i.e., other agents) obscure everything
		beyond them. In discrete setting, where only the cell state matters, it becomes
		necessary to precisely define what is visible. 
		We here exclusively work under a visibility range of $1$ step (see \cref{sub:roadframe.L1}. 
		Under this assumption, an agent observes only its immediate $4$ neighborhood, and no 
		grid cells inside the visibility range are obscured by \enquote{closer} objects.
		
		Note, that, it is impossible to execute any cooperative strategy under a $0$-visibility
		regime, since even immediate neighbors are beyond the horizon. We of course want to model
		physical reality, and collisions must be avoided. This imposes a minimum visibility range
		requirement of $1$ cell, which is also employed by algorithms in this work.
		
		In our model, the actions of an agent depend not only on the neighborhood, but also on the 
		\enquote{internal agent state}. The facility that enables holding the agent state is the agent memory, 
		and measure it in bits. Memory allows different agent actions given the same visible
		surroundings. An agent at state $X$ in a given neighborhood could remain motionless, while
		at state $Y$ it could make a simple move from its current row to the row with one lower index, for example!
		
		In physical reality all the vehicles move in the same direction on the road.
		Hence, we assume, that their motion/orientation naturally provides them a common North, 
		i.e., a preferred direction. Due to the limited visibility
		agents are unable to assess the dimensions of the grid frame, or their current location within the frame, 
		i.e., their row and column
		indexes. However, the preferred direction allows to state facts along the following lines: 
		\enquote{the cell to the north of the current cell I am in, i.e., the cell in front of me has
		a lower grid row index, but the same grid column index}.
		
		In our multi-agent paradigm, agents do not communicate with each other. In order, to prevent possible collisions,
		we assume time	synchronization, i.e. that all agents sense and reset their neighborhood at the same moment,
		once in an arbitrary defined unit time interval. 
		Then all agents act according to their current location, neighborhood occupancy pattern and state.

		Below we summarize agent capabilities and knowledge about the world:
		\begin{itemize}
			\item Sensing ($L_1$-visibility $r$).
			\item Memory ($b$ bits, hence $2^b$ states).
			\item Common North (well defined forward, backward, right and left).
			\item Synchronized operation (all agents operate in synchronized unit intervals).
		\end{itemize}
		Further we assume that agents
		\begin{itemize}
			\item \textbf{Do not} know their location on the grid.
			\item \textbf{Do not} know the frame dimensions.
			\item \textbf{Do not} have explicit communication capabilities (i.e., can not communicate 
			to surrounding agents or with any central controller).
		\end{itemize}
		
		Moreover, we assume all agents to be identical and anonymous, i.e. all the agents execute the same
		algorithm, and neighbors can not be distinguished one from the other.
		
		\subsection{The agent position on a grid}
			As mentioned previously, at the beginning of a unit time interval, each agent senses its surroundings. 
			The sensors are capable of detecting other agents, 
			the edges of the grid (i.e., the frame boundaries) and empty grid cells within the agent's visibility range. 
			Note, that detecting boundaries in reality is trivial near road shoulders, but is not at all obvious
			on the front and back boundaries of the frame. We assume, that these boundaries detection is based
			on some infrastructure provided support, for example a laser pulses from one side of the road to the other
			at predefined locations. A vehicle then register a laser pulse or the absence thereof and decides if it 
			moves near the frame front or the frame back, or not near the frame boundary at all.
			
			We have $9$ different types of positions on the grid, according to sensor readings of a simple $4$-neighborhood 
			of an agent (see \cref{sub:grid.positions}). An agent can be inside a frame, in one of the $4$ corners of it or in one 
			of the $4$ boundary rows. For example, on \cref{sub:L1.neighborhood} the agent senses position $6$, 
			since only the West (left) sensor reading is a wall/non-grid cell).
			Occasionally, we shall refer the reader to \cref{sub:grid.positions}, otherwise we
			shall mention the position type in text.
			
		\subsection{A local algorithm}
		We denote exiting agents, i.e., agents modeling vehicles with destination set at the next 
		exit, by $1$; continue agents, i.e., agents corresponding to vehicles, that continue down 
		the road, by $-1$; and, finally, empty cells will be denoted by $0$. Hence, we can model
		a frame as a rectangular grid filled with three possible numbers 
		from the set $\{+1, -1, 0\}$, where road lanes become columns of the grid.
		
		\begin{definition}
			A grid \textit{configuration} $C$, is a particular state of an $n \times m$ grid,
			over the number set $\{1, -1, 0\}$.
		\end{definition}
		\begin{definition}
			An $n \times m$ grid configuration $\mathcal{C}$ is a
			\textit{target grid configuration} if all the $1$s are located in the right-most, $m$th,
			column of a grid.
		\end{definition}
		\begin{definition}
			A \textit{legal move} of an agent $A$ is an exchange of positions of an agent
			$A$	(represented by $1$ or $-1$) into one of the $4$ neighboring empty spaces ($0$). 
			A legal move results in an empty space \enquote{taking} the place of an agent $A$, 
			and an agent $A$ moving into that empty space.
		\end{definition}
		
		In particular, a legal move leaves an agent on the grid, i.e. agents do not fall
		off the grid. But, not every set of legal moves, executed at the same time tick, constitutes a 
		legal configuration change of the grid. Two standalone legal moves executed at the same time 
		could lead to a vehicle collision.
		Therefore,
		\begin{definition}
			A \textit{legal move sequence} is a set of legal vehicle moves, such that no two legal moves
			in the set, that overlap in time, overlap also in the spatial dimension.
		\end{definition}
		\begin{definition}
			A legal move sequence $S$ is said to \textit{transform} an $n \times m$ grid from configuration $C$ 
			to configuration $T$, if an execution of $S$ moves changes grid configuration from $C$ to $T$.
		\end{definition}
		
		We make the following assumptions to ensure that there exists a legal move sequence, which
		transforms a grid from an initial grid configuration to a target grid configuration.
		Naturally, legal moves assume that there is at least one empty grid cell (cell with value $0$). 
		Additionally, the number of exiting agents ($1$s) should not exceed the number of 
		available grid cells in the right-most grid column. We summarize:
		\begin{equation*}
			N_0 > 0
		\end{equation*}
		\begin{equation*}			
			N_1 \leq n\tag{$C_0$}
			\label{eq:C_0}
		\end{equation*}
		\begin{equation*}
			N_0 + N_1 + N_{-1} = n \cdot m,
		\end{equation*}
		where $N_0$ - is the number of $0$s (empty frame slots), 
		$N_1$ - is the number of $1$s (exiting vehicles) and 
		$N_{-1}$ - is the number of $-1$s (continue vehicles). 
		
		Consider an $n \times m$ grid, an initial configuration $C$ and a non-empty set of target configurations 
		$\mathcal{T} = \{T_1, T_2, \ldots, T_k\}$. 
		We shall ask if there exists a legal move sequence that transforms a grid from an initial 
		configuration $C$ to some target configuration $T_1, T_2, \ldots, T_k$?
		Further, we shall refer to this question as the \textbf{Exit the Road (ER) Problem}.

		In the further discussion we shall make use of the following list of notations:
		\begin{itemize}
			\item
				$a_{i, j}$ - a grid cell at position $(i, j)$.
			\item
				$a_{i,j}(t)$ - an agent occupying grid cell $a_{i,j}$ at time $t$.
			\item
				$\mathcal{N}(a_{i, j}(t))$ - a visible \textit{neighborhood} of an agent $a_{i,j}(t)$ at time $t$.
			\item
				$M_{i,j}(t)$ - an agent $a_{i, j}(t)$ $b$-bit memory state.
		\end{itemize}
		And more generally, for agent $A$, $\mathcal{N}(A)$ and $M(A)$ will denote the visible neighborhood
		and the memory state of agent $A$, where agent implicit location $a_{i, j}$, and time $t$ are obvious from the 
		context.
		
		\begin{definition}
			Let $\mathcal{C}$ be an $n \times m$ grid configuration at time $t$.
			A \textit{local rule of exchanges} is a function that defines the location and the 
			memory state of every agent $A \in \mathcal{C}$ at the next time tick $t+1$ given $\mathcal{N}(A)$ and $M(A)$.
		\end{definition}
		
		In this paper we develop a particular local rule of exchanges $\mathcal{A}$. We show that an iterative execution
		of $\mathcal{A}$ by the agents on the grid $G$ at times $1, 2, 3, \ldots$ in a distributed manner eventually 
		solves the Exit the Road Problem, i.e., transforms a grid to a target configuration.
	
	\section{Sorting a multi-column grid}\label{sec:Multi-lane Algorithm A}
		We here propose a novel distributed algorithm that, if applied, transforms an $n \times m$ grid 
		from any possible initial configuration $IC$ to a target configuration.
		
		We consider a specific restricted implementation of an agent. An agent is missing any communication
		capabilities, and additionally does not posses the ability to distinguish between $1$ and $-1$ neighbor agents. 
		Therefore, from an agent point of view, grid neighbors in the $\mathcal{N}(a_{i,j}(t))$ appear in 
		three different types. Those are either other 
		\textit{agent} (indistinguishable) cells, \textit{empty} cells or \textit{border} cells.
		
		We assume that agents act under the LOOK - COMPUTE - MOVE paradigm at discrete times $1, 2, 3, \ldots$.
		No agent occupies more than one cell on the grid at a time, i.e. every move from a cell 
		to a neighbor cell could be accomplished in a single time tick.
		
		We adopt the following convention: North-to-South is the direction of increasing row indices, 
		while West-to-East is the direction of increasing column indices. Therefore, the exit lane 
		corresponds to the eastern-most column of the grid, and the front frame row to the northern row 
		of the grid.
		The exact local rules are listed in \cref{sec:Sorting.Algorithm.Exit} and in
		\cref{sec:Sorting.Algorithm.Continue}; below we provide the main properties and characteristics of 
		Algorithm $\mathcal{A}$.
		
\begin{figure}
			\centering
			\begin{subfigure}{.32\textwidth}
				\phantom{\resizebox{\textwidth}{!}{	\begin{tikzpicture}
	
		\pgfmathtruncatemacro{\xsize}{6}
		\pgfmathtruncatemacro{\ysize}{6}

		\pgfmathsetmacro{\arrowoffset}{0.07}
		\draw[dashed, very thin] (0,0) grid (\xsize, \ysize);
		\draw[thick] (0, 0) --
			(0, \ysize) --
			(\xsize, \ysize) --
			(\xsize, 0) --
			cycle;

		\pgfmathtruncatemacro{\xinnersize}{\xsize - 2}
		\pgfmathtruncatemacro{\yinnersize}{\ysize - 2}
		
		\begin{scope}[shift={(0.5, 0.5)}]
			\foreach \col in {0, ..., \xinnersize}
			{
				\draw[-latex] (\col + \arrowoffset, \yinnersize - \ifnum\col=0 -\arrowoffset+1 \else \arrowoffset \fi) -- (\col + \arrowoffset, 0); 
				\draw[-latex] (\col + 1 - \arrowoffset, 0) -- (\col + 1 - \arrowoffset, \yinnersize - \ifnum\col=\xinnersize -1+\arrowoffset \else \arrowoffset \fi); 

				\ifnum\col<\xinnersize
					\draw[-latex] (\col + \arrowoffset, 0) -- (\col + 1 - \arrowoffset, 0); 
				\fi
				\draw[-latex] (\col + 1 - 0.5 * \arrowoffset, \yinnersize + 1 + 2 * \arrowoffset) -- (\col + 0.5 * \arrowoffset, \yinnersize + 1 + 2 * \arrowoffset); 
			}
		\end{scope}
	\end{tikzpicture}}}
			\end{subfigure}
			\begin{subfigure}{.32\textwidth}
				\resizebox{\textwidth}{!}{	\begin{tikzpicture}
	
		\pgfmathtruncatemacro{\xsize}{6}
		\pgfmathtruncatemacro{\ysize}{6}

		\pgfmathsetmacro{\arrowoffset}{0.07}
		\draw[dashed, very thin] (0,0) grid (\xsize, \ysize);
		\draw[thick] (0, 0) --
			(0, \ysize) --
			(\xsize, \ysize) --
			(\xsize, 0) --
			cycle;
			
		\pgfmathtruncatemacro{\xinnersize}{\xsize - 2}
		\pgfmathtruncatemacro{\yinnersize}{\ysize - 2}
		
		\begin{scope}[shift={(0.5, 0.5)}]
			\foreach \col in {1, ..., \xinnersize}
			{
				\draw[-latex] (\col, \yinnersize + 1 - \arrowoffset) -- (\col, \yinnersize + \arrowoffset); 

				\foreach \row in {1, ..., \yinnersize}
				{
					\draw[-latex] (\col - 1 + \arrowoffset, \row) -- (\col - \arrowoffset, \row); 
				}
			}
		\end{scope}
	\end{tikzpicture}}
				\caption{}
				\label{sub:emptyspace.exiting}
			\end{subfigure}
			\begin{subfigure}{.32\textwidth}
				\resizebox{\textwidth}{!}{	\begin{tikzpicture}
	
		\pgfmathtruncatemacro{\xsize}{6}
		\pgfmathtruncatemacro{\ysize}{6}

		\pgfmathsetmacro{\arrowoffset}{0.07}
		\draw[dashed, very thin] (0,0) grid (\xsize, \ysize);
		\draw[thick] (0, 0) --
			(0, \ysize) --
			(\xsize, \ysize) --
			(\xsize, 0) --
			cycle;
			
		\pgfmathtruncatemacro{\xinnersize}{\xsize - 2}
		\pgfmathtruncatemacro{\yinnersize}{\ysize - 2}
		
		\begin{scope}[shift={(0.5, 0.5)}]
			\foreach \col in {1, ..., \xinnersize}
			{
				\draw[-latex] (\col + \arrowoffset, \yinnersize + 1 - \arrowoffset) -- (\col + \arrowoffset, \yinnersize + \arrowoffset); 
				\draw[-latex] (\col - \arrowoffset, \yinnersize + \arrowoffset) -- (\col - \arrowoffset, \yinnersize + 1 - \arrowoffset); 

			}
			\foreach \row in {0, ..., \yinnersize}
			{
				\draw[-latex] (\xinnersize + \arrowoffset, \row) -- (\xinnersize + 1 - \arrowoffset, \row); 
			}
		\end{scope}
	\end{tikzpicture}}
				\caption{}
				\label{sub:emptyspace.continue}
			\end{subfigure}
			\begin{subfigure}{.32\textwidth}
				\resizebox{\textwidth}{!}{	\begin{tikzpicture}
	
		\pgfmathtruncatemacro{\xsize}{6}
		\pgfmathtruncatemacro{\ysize}{6}

		\pgfmathsetmacro{\arrowoffset}{0.07}
		\draw[dashed, very thin] (0,0) grid (\xsize, \ysize);
		\draw[thick] (0, 0) --
			(0, \ysize) --
			(\xsize, \ysize) --
			(\xsize, 0) --
			cycle;

		\pgfmathtruncatemacro{\xinnersize}{\xsize - 2}
		\pgfmathtruncatemacro{\yinnersize}{\ysize - 2}
		
		\begin{scope}[shift={(0.5, 0.5)}]
			\foreach \col in {0, ..., \xinnersize}
			{
				\draw[-latex] (\col + \arrowoffset, \yinnersize - \ifnum\col=0 -\arrowoffset+1 \else \arrowoffset \fi) -- (\col + \arrowoffset, 0); 
				\draw[-latex] (\col + 1 - \arrowoffset, 0) -- (\col + 1 - \arrowoffset, \yinnersize - \ifnum\col=\xinnersize -1+\arrowoffset \else \arrowoffset \fi); 

				\ifnum\col<\xinnersize
					\draw[-latex] (\col + \arrowoffset, 0) -- (\col + 1 - \arrowoffset, 0); 
				\fi
				\draw[-latex] (\col + 1 - 0.5 * \arrowoffset, \yinnersize + 1 + 2 * \arrowoffset) -- (\col + 0.5 * \arrowoffset, \yinnersize + 1 + 2 * \arrowoffset); 
			}
		\end{scope}
	\end{tikzpicture}}
				\caption{}
				\label{sub:emptyspace.common}
			\end{subfigure}
			\begin{subfigure}{.32\textwidth}
				\resizebox{\textwidth}{!}{	\begin{tikzpicture}
	
		\pgfmathtruncatemacro{\xsize}{6}
		\pgfmathtruncatemacro{\ysize}{6}

		\pgfmathsetmacro{\arrowoffset}{0.07}
		\draw[dashed, very thin] (0,0) grid (\xsize, \ysize);
		\draw[thick] (0, 0) --
			(0, \ysize) --
			(\xsize, \ysize) --
			(\xsize, 0) --
			cycle;
			
		\pgfmathtruncatemacro{\xinnersize}{\xsize - 2}
		\pgfmathtruncatemacro{\yinnersize}{\ysize - 2}
		
		\begin{scope}[shift={(0.5, 0.5)}]
			\foreach \col in {0, ..., \xinnersize}
			{
				\draw[-latex] (\col + \arrowoffset, \yinnersize - \ifnum\col=0 -\arrowoffset+1 \else \arrowoffset \fi) -- (\col + \arrowoffset, 0); 
				\draw[-latex] (\col + 1 - \arrowoffset, 0) -- (\col + 1 - \arrowoffset, \yinnersize - \ifnum\col=\xinnersize -1+\arrowoffset \else \arrowoffset \fi); 

				\ifnum\col<\xinnersize
					\draw[-latex] (\col + \arrowoffset, 0) -- (\col + 1 - \arrowoffset, 0); 
				\fi
				\draw[-latex] (\col + 1 - 0.5 * \arrowoffset, \yinnersize + 1 + 2 * \arrowoffset) -- (\col + 0.5 * \arrowoffset, \yinnersize + 1 + 2 * \arrowoffset); 
			}
		\end{scope}
		\begin{scope}[shift={(0.5, 0.5)}]
			\foreach \col in {1, ..., \xinnersize}
			{
				\draw[-latex] (\col, \yinnersize + 1 - \arrowoffset) -- (\col, \yinnersize + \arrowoffset); 

				\foreach \row in {1, ..., \yinnersize}
				{
					\draw[-latex] (\col - 1 + \arrowoffset, \row) -- (\col - \arrowoffset, \row); 
				}
			}
		\end{scope}
	\end{tikzpicture}}
				\caption{}
				\label{sub:emptyspace.common+exiting}
			\end{subfigure}
			\begin{subfigure}{.32\textwidth}
				\resizebox{\textwidth}{!}{	\begin{tikzpicture}
	
		\pgfmathtruncatemacro{\xsize}{6}
		\pgfmathtruncatemacro{\ysize}{6}

		\pgfmathsetmacro{\arrowoffset}{0.07}
		\draw[dashed, very thin] (0,0) grid (\xsize, \ysize);
		\draw[thick] (0, 0) --
			(0, \ysize) --
			(\xsize, \ysize) --
			(\xsize, 0) --
			cycle;
			
		\pgfmathtruncatemacro{\xinnersize}{\xsize - 2}
		\pgfmathtruncatemacro{\yinnersize}{\ysize - 2}
		
		\begin{scope}[shift={(0.5, 0.5)}]
			\foreach \col in {0, ..., \xinnersize}
			{
				\draw[-latex] (\col + \arrowoffset, \yinnersize - \ifnum\col=0 -\arrowoffset+1 \else \arrowoffset \fi) -- (\col + \arrowoffset, 0); 
				\draw[-latex] (\col + 1 - \arrowoffset, 0) -- (\col + 1 - \arrowoffset, \yinnersize - \ifnum\col=\xinnersize -1+\arrowoffset \else \arrowoffset \fi); 

				\ifnum\col<\xinnersize
					\draw[-latex] (\col + \arrowoffset, 0) -- (\col + 1 - \arrowoffset, 0); 
				\fi
				\draw[-latex] (\col + 1 - 0.5 * \arrowoffset, \yinnersize + 1 + 2 * \arrowoffset) -- (\col + 0.5 * \arrowoffset, \yinnersize + 1 + 2 * \arrowoffset); 
			}
		\end{scope}
		\begin{scope}[shift={(0.5, 0.5)}]
			\foreach \col in {1, ..., \xinnersize}
			{
				\draw[-latex] (\col + \arrowoffset, \yinnersize + 1 - \arrowoffset) -- (\col + \arrowoffset, \yinnersize + \arrowoffset); 
				\draw[-latex] (\col - \arrowoffset, \yinnersize + \arrowoffset) -- (\col - \arrowoffset, \yinnersize + 1 - \arrowoffset); 

			}
			\foreach \row in {0, ..., \yinnersize}
			{
				\draw[-latex] (\xinnersize + \arrowoffset, \row) -- (\xinnersize + 1 - \arrowoffset, \row); 
			}
		\end{scope}
	\end{tikzpicture}}
				\caption{}
				\label{sub:emptyspace.common+continue}
			\end{subfigure}
			\caption{Empty space \enquote{movement} visualization. (\subref{sub:emptyspace.common}) A
			common pattern and \enquote{moves inspired by} (\subref{sub:emptyspace.exiting}) exiting and 
			(\subref{sub:emptyspace.continue}) continue agent only. Overall, a common pattern fused
			with a specific \enquote{movement} pattern produces a full empty spaces \enquote{movement} 
			pattern caused by (\subref{sub:emptyspace.common+exiting})
			exiting and	(\subref{sub:emptyspace.common+continue}) continue agents.}
		\end{figure}
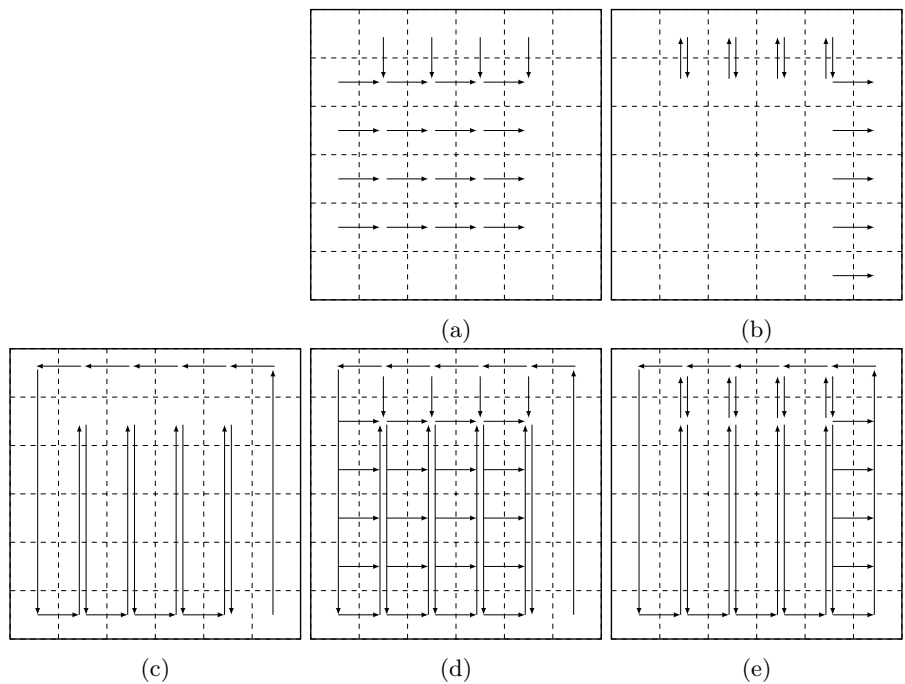

		\begin{itemize}
			\item
				An agent possesses a $3$ bit memory.
			\item
				An agent utilizes its $3$ bits in the following way: $2$ bits are used as the modulo $4$ timer, i.e.
				an agent counts times $0, 1, 2, 3, 0, 1, \ldots$. We follow the Biham et al. approach
				(see \cite{biham1992self}) to ensure
				that agents do not move in colliding directions at the same time. An additional bit $d$ is 
				utilized to store an agent \enquote{heading} direction.
			\item
				An agent memory could be initialized in an arbitrary way, except, that 
				timer bits should be initialized to the same value by all the agents, i.e.,
				agent timers are synchronized.
				Therefore, we canonically initialize all the memory bits to $0$.
			\item
				An agent senses an $L_1$ neighborhood of radius $1$, i.e. the standard \\ 
				$4$-neighborhood.
			\item
				An agent strives to continue the movement in the direction encoded by $d$.
			\item
				An agent movement direction could change, if an agent is unable to proceed in the direction
				encoded by $d$. However, a movement direction change occurs after a cool-down period.
			\item
				An exiting agent does not leave the last column $m$ (see the absence of 
				such \enquote{movement} patterns on \cref{sub:emptyspace.exiting}).
			\item
				Therefore, a continue agent must leave column $m$ westward (see
				\cref{sub:emptyspace.continue}).
			\item 
				Agents of both types move in a clockwise manner on the boundary of the grid
				(\cref{sub:emptyspace.common}, i.e. empty spaces ($0$) \enquote{move} in a 
				counter-clockwise direction).
			\item
				Agents of both types \enquote{enter} non-boundary columns, and traverse them
				in North, then South directions (\cref{sub:emptyspace.common}).
			\item
				An exiting agent exploits the opportunity to move closer to the first column and goes
				westward, in the interior columns of the grid. This behavior ensures, that an
				exiting agent is not stuck indefinitely in the middle of a column 
				(distinctive right arrow \enquote{movements}
				of empty spaces	on \cref{sub:emptyspace.exiting}).
		\end{itemize}
		
		The proposed algorithm $\mathcal{A}$ is a local rule of exchanges. Agents 
		apply $\mathcal{A}$ simultaneously and independently at discrete times $t = 1, 2, 3, \ldots$. 
		Moreover, the algorithm does not require agents to communicate their intentions
		tactically (signaling a lane change) or strategically (designating the type of an agent).
		Therefore, it is enough to sense only the presence of other agents, or the absence thereof. 
		However, an infrastructure support is required to appropriately implement the algorithm 
		on the road. Specifically, a moving frame boundaries should be marked in some way.
		
		We show that Algorithm $\mathcal{A}$ brings a general $n \times m$ grid to a target configuration.
		On the other hand, Algorithm $\mathcal{A}$ pose additional requirements and possesses certain
		shortcomings and limitations. First of all, the algorithm is inapplicable to two-column grids, 
		and some other algorithm should be applied in that specific case. 
		
		Additionally, continue agents demonstrate a somewhat \enquote{unreasonable} behavior, in particular, 
		from time to time a continue agent moves to the first row on the grid, then, immediately, 
		in the next time tick, moves back to its previous position (see the top row down/up movement arrows
		on \cref{sub:emptyspace.continue}). Although, the behavior may seem 
		irrational, those agents have no way to avoid it. 
		Agents with a limited visibility range can not identify the current row, unless it 
		is a row on the edges of the grid. Therefore, agents, located on the interior grid cells
		can detect an edge row (first or last) only after making a step into an edge row.
		It is when an agent understands it should have not leave its previous position 
		and decides to return back.
		
		On the other hand, exiting agents exploit the above \enquote{eccentric} behavior, and do \textbf{not}
		return back into the interior cell (see one-way down arrows on \cref{sub:emptyspace.exiting}). 
		Instead, they continue the movement in the first row, which is a significant shortcut for them
		on the way to the right-most exit lane.
		
		Finally, we assume a slightly more severe constraints 
		\begin{equation*}
			N_0 > 0
		\end{equation*}
		\begin{equation*}			
			N_1 < n\tag{$C_1$}%
			\label{eq:C_1}
		\end{equation*}
		\begin{equation*}
			N_0 + N_1 + N_{-1} = n \cdot m,
		\end{equation*}
		i.e., for the algorithm to work, the number of exiting agents should be strictly 
		less then the number of cells in the right-most grid column.
		
		We claim that the proposed Algorithm solves the general ER Problem.
		\begin{theorem}[Algorithm $\mathcal{A}$ Soundness Theorem]\label{thm:The.A.Theorem}
			Let $\mathcal{IC}$ be an initial grid configuration of an $n \times m$ grid $G$, where
			$m \geq 3, n \geq 2$. Suppose $\mathcal{IC}$ meets \eqref{eq:C_1} criterion. Then Algorithm 
			$\mathcal{A}$ solves the ER Problem and transforms $G$ from $\mathcal{IC}$ 
			to a target configuration.
		\end{theorem}

	\subsection{Time complexity}
	In this section we provide an analyzes of the proposed Algorithm time complexity as
	a function of various parameters.
	
	Recall that we are given an $n \times m$ grid $G$ with $N_0$ - empty cells and $N_1$ - cells 
	occupied by exiting agents. An Algorithm $\mathcal{A}$
	takes an initial grid configuration $\mathcal{C}$ and iteratively transforms it
	to a target configuration $\mathcal{T}$. Then, the number of time ticks it takes $\mathcal{A}$ 
	to execute the transformation - is the number we are looking for. However, a rigorous time 
	complexity analysis is complicated due to the intricacy of agent interactions with empty slots. 
	
	\begin{definition}
		A \textit{$(n, m, N_0, N_1)$ problem} is an ER Problem, where an initial configuration 
		$\mathcal{C}$ of an $n \times m$ grid is parameterized by $N_0$ empty spaces and $N_1$ 
		exiting agents.
	\end{definition}
	
	We informally define an \textit{empty space cycle} - a period of time it takes a designated
	empty space to \enquote{visit} a specific cell on the grid, e.g. $a_{1, 1}$.
	In the context of a single empty space initial configuration the designated empty space
	tracking is an unambiguous task.
	
	Below we provide a worst case time complexity analysis for this uncomplicated $N_0 = 1$ case. 
	
	\begin{lemma}
		Consider a $(n, m, 1, 1)$ problem.
		The worst case time complexity of Algorithm $\mathcal{A}$ is $\Theta(n^2m + m^2n)$.
	\end{lemma}
	\begin{proof}
		We have noted in \cref{observation: west-east exiting movements} that an exiting
		vehicle makes at most $2m$ West/East moves until it reaches the right-most lane. Additionally,
		an exiting vehicle executes at most $n$ North moves in the left-most column of the grid.
		
		By our informal definition an empty space cycle is a period of time until $a_{1, 1}$ becomes 
		empty once again. Let us ignore an exiting agent shortcut route, i.e. an opportunity of
		an exiting agent to immediately move into the first row from the current column (second row).
		Then, during every empty space cycle an exiting vehicle, either makes one East/West move, 
		or it makes one North move in the first column. Therefore, at most $2m + n$ cycles are required
		to bring an exiting vehicle	to the right-most column of the grid. 
		
		Each empty space cycle takes time proportional to the parameters $m, n$. The time is mostly
		spent on internal columns double \enquote{traversals}. Moreover, each empty space \enquote{move}
		takes a full four-time-ticks round to be executed (with a rare exception in the
		last column, or the u-turns at $a_{2, i}$ by continue agents, which are all require less time). 
		Hence, we can bound the time needed for a full empty space cycle by $8mn$ ticks from above.
		
		Therefore, in total, it takes at most $16m^2n + 8mn^2 = \Theta(n^2m + m^2n)$ time ticks, as claimed.
	\end{proof}
	
	\begin{lemma} \label{lemma:moving out of internal columns}
		Let $T_e$ be the first time tick, when internal columns have become empty of the exiting agents
		in the $(n, m, 1, N_1)$ problem,
		i.e. all exiting agents are located in the first/last column or the first row after $T_e$ ticks.
		Then $T_e = \Theta(m + N_1)$ empty space cycles or $\Theta((m + N_1)\cdot mn)$ time ticks.
	\end{lemma}
	
\begin{figure}
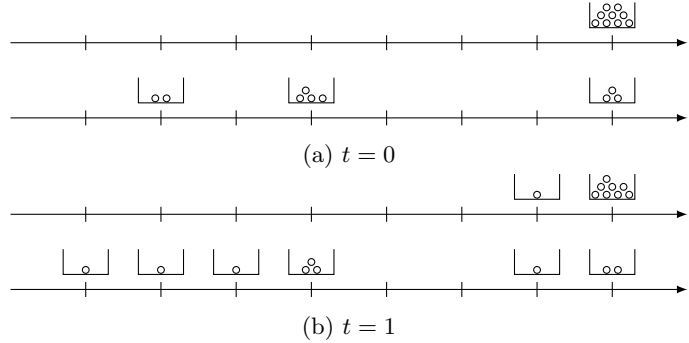

		\centering
		\subfloat[$t=0$]{%
				\begin{tikzpicture}
		\pgfmathsetmacro{\sidetick}{0.1}
		\pgfmathsetmacro{\basey}{0.2}
		\pgfmathsetmacro{\scalex}{0.15}
		\pgfmathtruncatemacro{\lasttick}{7}
		
		\begin{scope}[shift={(-\scalex * 2, \basey)}]
			\begin{scope}[shift={(\lasttick, 0)}]
				\begin{scope}[scale = \scalex]
					\pgfmathtruncatemacro{\balls}{9}
					\input{urns.n.balls}
				\end{scope}
			\end{scope}
		\end{scope}

		\draw[-latex] (-1, 0) -- (\lasttick+1, 0);
		\foreach \x in {0, ..., \lasttick}
		{
			\draw (\x, -\sidetick) -- (\x, \sidetick);
		}
		
		\begin{scope}[yshift = -1 cm]
			\begin{scope}[shift={(-\scalex * 2, \basey)}]
				\foreach \offset/\nballs in {1/2, 3/4, 7/3}
				{
					\begin{scope}[shift={(\offset, 0)}]
						\begin{scope}[scale = \scalex]
							\pgfmathtruncatemacro{\balls}{\nballs}
							\input{urns.n.balls}
						\end{scope}
					\end{scope}
				}
			\end{scope}

			\draw[-latex] (-1, 0) -- (\lasttick+1, 0);
			\foreach \x in {0, ..., \lasttick}
			{
				\draw (\x, -\sidetick) -- (\x, \sidetick);
			}
		\end{scope}
	\end{tikzpicture}
		}\par\vfill
		\subfloat[$t=1$]{%
				\begin{tikzpicture}
		\pgfmathsetmacro{\sidetick}{0.1}
		\pgfmathsetmacro{\basey}{0.2}
		\pgfmathsetmacro{\scalex}{0.15}
		\pgfmathtruncatemacro{\lasttick}{7}
		
		\begin{scope}[shift={(-\scalex * 2, \basey)}]
			\begin{scope}[shift={(\lasttick, 0)}]
				\foreach \offset/\nballs in {-1/1, 0/8}
				{
					\begin{scope}[shift={(\offset, 0)}]
						\begin{scope}[scale = \scalex]
							\pgfmathtruncatemacro{\balls}{\nballs}
							\input{urns.n.balls}
						\end{scope}
					\end{scope}
				}
			\end{scope}
		\end{scope}

		\draw[-latex] (-1, 0) -- (\lasttick+1, 0);
		\foreach \x in {0, ..., \lasttick}
		{
			\draw (\x, -\sidetick) -- (\x, \sidetick);
		}
		
		\begin{scope}[yshift = -1 cm]
			\begin{scope}[shift={(-\scalex * 2, \basey)}]
				\foreach \offset/\nballs in {0/1, 1/1, 2/1, 3/3, 6/1, 7/2}
				{
					\begin{scope}[shift={(\offset, 0)}]
						\begin{scope}[scale = \scalex]
							\pgfmathtruncatemacro{\balls}{\nballs}
							\input{urns.n.balls}
						\end{scope}
					\end{scope}
				}
			\end{scope}

			\draw[-latex] (-1, 0) -- (\lasttick+1, 0);
			\foreach \x in {0, ..., \lasttick}
			{
				\draw (\x, -\sidetick) -- (\x, \sidetick);
			}
		\end{scope}
	\end{tikzpicture}
		}
		\caption{Two different ball placements depicted side-by-side at $t=0$. A single 
		ball-relocation iteration depicted (after balls moved at $t=1$).}
		\label{fig:side-by-side simulations}
	\end{figure}

	In order to prove \cref{lemma:moving out of internal columns} we consider the following problem.
	Suppose, we put $M$ boxes indexed by positive integers $1, 2, \ldots, M$ and a sink box indexed 
	$0$. Then we arbitrary put $N$ identical balls into the boxes, including sink box. 
	At discrete time ticks $t=1, 2, 3, \ldots$ we pick a single ball from each non-empty 
	box $k$ and move it to box $k-1$ (see a one time tick simulation of two ball configurations 
	in \cref{fig:side-by-side simulations}). Balls in the sink box remain in the sink box forever. 
	Denote this ball movement protocol as $\mathcal{BMP}_1$.
	
	Question: When will all the balls disappear in the sink?
	
	The answer depends on the initial distribution of the balls. However, we are 
	more interested in the upper bound on the time, and we claim it to be $O(M + N)$.
	\begin{proof}
		Put all the balls in the box indexed $M$. We remove one ball at a time, hence
		the last ball will be removed at $t = N$. A standalone ball can \enquote{travel}
		at most one box at a tick. Therefore, $M$ ticks are required to move the last ball
		from box $M$ to box $0$. Note, that throughout the execution, all the balls that have
		already left the $M$ box are alone in their current location (except for balls in 
		the sink). Thus, all the balls are constantly moving towards the sink at box $0$. 
		Finally, the last ball will disappear in sink at $t = M + N$.
		
		We claim that an arbitrary placement of balls will lead to a faster
		ball disappearance in sink. Label the balls from $1$ to $N$, in such a manner, that
		a ball at the box $k$ will be labeled with a higher ordinal, than all the balls in
		the lower indexed boxes.
		
		For example, let $M=7$, $N=9$, such that two balls are at $1$ - we shall label them 
		$1$ and $2$; four balls are at $3$ - we shall label them $3,4,5$ and $6$, and the other 
		balls are at $7$ - and we shall label them $7,8,9$. See 
		\cref{fig:slow side-by-side simulations} at $t=0$.
		
		\begin{figure}
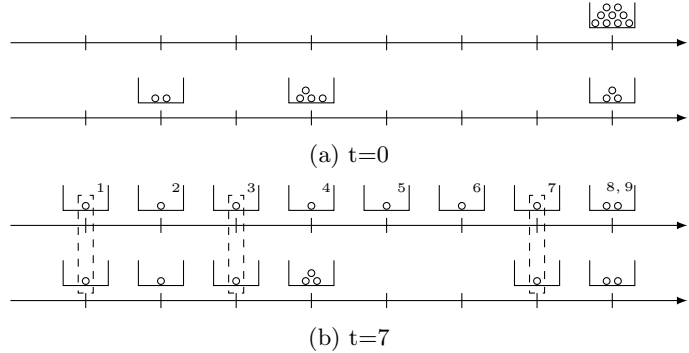

			\centering
			\subfloat[t=0]{%
					\begin{tikzpicture}
		\pgfmathsetmacro{\sidetick}{0.1}
		\pgfmathsetmacro{\basey}{0.2}
		\pgfmathsetmacro{\scalex}{0.15}
		\pgfmathtruncatemacro{\lasttick}{7}
		
		\begin{scope}[shift={(-\scalex * 2, \basey)}]
			\begin{scope}[shift={(\lasttick, 0)}]
				\begin{scope}[scale = \scalex]
					\pgfmathtruncatemacro{\balls}{9}
					\input{urns.n.balls}
				\end{scope}
			\end{scope}
		\end{scope}

		\draw[-latex] (-1, 0) -- (\lasttick+1, 0);
		\foreach \x in {0, ..., \lasttick}
		{
			\draw (\x, -\sidetick) -- (\x, \sidetick);
		}
		
		\begin{scope}[yshift = -1 cm]
			\begin{scope}[shift={(-\scalex * 2, \basey)}]
				\foreach \offset/\nballs in {1/2, 3/4, 7/3}
				{
					\begin{scope}[shift={(\offset, 0)}]
						\begin{scope}[scale = \scalex]
							\pgfmathtruncatemacro{\balls}{\nballs}
							\input{urns.n.balls}
						\end{scope}
					\end{scope}
				}
			\end{scope}

			\draw[-latex] (-1, 0) -- (\lasttick+1, 0);
			\foreach \x in {0, ..., \lasttick}
			{
				\draw (\x, -\sidetick) -- (\x, \sidetick);
			}
		\end{scope}
	\end{tikzpicture}
			}\par\vfill
			\subfloat[t=7]{%
					\begin{tikzpicture}
		\pgfmathsetmacro{\sidetick}{0.1}
		\pgfmathsetmacro{\basey}{0.2}
		\pgfmathsetmacro{\scalex}{0.15}
		\pgfmathtruncatemacro{\lasttick}{7}
		
		\begin{scope}[shift={(-\scalex * 2, \basey)}]
			\foreach \offset/\nballs in {0/1, 1/1, 2/1, 3/1, 4/1, 5/1, 6/1, 7/2}
			{
				\begin{scope}[shift={(\offset, 0)}]
					\begin{scope}[scale = \scalex]
						\pgfmathtruncatemacro{\balls}{\nballs}
						\input{urns.n.balls}
					\end{scope}
				\end{scope}
			}
		\end{scope}

		\draw[-latex] (-1, 0) -- (\lasttick+1, 0);
		\foreach \x in {0, ..., \lasttick}
		{
			\draw (\x, -\sidetick) -- (\x, \sidetick);
		}
		
		\begin{scope}[yshift = -1 cm]
			\begin{scope}[shift={(-\scalex * 2, \basey)}]
				\foreach \offset/\nballs in {0/1, 1/1, 2/1, 3/3, 6/1, 7/2}
				{
					\begin{scope}[shift={(\offset, 0)}]
						\begin{scope}[scale = \scalex]
							\pgfmathtruncatemacro{\balls}{\nballs}
							\input{urns.n.balls}
						\end{scope}
					\end{scope}
				}
			\end{scope}

			\draw[-latex] (-1, 0) -- (\lasttick+1, 0);
			\foreach \x in {0, ..., \lasttick}
			{
				\draw (\x, -\sidetick) -- (\x, \sidetick);
			}
		\end{scope}
		
		\pgfmathsetmacro{\dashedwidth}{0.2}
		\begin{scope}[yshift=\basey cm]
			\foreach \x in {0, 2, 6}
			{
				\begin{scope}[xshift=\x cm]
					\draw[dashed] (-\dashedwidth / 2, \basey) -- (-\dashedwidth / 2, -1 - \basey / 2) -- 
						(\dashedwidth / 2, -1 - \basey / 2) -- (\dashedwidth / 2, \basey) -- cycle;
				\end{scope}
			}
		\end{scope}
		
		\begin{scope}[xshift= -1 cm]
			\foreach \num in {1, ..., 7}
			{
				\node at (\num + 0.2, 0.5) {\tiny $\num$};
			}
			\node at (8 + 0.1, 0.5) {\tiny $8,9$};
		\end{scope}
	\end{tikzpicture}
			}
			\caption{Two simulations run according to the slowest possible mode.}
			\label{fig:slow side-by-side simulations}
		\end{figure}
		
		We compare ball disappearance time versus the \enquote{All in $M$ initially} 
		scenario duration. Denote the former original scenario as $\mathcal{O}$ and
		the latter \enquote{All in $M$ initially} scenario as $\mathcal{P}$.
		In $\mathcal{P}$ we always remove the lowest indexed 
		ball from the box. However, in the $\mathcal{O}$ scenario we proceed in the following
		way: we remove the ball with label $L$ from the box $k$, at the same time when
		the ball with label $L$ is removed from the box $k$ in the $\mathcal{P}$. We denote
		this changed protocol as $\mathcal{BMP}_2$.
		
		Since all the balls in $\mathcal{P}$ do eventually end up in sink, we ensure that 
		all the balls will end up in sink in $\mathcal{O}$ also under the protocol $\mathcal{BMP}_2$.
		However, an actual $\mathcal{BMP}_1$ run on $\mathcal{O}$ will be at least as fast as 
		$\mathcal{BMP}_2$. Hence, the time required to move	all the $\mathcal{O}$ balls into 
		the sink at $0$ under $\mathcal{BMP}_1$ is at most $N + M$.
	\end{proof}
	
	Let get back to the ER Problem and the Algorithm $\mathcal{A}$ time complexity analysis.
	
	An empty space \enquote{executes} a counter-clockwise walk on the grid. It \enquote{traverses} 
	internal columns North then South, then \enquote{moves} East. East movement from column $i$
	to column $i+1$ originates from three different agent movement rules:
	\begin{itemize}
		\item
			Exiting vehicles in the internal column $i+1$ \enquote{Rush West} if neighbored by
			an empty space from the West.
		\item
			Continue vehicles in the last column $m$ execute the rule \enquote{Leave the Exit Lane}
			under the same conditions.			
		\item
			Both types of agents leave $a_{n, i+1}$ West.
	\end{itemize}
	
	The second option is irrelevant in the context of exiting vehicles. Hence, we shall concentrate 
	on the two remaining cases.
	
	The last type of agent Westward movement is not initiated upon empty space 
	\enquote{arrival} to $a_{n, i}$ for the first time. Only the second encounter triggers an
	appropriate agent movement rule. This allows execution of the following scenario: an empty
	space \enquote{arrives} from $a_{n, i-1}$ to $a_{n, i}$, then \enquote{enters} column $i$ and 
	\enquote{traverses} it in the \enquote{regular} North-South fashion. Finally, an empty space
	\enquote{moves} to $a_{n, i+1}$ on the second arrival to $a_{n, i}$.
	
	However, consider the former case. During a North-South \enquote{traversal} an empty space 
	is \enquote{rushed} East by an exiting vehicle from column $i+1$. Then, an agent in the 
	$a_{n, i+1}$ cell is left waiting for its second encounter with an empty space in the following
	empty space cycles. Moreover, the next East \enquote{move} of an empty space from
	$a_{n, i-1}$ to $a_{n, i}$ could lead to a different \enquote{movement} pattern.
	Specifically, an empty space could \enquote{continue} to $a_{n, i+1}$. Though, an empty
	space will repeat a regular column $i$ North-South pattern in the very next cycle. It follows,
	that at least once per two consecutive empty space cycles, a group of column $i+1$ 
	exiting vehicles decreases by $1$.

	Then, a ball-box analogy can be applied. We shall set a \enquote{ball-box time tick} to count a 
	consecutive empty space cycle pairs. Since more than one exiting agent can leave a column (a box)
	in the same time tick, the ball-box model application is not immediate. However,
	removing a single ball restriction could only decrease the \enquote{All in the Sink}
	completion time. Thus, a complexity of time required for all the exiting vehicles to leave the 
	internal columns (except, maybe, second) is still $\Theta(m + N_1)$ empty space cycles.
	
	An exiting agent visible behavior in the second column differs slightly from 
	the behavior in	the other internal columns. We have previously noted, that an exiting
	agent in the internal column moves West once neighbored by an empty space. And that a vertical
	\enquote{movement} of an empty space in the internal column followed by a West movement
	of an exiting agent	is possible at most once per clock cycle. However, 
	in the first column an empty space can \enquote{move} South more than once during
	the same $4$ successive ticks. In fact, in the first column an exiting agent can move North
	$3$ out of $4$ ticks in one	clock cycle. Therefore, an empty space 
	could \enquote{pass} in the first column near an exiting vehicle in the second column 
	without that said exiting agent moving West in the same clock cycle.
	
	This second column complication is, however, caused by \textit{another} exiting agent moving
	North in the first column. Let $A$ be an exiting vehicle, that \enquote{misses} a neighbor 
	empty space. Denote the $A$ row index by $k$. In the simplest, single empty space, model, 
	the maximum number of \enquote{new} exiting vehicles that could occupy $a_{k, 1}$ during
	the same empty space cycle is $1$.
	Moreover, continue vehicles can not prevent $A$'s move into $a_{k,1}$ during the clock
	cycle when an empty space \enquote{moved} into $a_{k, 1}$
	
	Since there are at most $N_1$ exiting agents we conclude
	that in at most $Theta(m + N_1)$ empty space cycles all the exiting agents leave the internal 
	columns. 
	Thus we have established the correctness of \cref{lemma:moving out of internal columns}.

	We summarize the statements above and we come up with a specific bound of $2m + 3N_1$ cycles. 
	Recall that $2$ comes from the movement rule of agents at position $5$ (i.e. empty 
	space double encounter), and one additional $N_1$ was added as a second column complication 
	payment. However, this last summand will be later counted towards the cycles required for
	exiting vehicles to leave the first column.
	
	Suppose, at $T$ all exiting agents have already left the internal columns. Then an exiting 
	agent in the first row or the first column moves exactly one step clockwise in 
	the first row/first column during a single empty space cycle. Hence, in at most $(m + n)$
	empty space cycles all the exiting vehicles will move into the last column. The Algorithm
	$\mathcal{A}$ will stop then upon reaching a target configuration. We conclude, that the 
	time complexity of the original Algorithm $\mathcal{A}$ is $\Theta(n + m + N_1)$ empty
	space cycles. Specifically, the upper bound is $n + 3m + 2N_1$ cycles

	Finally, we shall analyze the most general $(n, m, N_0, N_1)$ problem, where $N_0$ is
	the number of empty spaces on the grid. Initially, we shall assume, that empty spaces do not 
	\enquote{interfere} with each other, i.e. empty spaces do not \enquote{wait}, because
	neighbor agents interact with \textit{other} empty spaces. Then 
	we get a reciprocity: $N_0$ empty spaces pass through designated cell $a_{i, j}$ during 
	one empty space cycle. Hence each exiting agent observes $N_0$ empty spaces during the 
	same period of time which is equivalent to observing a single empty space, but $N_0$ times.
	
	In reality an \enquote{empty space cycle} is not length-constant over time. But 
	depends on the actual \enquote{interference} of empty spaces during the execution of
	the Algorithm. Moreover, due to the certain positions agent movement rules the
	cycles are even shorter comparatively to the expected $1 / N_0$ rate. Indeed, if one empty
	space \enquote{moves} into an internal column $i$ from the position $5$, then the next empty
	space \enquote{skips} to the next $i+1$ column instead of entering the column $i$,
	as the \enquote{previous} empty space.
	
	Thus, an overall time upper bound is as follows:
	\begin{equation*}
		T \leq \frac{3m + n + 2N_1}{N_0}\cdot 8mn,
	\end{equation*}
	where $8mn$ is an upper bound on the \enquote{empty space cycle} length.
	
	\subsubsection{Average case time complexity}
	In some scenarios, not only can we bound the worst-case time complexity, but we can also estimate 
	an average case time complexity. 
	
	Consider a special case $N_0 = N_1 = 1$, an 
	\enquote{empty space cycle} is bounded by $8mn$. We can also estimate a number of 
	cycles required to move an exiting vehicle from \textit{any} cell of the grid to the 
	right-most column of the grid.
	
	We shall analyze a number of exiting agent initial positions:
	\begin{enumerate}
		\item
			The right-most column $m$. Agents, starting in this column, need $0$ cycles to finish
			the algorithm. There are $n$ distinct starting positions in this column.
		\item
			The first row. Agents, starting in the first row, are expected to make a number 
			of steps to the right, and the number of steps is exactly the distance between the agent
			initial column and the $m$th column.
			However, an agent is only able to make a single step during one full empty space cycle. 
			Hence, e.g. it will take a whole $(m - k)$ empty space cycles for an agent starting at cell $a_{1, k}$
			to reach the right-most column.
			
			There are $(m-1)$ distinct starting positions in the first row 
			(not already mentioned in the previous bullet).
		\item
			The second row. An agent starting in the second row has a \textit{shortcut} opportunity
			to move into the first row. And then to behave like a \textit{first row} agent from
			the previous bullet. In other words, an additional empty space
			cycle will be required to complete move to the last column for an agent starting at cell 
			$a_{2, k}$ versus an agent starting at $a_{1, k}$.
			
			However, in the worst case, the first encounter of an empty space will end up
			in a West movement, i.e. an Algorithm will require additional two empty space cycles 
			to accomplish the task in comparison to the optimistic scenario of an \enquote{immediate
			shortcut}, or three cycles in comparison to an agent at $a_{1, k}$.
		\item
			The first column. An agent, starting in the first column, will need to make all 
			the North steps to reach the first row. Then an additional $(m-1)$ East steps in the 
			first row. The number of required steps in the first column is a function of a starting row:
			agents starting at $a_{k, 1}$ will be required to make $(k-1)$ steps North.
		\item
			The last and the most numerous group of agents, are agents starting in the general
			internal column position. An agent starting at $a_{i, j}$ will need to move West $(j-1)$
			steps, then $(i - 1)$ steps North, before, finally, proceed $(m-1)$ East steps in the
			first row.
	\end{enumerate}

	To estimate the bound on the average completion time we sum up the completion times for all 
	the possible initial placements of an exiting vehicle. The below times are expressed 
	in the empty space cycles.
	
	First group can be skipped.
	Agents from the second group will need a total number of cycles equal to:
	\begin{equation*}
		\sum\limits_{j=1}^{m-1} (m - j) = \frac{m \cdot (m-1)}{2},
	\end{equation*}

	While, agents from the third group in the optimistic scenario
	\begin{equation*}
		\sum\limits_{j=1}^{m-1} (m - j + 1) = \frac{(m-1) \cdot (m+2)}{2},
	\end{equation*}
	and in the pessimistic case:
	\begin{equation*}
		m + \sum\limits_{j=2}^{m-1} (m - j + 3) = \frac{(m-1) \cdot (m+4)}{2} + (m - 3),
	\end{equation*}
	where $m$ cycles are paid for a North-then-East movement of an agent initially
	located at $a_{2, 1}$.
	
	In the forth group, we pay cycles according to:
	\begin{equation*}
		\sum\limits_{i=3}^{n} ((i - 1) + (m-1)) = (m-1)(n-2) + \frac{n(n-1)}{2} - 1.
	\end{equation*}
	
	In the last group, the calculate payment in two steps. An agent should move West from
	$a_{i, j}$ to $a_{i, 1}$, total in
	\begin{equation*}
		(n-2) \cdot\sum\limits_{j=2}^{m-1} (j - 1) = (n-2) \cdot \frac{(m-1) \cdot (m-2)}{2}.
	\end{equation*}
	From $a_{i, 1}$ the cost was already calculated, hence this is $m-2$ copies of the forth
	group:
	\begin{equation*}
		(m-2)\left[(m-1)(n-2) + \frac{n(n-1)}{2} - 1\right].
	\end{equation*}
	
	The total number of initial positions is equal to the number of cells on the grid. 
	Though in the worst case we shall additionally pay
	a whole \enquote{empty space cycle} due to the random initial placement of an empty space.
	After summing previous terms and multiplying by the cycle length we get an upper bound
	on the average time:
	\begin{equation*}
		\bar{T} \leq 4(3m + n)mn = O(m^2n+mn^2)
	\end{equation*}
	
	On the other hand, the second and the third groups in the optimistic 
	scenario require $m^2 - 1$ cycles.
	While, two last groups total time could be bounded from below by 
	$\sfrac{1}{2}(m-1)(n-2)(m + \sfrac{1}{2}n - 1)$
	cycles. Therefore, $\mathbb{E}(T) = \Omega(m^2n+mn^2)$. Hence, 
	\begin{equation*}
		\mathbb{E}(T) = \Theta(m^2n+mn^2).
	\end{equation*}
	We conclude that the average and the worst cases of the $(m, n, 1, 1)$ problem have the same 
	time complexity of $\Theta(m^2n+mn^2)$ ticks.
	
	In the general case, $N_1 > 1$, the Algorithm completion time is intuitively determined by
	an agent in the internal column with the highest index and/or highest row index. 
	For an arbitrary initial configuration
	the probability that an agent is located in the bottom half could be bounded from below by:
	\begin{equation*}
		\mathbb{P}(\text{agent is in the grid bottom half}) = 
		\frac{\sfrac{n}{2}\cdot (m-1)}{n \cdot m} = \frac{m-1}{2m} \geq \frac{1}{4}.
	\end{equation*}
	Therefore, the probability that no agent is located in the bottom half area is:
	\begin{equation*}
		\mathbb{P}(\text{grid bottom half is agent free}) \leq (1 - \sfrac{1}{4})^{N_1} \to 0,
	\end{equation*}
	and the complementary event probability is high, in particular at least $\sfrac{1}{4}$.
	Hence, in at least $\sfrac{1}{4}$ cases the time it takes the Algorithm to solve the problem
	is $\Theta(m^2n+n^2m)$, as in the worst case. It follows, that in the average case
	\begin{equation*}
		\mathbb{E}(T) = \Theta(m^2n+mn^2).
	\end{equation*}
	
	Moreover, our previous observations hold, therefore
	\begin{equation*}
		\mathbb{E}(T) = \Theta\left(\frac{1}{N_0}\cdot (m + n + N_1)\cdot mn\right).
	\end{equation*}

	\subsection{Simulation results}
	Alongside, the theoretical complexity, we are extremely keen to investigate \enquote{an average 
	case} behavior of the Algorithm in the settings of practical interest. The number of lanes
	on the real roads is not unbounded or extremely high. In fact, $3 \leq m \leq 6$ cover the vast
	majority of the modern infrastructure.
	
	We have conducted a series of simulations to estimate an average Algorithm completion time.
	Note, that a single time tick in a simulation represents roughly one second in the real-world 
	environment required to	execute a lane change or a \enquote{move}-forward/backward maneuver 
	by a real vehicle on the road.
	
	We have run the simulations on the proprietary software implemented in C++. The agents
	were implemented in the OOP paradigm with a full memory separation between the grid state and
	the agent state. Every time tick an agent sensed the immediate neighborhood and
	decided to act based on the visible neighborhood and the agent internal state, i.e. according
	to the Algorithm $\mathcal{A}$. Exactly by the Algorithm, an internal state change kept secret,
	while a \enquote{move} decision was communicated to the grid. After all agent move decisions
	were collected, the grid moved all the agents at once.
	
	Each test run started with a random initial configuration parametrized by a tuple 
	$(n, m, N_0, N_1)$. Then the test was executed until a target configuration was reached. 
	Completion times were saved and processed later, with the graphical results presented below.
	If not otherwise stated, we have set $N_1 = n - 1$ throughout the tests.
	
	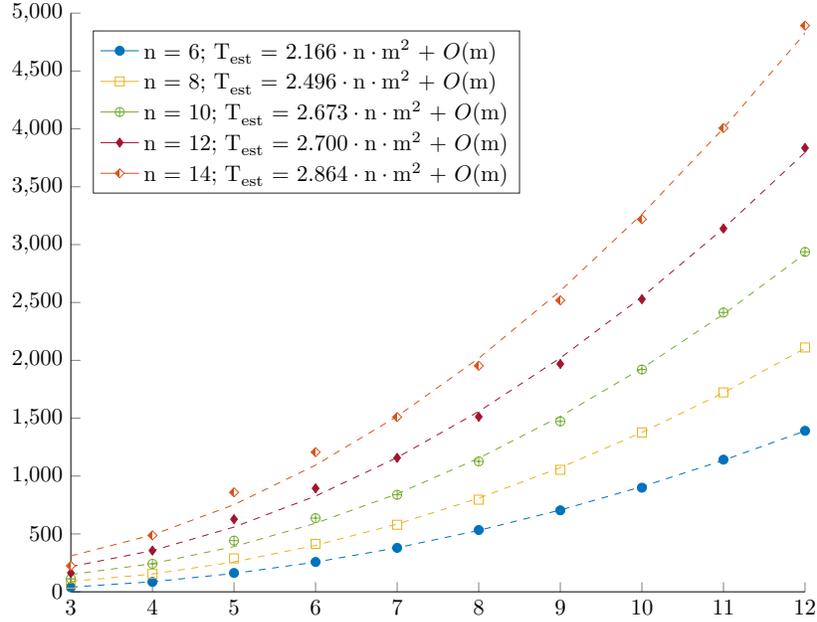
\begin{figure}
		\centering
		\scalebox{0.85}{
%
%
\definecolor{mycolor1}{rgb}{0.00000,0.44700,0.74100}%
\definecolor{mycolor2}{rgb}{0.92900,0.69400,0.12500}%
\definecolor{mycolor3}{rgb}{0.46600,0.67400,0.18800}%
\definecolor{mycolor4}{rgb}{0.63500,0.07800,0.18400}%
\definecolor{mycolor5}{rgb}{0.85000,0.32500,0.09800}%
\begin{tikzpicture}

\begin{axis}[%
width=4.521in,
height=3.566in,
at={(0.758in,0.481in)},
scale only axis,
xmin=3,
xmax=12,
xlabel style={font=\color{white!15!black}},
ymin=0,
ymax=5000,
ylabel style={font=\color{white!15!black}},
axis background/.style={fill=white},
title style={font=\bfseries},
axis x line*=bottom,
axis y line*=left,
legend style={at={(0.03,0.97)}, anchor=north west, legend cell align=left, align=left, draw=white!15!black}
]
\addplot [color=mycolor1, draw=none, mark=*, mark options={solid, mycolor1}]
  table[row sep=crcr]{%
3	38.922\\
4	85.334\\
5	162.84\\
6	257.966\\
7	379.72\\
8	533.968\\
9	704.074\\
10	899.646\\
11	1141.9\\
12	1391.172\\
};
\addlegendentry{$\text{n =  6; T}_{\text{est}}\text{ = 2.166} \cdot \text{n} \cdot \text{m}^\text{2}\text{ + }O(\text{m})$}

\addplot [color=mycolor1, dashed, forget plot]
  table[row sep=crcr]{%
3	40.73536363636\\
4	86.7016909090896\\
5	158.665548484849\\
6	256.626936363638\\
7	380.585854545457\\
8	530.542303030305\\
9	706.496281818184\\
10	908.447790909092\\
11	1136.39683030303\\
12	1390.3434\\
};
\addplot [color=mycolor2, draw=none, mark=square, mark options={solid, mycolor2}]
  table[row sep=crcr]{%
3	70.482\\
4	155.078\\
5	288.784\\
6	413.962\\
7	579.478\\
8	796.068\\
9	1054.262\\
10	1377.316\\
11	1722.83\\
12	2111.108\\
};
\addlegendentry{$\text{n =  8; T}_{\text{est}}\text{ = 2.496} \cdot \text{n} \cdot \text{m}^\text{2}\text{ + }O(\text{m})$}

\addplot [color=mycolor2, dashed, forget plot]
  table[row sep=crcr]{%
3	89.680945454544\\
4	153.669115151514\\
5	257.599709090909\\
6	401.472727272727\\
7	585.28816969697\\
8	809.046036363637\\
9	1072.74632727273\\
10	1376.38904242424\\
11	1719.97418181818\\
12	2103.50174545455\\
};
\addplot [color=mycolor3, draw=none, mark=oplus, mark options={solid, mycolor3}]
  table[row sep=crcr]{%
3	112.686\\
4	241.676\\
5	442.154\\
6	636.586\\
7	838.246\\
8	1126.738\\
9	1472.368\\
10	1920.808\\
11	2413.222\\
12	2936.758\\
};
\addlegendentry{$\text{n = 10; T}_{\text{est}}\text{ = 2.673} \cdot \text{n} \cdot \text{m}^\text{2}\text{ + }O(\text{m})$}

\addplot [color=mycolor3, dashed, forget plot]
  table[row sep=crcr]{%
3	149.249345454545\\
4	243.328606060606\\
5	390.867715151515\\
6	591.866672727273\\
7	846.325478787879\\
8	1154.24413333333\\
9	1515.62263636364\\
10	1930.46098787879\\
11	2398.75918787879\\
12	2920.51723636364\\
};
\addplot [color=mycolor4, draw=none, mark=diamond*, mark options={solid, mycolor4}]
  table[row sep=crcr]{%
3	162.898\\
4	355.852\\
5	627.17\\
6	893.99\\
7	1157.482\\
8	1512.326\\
9	1968.836\\
10	2528.112\\
11	3138.58\\
12	3835.148\\
};
\addlegendentry{$\text{n = 12; T}_{\text{est}}\text{ = 2.700} \cdot \text{n} \cdot \text{m}^\text{2}\text{ + }O(\text{m})$}

\addplot [color=mycolor4, dashed, forget plot]
  table[row sep=crcr]{%
3	217.40690909092\\
4	355.847503030307\\
5	559.092248484848\\
6	827.141145454541\\
7	1159.99419393939\\
8	1557.65139393939\\
9	2020.11274545454\\
10	2547.37824848485\\
11	3139.44790303031\\
12	3796.32170909092\\
};
\addplot [color=mycolor5, draw=none, mark=halfsquare left*, mark options={solid, mycolor5}]
  table[row sep=crcr]{%
3	224.602\\
4	487.724\\
5	859.618\\
6	1206.27\\
7	1510.346\\
8	1952.838\\
9	2518.19\\
10	3218.688\\
11	4006.49\\
12	4892.054\\
};
\addlegendentry{$\text{n = 14; T}_{\text{est}}\text{ = 2.864} \cdot \text{n} \cdot \text{m}^\text{2}\text{ + }O(\text{m})$}

\addplot [color=mycolor5, dashed, forget plot]
  table[row sep=crcr]{%
3	310.260309090919\\
4	491.425896969701\\
5	752.772863636363\\
6	1094.30120909091\\
7	1516.01093333333\\
8	2017.90203636363\\
9	2599.97451818181\\
10	3262.22837878788\\
11	4004.66361818182\\
12	4827.28023636364\\
};
\end{axis}
\end{tikzpicture}%
		}
		\caption{The average completion time on an $n \times m$ grid vs. the number of lanes ($m$). The number
		of empty spaces $N_0 = 5$. The best fit second degree polynomial is depicted by a dash line for every
		data series.}
		\label{fig:simulations.m}
	\end{figure}

	\begin{figure}
		\centering
		\begin{subfigure}{.48\textwidth}
			\resizebox{\textwidth}{!}{
%
%
\definecolor{mycolor1}{rgb}{0.00000,0.44700,0.74100}%
\begin{tikzpicture}

\begin{axis}[%
width=4.521in,
height=3.566in,
at={(0.758in,0.481in)},
scale only axis,
xmin=3,
xmax=21.5,
xlabel style={font=\color{white!15!black}},
ymin=0,
ymax=120000,
ylabel style={font=\color{white!15!black}},
ytick={20000, 40000, 60000, 80000, 100000, 120000},
scaled y ticks=true,
axis background/.style={fill=white},
title style={font=\bfseries},
axis x line*=bottom,
axis y line*=left,
legend style={at={(0.03,0.97)}, anchor=north west, legend cell align=left, align=left, draw=white!15!black}
]
\addplot [color=mycolor1, draw=none, mark=*, mark options={solid, mycolor1}, line width=2pt]
  table[row sep=crcr]{%
3	644.675\\
6	3060.656\\
9	7860.8345\\
12	13998.8505\\
15	20962.2475\\
18	29140.732\\
21	38404.74\\
};

\addplot [color=mycolor1, loosely dotted, line width=2pt]
  table[row sep=crcr]{%
3	8784\\
6	20160\\
9	34128\\
12	50688\\
15	69840\\
18	91584\\
21	115920\\
};
\addplot [color=mycolor1, draw=none, line width=2pt, forget plot]
 plot [error bars/.cd, y dir = both, y explicit, error bar style={line width=2pt}, error mark options={line width=1pt,mark size=4pt,rotate=90}]
 table[row sep=crcr, y error plus index=2, y error minus index=3]{%
3	644.675	844	364\\
6	3060.656	4664	2005\\
9	7860.8345	10020	4940\\
12	13998.8505	16776	8410\\
15	20962.2475	25080	13824\\
18	29140.732	34372	20614\\
21	38404.74	43160	29396\\
};
\addplot [color=mycolor1, dashed, line width=2pt, forget plot]
  table[row sep=crcr]{%
3	308.441916666572\\
6	3543.12221428571\\
9	8034.52696428576\\
12	13782.6561666667\\
15	20787.5098214286\\
18	29049.0879285714\\
21	38567.3904880952\\
};
\end{axis}
\end{tikzpicture}%
			}
			\caption{}
			\label{subfig:n_0=3 n=18}
		\end{subfigure}
		\quad
		\begin{subfigure}{.48\textwidth}
			\resizebox{\textwidth}{!}{
%
%
\definecolor{mycolor1}{rgb}{0.00000,0.44700,0.74100}%
\begin{tikzpicture}

\begin{axis}[%
width=4.521in,
height=3.566in,
at={(0.758in,0.481in)},
scale only axis,
xmin=3,
xmax=21.5,
xlabel style={font=\color{white!15!black}},
ymin=0,
ymax=18000,
ylabel style={font=\color{white!15!black}},
axis background/.style={fill=white},
title style={font=\bfseries},
axis x line*=bottom,
axis y line*=left,
legend style={at={(0.03,0.97)}, anchor=north west, legend cell align=left, align=left, draw=white!15!black}
]
\addplot [color=mycolor1, draw=none, mark=*, mark options={solid, mycolor1}, line width=2pt]
  table[row sep=crcr]{%
3	92.795\\
6	459.7155\\
9	988.7515\\
12	1614.389\\
15	2424.197\\
18	3219.723\\
21	3998.259\\
};

\addplot [color=mycolor1, loosely dotted, line width=2pt]
  table[row sep=crcr]{%
3	1254.85714285714\\
6	2880\\
9	4875.42857142857\\
12	7241.14285714286\\
15	9977.14285714286\\
18	13083.4285714286\\
21	16560\\
};
\addplot [color=mycolor1, draw=none, line width=2pt, forget plot]
 plot [error bars/.cd, y dir = both, y explicit, error bar style={line width=2pt}, error mark options={line width=1pt,mark size=4pt,rotate=90}]
 table[row sep=crcr, y error plus index=2, y error minus index=3]{%
3	92.795	152	37\\
6	459.7155	624	304\\
9	988.7515	1280	729\\
12	1614.389	2040	1218\\
15	2424.197	2930	1884\\
18	3219.723	3836	2565\\
21	3998.259	5012	3306\\
};
\addplot [color=mycolor1, dashed, line width=2pt, forget plot]
  table[row sep=crcr]{%
3	51.447976190482\\
6	494.55767857143\\
9	1027.16425\\
12	1649.26769047619\\
15	2360.868\\
18	3161.96517857143\\
21	4052.55922619048\\
};
\end{axis}
\end{tikzpicture}%
			}
			\caption{}
			\label{subfig:n_0=21 n=18}
		\end{subfigure}
		\caption{The completion time on an $n \times m$ grid vs. $m$ (lane number)
		($n = 18$, (\subref{subfig:n_0=3 n=18}) $N_0 = 3$ and (\subref{subfig:n_0=21 n=18}) $N_0 = 21$ ).
		Minimum and maximum times are depicted along the averages
		as on \cref{fig:simulations.m}. The best fit second degree polynomials 
		(\subref{subfig:n_0=3 n=18}) $T_{est} = \frac{11.637}{N_0} \cdot n \cdot m^2 + O(m)$ and
		(\subref{subfig:n_0=21 n=18}) $T_{est} = \frac{5.796}{N_0} \cdot n \cdot m^2 + O(m)$ are depicted 
		by a dash line. 
		Loosely dotted line is the time upper bound $T_{ub} = \dfrac{24}{N_0} \cdot n \cdot m^2 + O(m)$ 
		(see \eqref{equation:TimeUpperBound}). }
		\label{fig:simulations.mm}
	\end{figure}
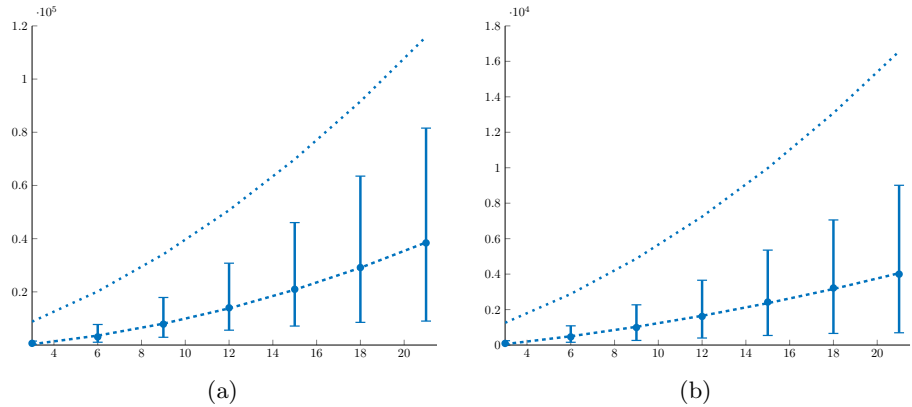

	\begin{figure}
		\centering
		\scalebox{0.85}{
%
%
\definecolor{mycolor1}{rgb}{0.00000,0.44700,0.74100}%
\definecolor{mycolor2}{rgb}{0.92900,0.69400,0.12500}%
\definecolor{mycolor3}{rgb}{0.46600,0.67400,0.18800}%
\definecolor{mycolor4}{rgb}{0.63500,0.07800,0.18400}%
\definecolor{mycolor5}{rgb}{0.85000,0.32500,0.09800}%
\begin{tikzpicture}

\begin{axis}[%
width=4.521in,
height=3.566in,
at={(0.758in,0.481in)},
scale only axis,
xmin=6,
xmax=15,
xlabel style={font=\color{white!15!black}},
ymin=0,
ymax=1800,
ylabel style={font=\color{white!15!black}},
axis background/.style={fill=white},
title style={font=\bfseries},
axis x line*=bottom,
axis y line*=left,
legend style={at={(0.03,0.97)}, anchor=north west, legend cell align=left, align=left, draw=white!15!black}
]
\addplot [color=mycolor1, draw=none, mark=*, mark options={solid, mycolor1}]
  table[row sep=crcr]{%
6	39.87\\
7	54.6\\
8	71.374\\
9	90.742\\
10	113.546\\
11	136.656\\
12	163.044\\
13	190.74\\
14	224.654\\
15	255.43\\
};
\addlegendentry{$\text{m = 3; T}_{\text{est}}\text{ = 0.386 }\cdot \text{m} \cdot \text{n}^\text{2}\text{ + O(n)}$}

\addplot [color=mycolor1, dashed, forget plot]
  table[row sep=crcr]{%
6	39.7778545454546\\
7	54.548303030303\\
8	71.6371454545454\\
9	91.0443818181818\\
10	112.770012121212\\
11	136.814036363636\\
12	163.176454545455\\
13	191.857266666667\\
14	222.856472727273\\
15	256.174072727273\\
};
\addplot [color=mycolor2, draw=none, mark=square, mark options={solid, mycolor2}]
  table[row sep=crcr]{%
6	86.474\\
7	117.85\\
8	154.456\\
9	195.29\\
10	242.642\\
11	294.888\\
12	354.16\\
13	416.606\\
14	485.398\\
15	555.786\\
};
\addlegendentry{$\text{m = 4; T}_{\text{est}}\text{ = 0.654 }\cdot \text{m} \cdot \text{n}^\text{2}\text{ + O(n)}$}

\addplot [color=mycolor2, dashed, forget plot]
  table[row sep=crcr]{%
6	86.2083636363648\\
7	117.624303030303\\
8	154.271484848485\\
9	196.149909090909\\
10	243.259575757575\\
11	295.600484848484\\
12	353.172636363636\\
13	415.97603030303\\
14	484.010666666667\\
15	557.276545454546\\
};
\addplot [color=mycolor3, draw=none, mark=oplus, mark options={solid, mycolor3}]
  table[row sep=crcr]{%
6	161.644\\
7	221.38\\
8	288.6\\
9	362.958\\
10	444.712\\
11	533.07\\
12	636.62\\
13	741.686\\
14	854.076\\
15	975.69\\
};
\addlegendentry{$\text{m = 5; T}_{\text{est}}\text{ = 0.784 }\cdot \text{m} \cdot \text{n}^\text{2}\text{ + O(n)}$}

\addplot [color=mycolor3, dashed, forget plot]
  table[row sep=crcr]{%
6	161.895054545448\\
7	221.023660606058\\
8	287.991418181819\\
9	362.79832727273\\
10	445.444387878792\\
11	535.929600000004\\
12	634.253963636367\\
13	740.41747878788\\
14	854.420145454544\\
15	976.261963636359\\
};
\addplot [color=mycolor4, draw=none, mark=diamond*, mark options={solid, mycolor4}]
  table[row sep=crcr]{%
6	256.79\\
7	330.746\\
8	417.06\\
9	522.352\\
10	635.976\\
11	762.868\\
12	900.878\\
13	1052.474\\
14	1214.118\\
15	1383.794\\
};
\addlegendentry{$\text{m = 6; T}_{\text{est}}\text{ = 1.009 }\cdot \text{m} \cdot \text{n}^\text{2}\text{ + O(n)}$}

\addplot [color=mycolor4, dashed, forget plot]
  table[row sep=crcr]{%
6	254.020327272734\\
7	331.432569696972\\
8	420.95566060606\\
9	522.589599999997\\
10	636.334387878784\\
11	762.19002424242\\
12	900.156509090906\\
13	1050.23384242424\\
14	1212.42202424243\\
15	1386.72105454546\\
};
\addplot [color=mycolor5, draw=none, mark=halfsquare left*, mark options={solid, mycolor5}]
  table[row sep=crcr]{%
6	377.246\\
7	472.358\\
8	581.354\\
9	704.34\\
10	833.164\\
11	982.628\\
12	1150.17\\
13	1323.426\\
14	1523.172\\
15	1732.586\\
};
\addlegendentry{$\text{m = 7; T}_{\text{est}}\text{ = 1.038 }\cdot \text{m} \cdot \text{n}^\text{2}\text{ + O(n)}$}

\addplot [color=mycolor5, dashed, forget plot]
  table[row sep=crcr]{%
6	380.223072727265\\
7	472.09096363636\\
8	578.493506060606\\
9	699.430700000003\\
10	834.902545454549\\
11	984.909042424247\\
12	1149.45019090909\\
13	1328.52599090909\\
14	1522.13644242424\\
15	1730.28154545454\\
};
\end{axis}
\end{tikzpicture}%
		}
		\caption{The average completion time vs. $n$ (row number) for $N_0 = 5$. 
		Dashed lines are the best fit second degree polynomial for each data series.}
		\label{fig:simulations.n}
	\end{figure}
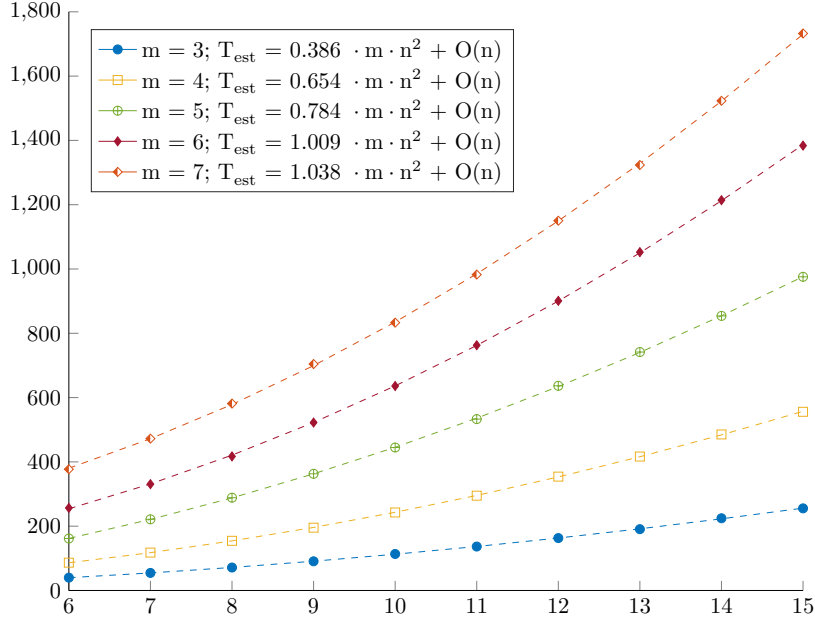
	

	In \cref{fig:simulations.m} and \cref{fig:simulations.n} we have measured the time against the
	number of grid columns and rows respectfully. In the previous section we have claimed that the time
	to solution is upper-bounded by:
	\begin{equation*}
		T \leq \frac{3m + n + 2N_1}{N_0}\cdot 8mn,
	\end{equation*}
	i.e.
	\begin{equation}
		T \leq \frac{1}{N_0} \cdot \left(24n\cdot m^2+8m\cdot n^2+16mn \cdot N_1\right).
		\label{equation:TimeUpperBound}
	\end{equation}
	Therefore, we have fitted a second degree polynomial line (dashed lines on \cref{fig:simulations.m} 
	and \cref{fig:simulations.n}). Both parameter graphs nicely support our findings from the
	above section.
	
	On the other hand, our time upper bound estimates seem way too cautious. We provide a number of plausible 
	explanations below:
	\begin{itemize}
		\item
			Exiting vehicles in the neighboring internal columns rush the same empty space without
			that empty space \enquote{finishing} double traversal of an internal column, i.e. in less time than
			needed for double traversal completion.
		\item
			A pair of empty space \enquote{moving} in close proximity could \enquote{save} each other half 
			of double traversals.
			Once the first empty space \enquote{enters} the column to double \enquote{pass} it, the second 
			empty space \enquote{skips} that same column. However, the second empty space in the pair 
			\enquote{traverses} the next right column twice. Then roles switch once again. Thus, in practice, 
			a single empty space in a pair \enquote{traverses} every second column twice, but also \enquote{skips}
			every second column. Therefore, we should expect $m$ - $n$-long traversals instead of $2m$. 
			Increasing the empty space numbers increases the chances for such \enquote{closely running} 
			empty space pair/triplet etc. The effect become prominent after all the first algorithm stage,
			i.e. after the exiting vehicles exodus from the internal columns.
			
			Indeed, in relative terms, \cref{subfig:n_0=21 n=18} featuring $N_0 = 21$ 
			(large number of empty spaces) is almost 
			twice faster than \cref{subfig:n_0=3 n=18} showing results corresponding to $N_0 = 3$ 
			(small number of empty spaces).
		\item
			Due to the practical limitations, the number of simulations that we have executed was not high
			enough to cover enough \enquote{bad} initial configurations. Thus the set of problem parameters 
			we have chosen to test, led to better times than we should have expect.
	\end{itemize}
	
	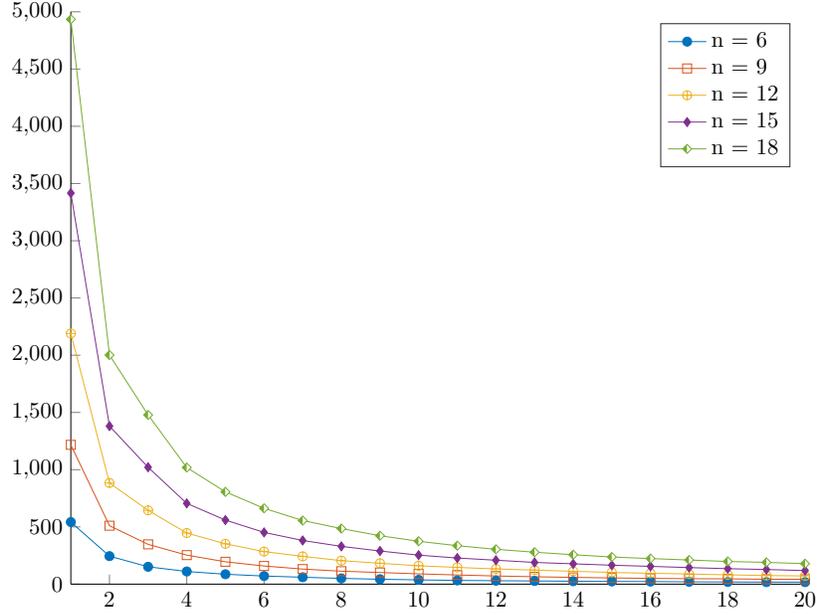
\begin{figure}
		\centering
		\scalebox{0.85}{
%
%
\definecolor{mycolor1}{rgb}{0.00000,0.44700,0.74100}%
\definecolor{mycolor2}{rgb}{0.85000,0.32500,0.09800}%
\definecolor{mycolor3}{rgb}{0.92900,0.69400,0.12500}%
\definecolor{mycolor4}{rgb}{0.49400,0.18400,0.55600}%
\definecolor{mycolor5}{rgb}{0.46600,0.67400,0.18800}%
\begin{tikzpicture}

\begin{axis}[%
width=4.521in,
height=3.527in,
at={(0.758in,0.519in)},
scale only axis,
xmin=1,
xmax=20,
xlabel style={font=\color{white!15!black}},
ymin=0,
ymax=5000,
ylabel style={font=\color{white!15!black}},
axis background/.style={fill=white},
title style={font=\bfseries},
axis x line*=bottom,
axis y line*=left,
legend style={legend cell align=left, align=left, draw=white!15!black}
]
\addplot [color=mycolor1, mark=*]
  table[row sep=crcr]{%
1	542.284\\
2	244.662\\
3	151.69\\
4	110.658\\
5	86.54\\
6	70.822\\
7	60.312\\
8	50.15\\
9	42.854\\
10	37.27\\
11	32.474\\
12	29.932\\
13	27.746\\
14	25.436\\
15	23.57\\
16	22.474\\
17	20.236\\
18	18.766\\
19	17.396\\
20	17.428\\
};
\addlegendentry{n =  6}

\addplot [color=mycolor2, mark=square]
  table[row sep=crcr]{%
1	1216.918\\
2	510.594\\
3	347.502\\
4	253.512\\
5	195.372\\
6	158.51\\
7	131.702\\
8	113.078\\
9	101.696\\
10	90.038\\
11	80.108\\
12	71.374\\
13	64.486\\
14	58.926\\
15	54.28\\
16	50.298\\
17	47.39\\
18	46.22\\
19	42.778\\
20	42.268\\
};
\addlegendentry{n =  9}

\addplot [color=mycolor3, mark=oplus]
  table[row sep=crcr]{%
1	2189.89\\
2	883.856\\
3	644.514\\
4	446.526\\
5	352.494\\
6	284.566\\
7	242.686\\
8	205.386\\
9	182.358\\
10	160.584\\
11	146.08\\
12	132.904\\
13	122.696\\
14	112.134\\
15	102.302\\
16	94.7\\
17	88.546\\
18	81.26\\
19	75.444\\
20	71.766\\
};
\addlegendentry{n = 12}

\addplot [color=mycolor4, mark=diamond*]
  table[row sep=crcr]{%
1	3416.102\\
2	1380.54\\
3	1020.878\\
4	705.474\\
5	559.888\\
6	451.902\\
7	381.442\\
8	330\\
9	289.546\\
10	253.454\\
11	228.256\\
12	209.09\\
13	189.626\\
14	177.558\\
15	165.478\\
16	155.166\\
17	144.684\\
18	135.096\\
19	127.962\\
20	118.554\\
};
\addlegendentry{n = 15}

\addplot [color=mycolor5, mark=halfsquare left*]
  table[row sep=crcr]{%
1	4934.642\\
2	2001.442\\
3	1478.09\\
4	1018.862\\
5	806.188\\
6	662.42\\
7	555.694\\
8	485.338\\
9	422.96\\
10	374.482\\
11	335.79\\
12	304.394\\
13	278.876\\
14	257.018\\
15	237.106\\
16	223.422\\
17	210.852\\
18	197.942\\
19	190.152\\
20	178.806\\
};
\addlegendentry{n = 18}

\end{axis}
\end{tikzpicture}%
		}
		\caption{An average completion time vs. an empty space number ($N_0$) on $n \times 4$ grids.}
		\label{fig:Normal.K.1}
	\end{figure}
	\begin{figure}
		\centering
		\scalebox{0.85}{
%
%
\definecolor{mycolor1}{rgb}{0.00000,0.44700,0.74100}%
\definecolor{mycolor2}{rgb}{0.85000,0.32500,0.09800}%
\definecolor{mycolor3}{rgb}{0.92900,0.69400,0.12500}%
\definecolor{mycolor4}{rgb}{0.49400,0.18400,0.55600}%
\definecolor{mycolor5}{rgb}{0.46600,0.67400,0.18800}%
\begin{tikzpicture}

\begin{axis}[%
width=4.521in,
height=3.496in,
at={(0.758in,0.551in)},
scale only axis,
xmode=log,
xmin=1,
xmax=20,
xminorticks=true,
xlabel style={font=\color{white!15!black}},
ymode=log,
ymin=10,
ymax=10000,
yminorticks=true,
ylabel style={font=\color{white!15!black}},
axis background/.style={fill=white},
title style={font=\bfseries},
axis x line*=bottom,
axis y line*=left,
legend style={legend cell align=left, align=left, draw=white!15!black}
]
\addplot [color=mycolor1, draw=none, mark size=1.4pt, mark=*, mark options={solid, mycolor1}]
 plot [error bars/.cd, y dir = both, y explicit]
 table[row sep=crcr, y error plus index=2, y error minus index=3]{%
1	542.284	91.8059475634124	91.8059475634124\\
2	244.662	46.5257029831485	46.5257029831485\\
3	151.69	29.0484685836295	29.0484685836295\\
4	110.658	21.6869935070482	21.6869935070482\\
5	86.54	18.1984275565518	18.1984275565518\\
6	70.822	14.4302036560511	14.4302036560511\\
7	60.312	13.899966695942	13.899966695942\\
8	50.15	13.966302847585	13.966302847585\\
9	42.854	12.5302313384386	12.5302313384386\\
10	37.27	10.8862353566548	10.8862353566548\\
11	32.474	9.28664673685989	9.28664673685989\\
12	29.932	8.40479965939874	8.40479965939874\\
13	27.746	7.87461129237578	7.87461129237578\\
14	25.436	7.42834197712102	7.42834197712102\\
15	23.57	7.71077246610534	7.71077246610534\\
16	22.474	6.99794730853742	6.99794730853742\\
17	20.236	6.02834852774323	6.02834852774323\\
18	18.766	4.9274540695185	4.9274540695185\\
19	17.396	3.61293379141693	3.61293379141693\\
20	17.428	3.74872269495722	3.74872269495722\\
};
\addlegendentry{$\text{n =  6; }\beta\text{ = -1.170}$}

\addplot [color=mycolor1, forget plot]
  table[row sep=crcr]{%
1	540.813933127647\\
2	240.348046091738\\
3	149.558998025173\\
4	106.815264403481\\
5	82.2712133630353\\
6	66.4668765889655\\
7	55.4979628609517\\
8	47.4707445935744\\
9	41.3596849491935\\
10	36.5628993081708\\
11	32.7047538575175\\
12	29.5391500466537\\
13	26.8983771502206\\
14	24.6643551850822\\
15	22.7516332017034\\
16	21.0969059979684\\
17	19.6523150924083\\
18	18.3810343180678\\
19	17.2542820735733\\
20	16.2492511192267\\
};
\addplot [color=mycolor2, draw=none, mark size=1.4pt, mark=square, mark options={solid, mycolor2}]
 plot [error bars/.cd, y dir = both, y explicit]
 table[row sep=crcr, y error plus index=2, y error minus index=3]{%
1	1216.918	157.021255329495	157.021255329495\\
2	510.594	72.7084741169613	72.7084741169613\\
3	347.502	52.111622873184	52.111622873184\\
4	253.512	35.4519327409683	35.4519327409683\\
5	195.372	27.300260519799	27.300260519799\\
6	158.51	24.9037807690557	24.9037807690557\\
7	131.702	20.3927665080465	20.3927665080465\\
8	113.078	19.2342358219741	19.2342358219741\\
9	101.696	17.0216055091233	17.0216055091233\\
10	90.038	16.6114549896675	16.6114549896675\\
11	80.108	16.7870171229964	16.7870171229964\\
12	71.374	15.6792155288203	15.6792155288203\\
13	64.486	14.5769262039399	14.5769262039399\\
14	58.926	12.4998005595312	12.4998005595312\\
15	54.28	12.3905064181654	12.3905064181654\\
16	50.298	11.8788640375098	11.8788640375098\\
17	47.39	11.1093084153305	11.1093084153305\\
18	46.22	10.2392032192515	10.2392032192515\\
19	42.778	10.2592016471494	10.2592016471494\\
20	42.268	9.85568084354293	9.85568084354293\\
};
\addlegendentry{$\text{n =  9; }\beta\text{ = -1.137}$}

\addplot [color=mycolor2, forget plot]
  table[row sep=crcr]{%
1	1186.33165009797\\
2	539.392663787593\\
3	340.150989462989\\
4	245.247141239001\\
5	190.286342473329\\
6	154.657382934422\\
7	129.791272545265\\
8	111.507190074776\\
9	97.5298059552697\\
10	86.5180130198924\\
11	77.6316705687329\\
12	70.3184963062616\\
13	64.2009713257098\\
14	59.0125537228902\\
15	54.5598590986344\\
16	50.6992798185359\\
17	47.3220080480403\\
18	44.3441442606335\\
19	41.6999884037533\\
20	39.3373821768604\\
};
\addplot [color=mycolor3, draw=none, mark size=1.4pt, mark=oplus, mark options={solid, mycolor3}]
 plot [error bars/.cd, y dir = both, y explicit]
 table[row sep=crcr, y error plus index=2, y error minus index=3]{%
1	2189.89	216.129604431747	216.129604431747\\
2	883.856	100.44922776566	100.44922776566\\
3	644.514	77.2202028192551	77.2202028192551\\
4	446.526	53.7597017762389	53.7597017762389\\
5	352.494	43.3005755583846	43.3005755583846\\
6	284.566	36.9814128149206	36.9814128149206\\
7	242.686	30.7523465386834	30.7523465386834\\
8	205.386	27.2892361347407	27.2892361347407\\
9	182.358	25.6789535087903	25.6789535087903\\
10	160.584	22.8388741479235	22.8388741479235\\
11	146.08	19.7070122714056	19.7070122714056\\
12	132.904	17.682650713978	17.682650713978\\
13	122.696	19.1003988039239	19.1003988039239\\
14	112.134	18.1512252466855	18.1512252466855\\
15	102.302	18.8195416068666	18.8195416068666\\
16	94.7	17.9355895008077	17.9355895008077\\
17	88.546	17.3104357649582	17.3104357649582\\
18	81.26	15.1685387937756	15.1685387937756\\
19	75.444	14.1947118212162	14.1947118212162\\
20	71.766	13.7045566782547	13.7045566782547\\
};
\addlegendentry{$\text{n = 12; }\beta\text{ = -1.124}$}

\addplot [color=mycolor3, forget plot]
  table[row sep=crcr]{%
1	2113.34825126404\\
2	969.462597019813\\
3	614.548754994372\\
4	444.72449179081\\
5	346.049080528073\\
6	281.913798000772\\
7	237.055392521466\\
8	204.009803169902\\
9	178.707021920897\\
10	158.744135096714\\
11	142.613617075689\\
12	129.323164122182\\
13	118.193668134444\\
14	108.745132911224\\
15	100.628957616557\\
16	93.5860303574185\\
17	87.4198589147459\\
18	81.9788094430175\\
19	77.1440501155891\\
20	72.8211743523458\\
};
\addplot [color=mycolor4, draw=none, mark size=1.4pt, mark=diamond*, mark options={solid, mycolor4}]
 plot [error bars/.cd, y dir = both, y explicit]
 table[row sep=crcr, y error plus index=2, y error minus index=3]{%
1	3416.102	349.98500339472	349.98500339472\\
2	1380.54	140.347151898755	140.347151898755\\
3	1020.878	114.191060563536	114.191060563536\\
4	705.474	67.3897029829949	67.3897029829949\\
5	559.888	58.5761537451475	58.5761537451475\\
6	451.902	50.2093851856547	50.2093851856547\\
7	381.442	43.0644136549761	43.0644136549761\\
8	330	36.8861651348675	36.8861651348675\\
9	289.546	33.7797280043534	33.7797280043534\\
10	253.454	30.7002425688719	30.7002425688719\\
11	228.256	27.120869217158	27.120869217158\\
12	209.09	25.1282462085007	25.1282462085007\\
13	189.626	22.484510901083	22.484510901083\\
14	177.558	21.8556623315768	21.8556623315768\\
15	165.478	22.3760338259388	22.3760338259388\\
16	155.166	22.2374335502504	22.2374335502504\\
17	144.684	21.1134957647191	21.1134957647191\\
18	135.096	19.4829657883964	19.4829657883964\\
19	127.962	18.9997509740035	18.9997509740035\\
20	118.554	20.6523636479822	20.6523636479822\\
};
\addlegendentry{$\text{n = 15; }\beta\text{ = -1.106}$}

\addplot [color=mycolor4, forget plot]
  table[row sep=crcr]{%
1	3252.42921063111\\
2	1511.0371623061\\
3	964.988185776635\\
4	702.008608952027\\
5	548.48157855467\\
6	448.321213303795\\
7	378.048551161983\\
8	326.144253322424\\
9	286.309751383579\\
10	254.817551548063\\
11	229.324216279135\\
12	208.284322296055\\
13	190.638423927574\\
14	175.636539019674\\
15	162.733209285952\\
16	151.52246371171\\
17	141.696062694748\\
18	133.015861761757\\
19	125.295030138358\\
20	118.384987054725\\
};
\addplot [color=mycolor5, draw=none, mark size=1.4pt, mark=halfsquare left*, mark options={solid, mycolor5}]
 plot [error bars/.cd, y dir = both, y explicit]
 table[row sep=crcr, y error plus index=2, y error minus index=3]{%
1	4934.642	451.698630883784	451.698630883784\\
2	2001.442	209.162732659207	209.162732659207\\
3	1478.09	150.842476794871	150.842476794871\\
4	1018.862	89.0562349415465	89.0562349415465\\
5	806.188	86.4307422298617	86.4307422298617\\
6	662.42	68.2862917082963	68.2862917082963\\
7	555.694	56.1539214105263	56.1539214105263\\
8	485.338	50.3757002765888	50.3757002765888\\
9	422.96	41.7585194648444	41.7585194648444\\
10	374.482	39.8994939219843	39.8994939219843\\
11	335.79	35.7023137964592	35.7023137964592\\
12	304.394	34.7408782451233	34.7408782451233\\
13	278.876	28.6934007481334	28.6934007481334\\
14	257.018	27.7228929512628	27.7228929512628\\
15	237.106	26.3497130678403	26.3497130678403\\
16	223.422	24.1358030730274	24.1358030730274\\
17	210.852	23.048917571166	23.048917571166\\
18	197.942	21.9654181656161	21.9654181656161\\
19	190.152	22.9444937967954	22.9444937967954\\
20	178.806	21.6778278533824	21.6778278533824\\
};
\addlegendentry{$\text{n = 18; }\beta\text{ = -1.101}$}

\addplot [color=mycolor5, forget plot]
  table[row sep=crcr]{%
1	4714.12020095943\\
2	2197.43949891465\\
3	1406.08436064614\\
4	1024.31421888808\\
5	801.160035613185\\
6	655.432017253427\\
7	553.105763748629\\
8	477.47372318306\\
9	419.393894294692\\
10	373.452655460499\\
11	336.244641156616\\
12	305.523012177938\\
13	279.746788110904\\
14	257.824663039188\\
15	238.962637444233\\
16	222.569551536415\\
17	208.196412745075\\
18	195.496226154613\\
19	184.196687144327\\
20	174.081181874923\\
};
\end{axis}
\end{tikzpicture}%
		}
		\caption{An average completion time vs. an empty space number ($N_0$) on $n \times 4$ grids. 
		In a log X-log Y scale. The best fit line slope $\beta$ is shown.}
		\label{fig:LogLog.K.1}
	\end{figure}
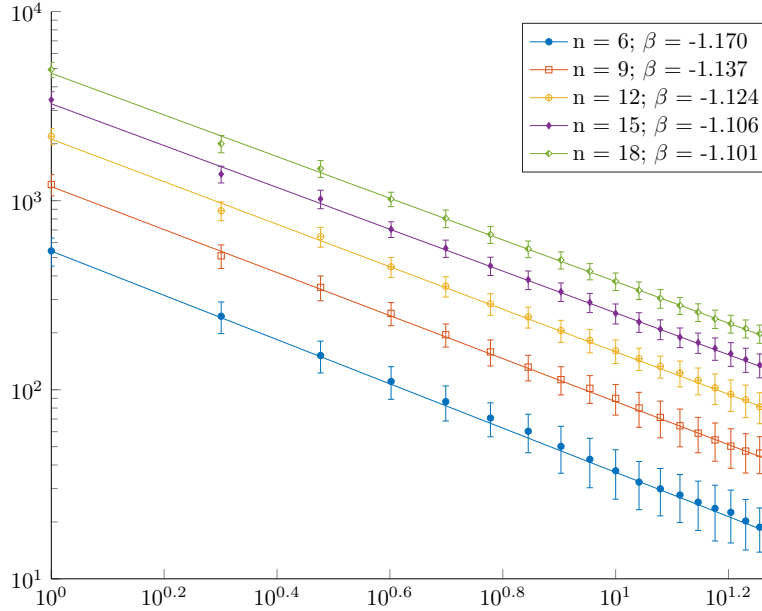
	
	\noindent
	We have assumed that 
	\begin{equation*}
		T(N_0) = c \;\cdot \frac{1}{N_0}.
	\end{equation*}
	Therefore, a dependency relation between $N_0$ and $T$ in the Log-Log plane should be a line with a 
	slope equal to $-1$, i.e. taking logarithms on both sides produces
	\begin{equation*}
		\log T(N_0) = -1 \cdot \log N_0 + \log c .
	\end{equation*}
	And after standard algebraic transformations
	\begin{equation*}
		\frac{\log T - \log c}{\log N_0} = -1
	\end{equation*}
	
	This reciprocity can be observed on \cref{fig:LogLog.K.1} (the slope $\beta$ is depicted 
	in the legend).
	
	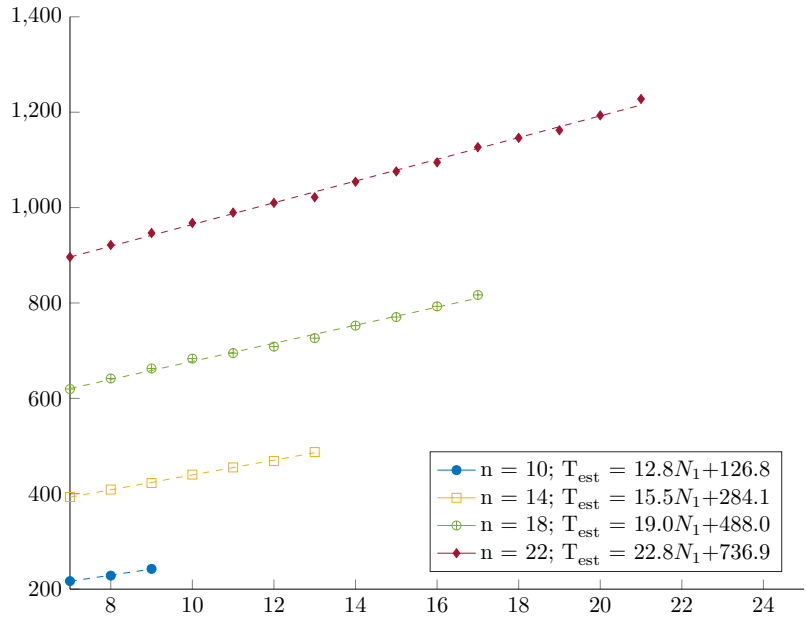
\begin{figure}
		\centering
		\scalebox{0.85}{
%
%
\definecolor{mycolor1}{rgb}{0.00000,0.44700,0.74100}%
\definecolor{mycolor2}{rgb}{0.92900,0.69400,0.12500}%
\definecolor{mycolor3}{rgb}{0.46600,0.67400,0.18800}%
\definecolor{mycolor4}{rgb}{0.63500,0.07800,0.18400}%
\begin{tikzpicture}

\begin{axis}[%
width=4.521in,
height=3.527in,
at={(0.758in,0.519in)},
scale only axis,
xmin=7,
xmax=25,
xlabel style={font=\color{white!15!black}},
ymin=200,
ymax=1400,
ylabel style={font=\color{white!15!black}},
axis background/.style={fill=white},
title style={font=\bfseries},
axis x line*=bottom,
axis y line*=left,
legend style={at={(0.97,0.03)}, anchor=south east, legend cell align=left, align=left, draw=white!15!black}
]
\addplot [color=mycolor1, draw=none, mark=*, mark options={solid, mycolor1}]
  table[row sep=crcr]{%
1	82.842\\
2	135.686\\
3	161.58\\
4	177.16\\
5	188.936\\
6	204.384\\
7	216.9\\
8	228.38\\
9	242.516\\
};
\addlegendentry{$\text{n = 10; T}_{\text{est}}\text{ = 12.8}N_1{ + 126.8}$}

\addplot [color=mycolor1, dashed, forget plot]
  table[row sep=crcr]{%
7	216.457333333333\\
8	229.265333333333\\
9	242.073333333333\\
};
\addplot [color=mycolor2, draw=none, mark=square, mark options={solid, mycolor2}]
  table[row sep=crcr]{%
1	181.69\\
2	255.99\\
3	298.506\\
4	333.202\\
5	357.77\\
6	376.562\\
7	393.148\\
8	408.84\\
9	422.608\\
10	440.154\\
11	455.224\\
12	468.552\\
13	487.432\\
};
\addlegendentry{$\text{n = 14; T}_{\text{est}}\text{ = 15.5}N_1{ + 284.1}$}

\addplot [color=mycolor2, dashed, forget plot]
  table[row sep=crcr]{%
7	392.827\\
8	408.358857142857\\
9	423.890714285714\\
10	439.422571428571\\
11	454.954428571429\\
12	470.486285714286\\
13	486.018142857143\\
};
\addplot [color=mycolor3, draw=none, mark=oplus, mark options={solid, mycolor3}]
  table[row sep=crcr]{%
1	271.33\\
2	402.53\\
3	493.126\\
4	540.488\\
5	567.678\\
6	590.266\\
7	619.5\\
8	641.7\\
9	662.688\\
10	683.362\\
11	694.83\\
12	708.528\\
13	726.296\\
14	752.584\\
15	770.5\\
16	792.914\\
17	816.864\\
};
\addlegendentry{$\text{n = 18; T}_{\text{est}}\text{ = 19.0}N_1{ + 488.0}$}

\addplot [color=mycolor3, dashed, forget plot]
  table[row sep=crcr]{%
7	620.659545454546\\
8	639.614290909091\\
9	658.569036363637\\
10	677.523781818182\\
11	696.478527272727\\
12	715.433272727273\\
13	734.388018181818\\
14	753.342763636364\\
15	772.297509090909\\
16	791.252254545454\\
17	810.207\\
};
\addplot [color=mycolor4, draw=none, mark=diamond*, mark options={solid, mycolor4}]
  table[row sep=crcr]{%
1	401.958\\
2	589.706\\
3	720.312\\
4	782.38\\
5	839.844\\
6	848.182\\
7	896.422\\
8	921.57\\
9	946.808\\
10	967.712\\
11	989.596\\
12	1009.926\\
13	1021.77\\
14	1054.156\\
15	1075.84\\
16	1094.902\\
17	1126.682\\
18	1146.16\\
19	1162.248\\
20	1193.368\\
21	1227.686\\
};
\addlegendentry{$\text{n = 22; T}_{\text{est}}\text{ = 22.8}N_1{ + 736.9}$}

\addplot [color=mycolor4, dashed, forget plot]
  table[row sep=crcr]{%
7	896.258699999999\\
8	919.029799999999\\
9	941.800899999999\\
10	964.571999999999\\
11	987.343099999999\\
12	1010.1142\\
13	1032.8853\\
14	1055.6564\\
15	1078.4275\\
16	1101.1986\\
17	1123.9697\\
18	1146.7408\\
19	1169.5119\\
20	1192.283\\
21	1215.0541\\
};
\end{axis}
\end{tikzpicture}%
		}
		\caption{An average completion time vs. $N_1$, an exiting agent number on $n \times 4$ grids. 
		($N_0 = 5$).}
			\label{fig:ones.N1.10}
	\end{figure}
	
	The last parameter that we have tested is $N_1$ - the exiting agent number. We have predicted,
	that a completion time is linear in $N_1$. Our findings support nicely the theoretical analysis
	(see in \cref{fig:ones.N1.10}). However, the Algorithm has been proven to complete the task in a limited 
	range of $N_1$ values (namely for $N_1 \in \{1, \ldots, n-1\}$).
		
\begin{figure}
		\centering
		\scalebox{0.85}{
%
%
\definecolor{mycolor1}{rgb}{0.00000,0.44700,0.74100}%
\definecolor{mycolor2}{rgb}{0.92900,0.69400,0.12500}%
\definecolor{mycolor3}{rgb}{0.46600,0.67400,0.18800}%
\definecolor{mycolor4}{rgb}{0.63500,0.07800,0.18400}%
\definecolor{mycolor5}{rgb}{0.85000,0.32500,0.09800}%
\begin{tikzpicture}

\begin{axis}[%
width=4.521in,
height=3.566in,
at={(0.758in,0.481in)},
scale only axis,
xmin=4,
xmax=80,
xlabel style={font=\color{white!15!black}},
ymin=0,
ymax=600,
ylabel style={font=\color{white!15!black}},
axis background/.style={fill=white},
title style={font=\bfseries},
axis x line*=bottom,
axis y line*=left,
legend style={at={(0.03,0.97)}, anchor=north west, legend cell align=left, align=left, draw=white!15!black}
]
\addplot [color=mycolor1, draw=none, mark=*, mark options={solid, mycolor1}]
  table[row sep=crcr]{%
4	16.632\\
8	34.34\\
12	55.014\\
16	73.09\\
20	90.776\\
24	112.036\\
28	132.518\\
32	153.51\\
36	171.666\\
40	190.648\\
44	210.138\\
48	231.05\\
52	252.108\\
56	271.542\\
60	292.424\\
64	311.9\\
68	330.128\\
72	350.298\\
76	369\\
80	390.754\\
};
\addlegendentry{$\text{m = 4; T}_{\text{est}}\text{ = 4.9n + -5.6}$}

\addplot [color=mycolor1, dashed, forget plot]
  table[row sep=crcr]{%
4	14.2075428571428\\
8	33.9729172932331\\
12	53.7382917293233\\
16	73.5036661654135\\
20	93.2690406015037\\
24	113.034415037594\\
28	132.799789473684\\
32	152.565163909774\\
36	172.330538345865\\
40	192.095912781955\\
44	211.861287218045\\
48	231.626661654135\\
52	251.392036090226\\
56	271.157410526316\\
60	290.922784962406\\
64	310.688159398496\\
68	330.453533834586\\
72	350.218908270677\\
76	369.984282706767\\
80	389.749657142857\\
};
\addplot [color=mycolor2, draw=none, mark=square, mark options={solid, mycolor2}]
  table[row sep=crcr]{%
4	21.752\\
8	41.802\\
12	63.336\\
16	86.266\\
20	106.41\\
24	127.396\\
28	150.634\\
32	173.048\\
36	195.992\\
40	217.968\\
44	238.88\\
48	261.824\\
52	284.88\\
56	306.572\\
60	329.674\\
64	352.514\\
68	372.778\\
72	398.458\\
76	420.834\\
80	442.106\\
};
\addlegendentry{$\text{m = 5; T}_{\text{est}}\text{ = 5.6n + -3.9}$}

\addplot [color=mycolor2, dashed, forget plot]
  table[row sep=crcr]{%
4	18.3809428571425\\
8	40.6204436090223\\
12	62.859944360902\\
16	85.0994451127817\\
20	107.338945864661\\
24	129.578446616541\\
28	151.817947368421\\
32	174.057448120301\\
36	196.29694887218\\
40	218.53644962406\\
44	240.77595037594\\
48	263.01545112782\\
52	285.254951879699\\
56	307.494452631579\\
60	329.733953383459\\
64	351.973454135338\\
68	374.212954887218\\
72	396.452455639098\\
76	418.691956390978\\
80	440.931457142857\\
};
\addplot [color=mycolor3, draw=none, mark=oplus, mark options={solid, mycolor3}]
  table[row sep=crcr]{%
4	28.54\\
8	49.406\\
12	72.106\\
16	95.55\\
20	117.462\\
24	142.914\\
28	165.852\\
32	189.414\\
36	215.37\\
40	239.978\\
44	261.532\\
48	286.492\\
52	309.896\\
56	335.27\\
60	356.482\\
64	381.54\\
68	406.156\\
72	430.814\\
76	453.408\\
80	478.016\\
};
\addlegendentry{$\text{m = 6; T}_{\text{est}}\text{ = 6.0n +  0.7}$}

\addplot [color=mycolor3, dashed, forget plot]
  table[row sep=crcr]{%
4	24.5470285714286\\
8	48.3641729323309\\
12	72.1813172932331\\
16	95.9984616541354\\
20	119.815606015038\\
24	143.63275037594\\
28	167.449894736842\\
32	191.267039097744\\
36	215.084183458647\\
40	238.901327819549\\
44	262.718472180451\\
48	286.535616541353\\
52	310.352760902256\\
56	334.169905263158\\
60	357.98704962406\\
64	381.804193984962\\
68	405.621338345865\\
72	429.438482706767\\
76	453.255627067669\\
80	477.072771428571\\
};
\addplot [color=mycolor4, draw=none, mark=diamond*, mark options={solid, mycolor4}]
  table[row sep=crcr]{%
4	35.936\\
8	55.408\\
12	78.948\\
16	102.94\\
20	125.826\\
24	153.906\\
28	176.96\\
32	204.628\\
36	229.242\\
40	252.56\\
44	279.274\\
48	303.218\\
52	327.866\\
56	353.502\\
60	380.386\\
64	404.484\\
68	428.512\\
72	453.026\\
76	476.676\\
80	501.386\\
};
\addlegendentry{$\text{m = 7; T}_{\text{est}}\text{ = 6.2n +  5.3}$}

\addplot [color=mycolor4, dashed, forget plot]
  table[row sep=crcr]{%
4	30.1420857142854\\
8	54.9938872180448\\
12	79.8456887218043\\
16	104.697490225564\\
20	129.549291729323\\
24	154.401093233083\\
28	179.252894736842\\
32	204.104696240601\\
36	228.956497744361\\
40	253.80829924812\\
44	278.66010075188\\
48	303.511902255639\\
52	328.363703759399\\
56	353.215505263158\\
60	378.067306766917\\
64	402.919108270677\\
68	427.770909774436\\
72	452.622711278196\\
76	477.474512781955\\
80	502.326314285714\\
};
\addplot [color=mycolor5, draw=none, mark=halfsquare left*, mark options={solid, mycolor5}]
  table[row sep=crcr]{%
4	40.432\\
8	62.556\\
12	86.366\\
16	110.658\\
20	137.434\\
24	162.832\\
28	189.41\\
32	213.208\\
36	239.304\\
40	266.258\\
44	292.156\\
48	316.11\\
52	344.41\\
56	370.902\\
60	394.412\\
64	421.772\\
68	447.608\\
72	474.72\\
76	498.792\\
80	523.882\\
};
\addlegendentry{$\text{m = 8; T}_{\text{est}}\text{ = 6.4n +  9.6}$}

\addplot [color=mycolor5, dashed, forget plot]
  table[row sep=crcr]{%
4	35.3615999999998\\
8	61.0773368421051\\
12	86.7930736842104\\
16	112.508810526316\\
20	138.224547368421\\
24	163.940284210526\\
28	189.656021052632\\
32	215.371757894737\\
36	241.087494736842\\
40	266.803231578947\\
44	292.518968421053\\
48	318.234705263158\\
52	343.950442105263\\
56	369.666178947368\\
60	395.381915789474\\
64	421.097652631579\\
68	446.813389473684\\
72	472.52912631579\\
76	498.244863157895\\
80	523.9606\\
};
\end{axis}
\end{tikzpicture}%
		}
		\caption{An average completion time vs. $n$ (rows) on $n \times m$ grid with a constant ratio $\rho = 0.6$ of 
			empty spaces.}
		\label{fig:FreeRatio.N.14}
	\end{figure}
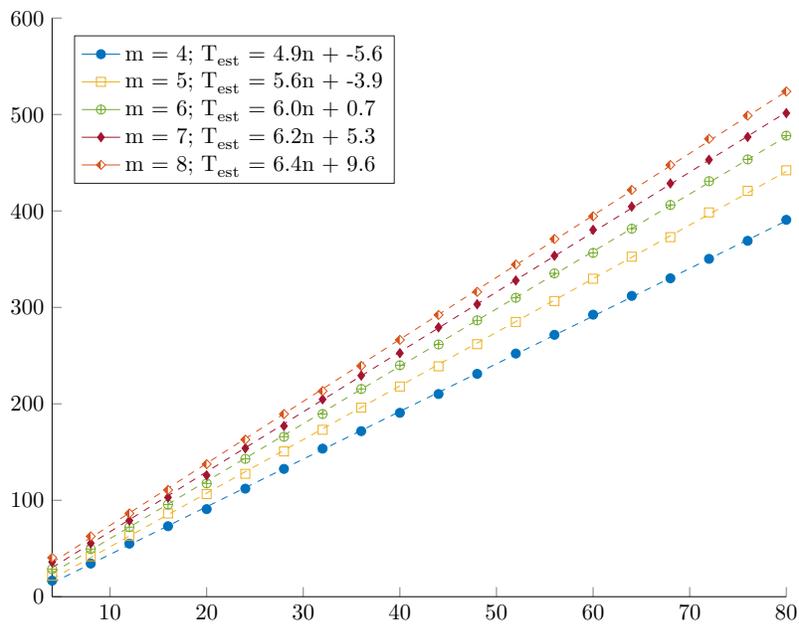

	An interesting question raises from our daily lives: \enquote{what is the algorithm behavior 
	not as a function of $m, n$ and $N_0$}, but rather \enquote{what is the algorithm behavior 
	on the road with a constant fraction of empty slots}? I.e. what happens when the road is
	half-empty, or the congestion is high?
	How well the time scales as a function of grid dimensions? We provide an answer in
	\cref{fig:FreeRatio.M.12} and \cref{fig:FreeRatio.N.14}. Denote the 
	ratio of empty spaces on grid as $\rho$, we have that $N_0 = \rho \cdot mn$, hence the upper bound 
	in this case is:
	\begin{equation*}
		T \leq \frac{3m + n + 2N_1}{\rho \cdot mn}\cdot 8mn.	
	\end{equation*}
	After simple algebraic reduction we get, that time is expected to be linearly dependent in $m$ and $n$
	for roads with a constant free spaces ratio.
	
	\begin{equation*}
		T \leq \frac{8}{\rho} (3m + n + 2N_1)
	\end{equation*}
	
	\begin{figure}
		\centering
		\scalebox{0.85}{
%
%
\definecolor{mycolor1}{rgb}{0.00000,0.44700,0.74100}%
\definecolor{mycolor2}{rgb}{0.92900,0.69400,0.12500}%
\definecolor{mycolor3}{rgb}{0.46600,0.67400,0.18800}%
\definecolor{mycolor4}{rgb}{0.63500,0.07800,0.18400}%
\begin{tikzpicture}

\begin{axis}[%
width=4.521in,
height=3.566in,
at={(0.758in,0.481in)},
scale only axis,
xmin=4,
xmax=80,
xlabel style={font=\color{white!15!black}},
ymin=0,
ymax=900,
ylabel style={font=\color{white!15!black}},
axis background/.style={fill=white},
title style={font=\bfseries},
axis x line*=bottom,
axis y line*=left,
legend style={at={(0.03,0.97)}, anchor=north west, legend cell align=left, align=left, draw=white!15!black}
]
\addplot [color=mycolor1, draw=none, mark=*, mark options={solid, mycolor1}]
  table[row sep=crcr]{%
4	16.99\\
8	40.544\\
12	69.742\\
16	92.58\\
20	113.342\\
24	139.644\\
28	155.592\\
32	182.296\\
36	199.912\\
40	216.792\\
44	236.878\\
48	259.022\\
52	281.12\\
56	298.344\\
60	311.04\\
64	330.848\\
68	341.184\\
72	367.072\\
76	385.168\\
80	391.784\\
};
\addlegendentry{$\text{n =  4; T}_{\text{est}}\text{ =  4.9m + 13.9}$}

\addplot [color=mycolor1, dashed, forget plot]
  table[row sep=crcr]{%
4	33.6559428571427\\
8	53.4284436090224\\
12	73.2009443609021\\
16	92.9734451127819\\
20	112.745945864662\\
24	132.518446616541\\
28	152.290947368421\\
32	172.063448120301\\
36	191.83594887218\\
40	211.60844962406\\
44	231.38095037594\\
48	251.15345112782\\
52	270.925951879699\\
56	290.698452631579\\
60	310.470953383459\\
64	330.243454135338\\
68	350.015954887218\\
72	369.788455639098\\
76	389.560956390978\\
80	409.333457142857\\
};
\addplot [color=mycolor2, draw=none, mark=square, mark options={solid, mycolor2}]
  table[row sep=crcr]{%
4	34.398\\
8	61.748\\
12	94.038\\
16	129.594\\
20	165.386\\
24	198.174\\
28	238.762\\
32	273.756\\
36	304.336\\
40	338.808\\
44	378.05\\
48	403.462\\
52	431.526\\
56	465.938\\
60	500.78\\
64	523.18\\
68	561.002\\
72	592.486\\
76	606.108\\
80	635.814\\
};
\addlegendentry{$\text{n =  8; T}_{\text{est}}\text{ =  8.1m +  6.3}$}

\addplot [color=mycolor2, dashed, forget plot]
  table[row sep=crcr]{%
4	38.7177428571427\\
8	71.1545383458645\\
12	103.591333834586\\
16	136.028129323308\\
20	168.46492481203\\
24	200.901720300752\\
28	233.338515789474\\
32	265.775311278195\\
36	298.212106766917\\
40	330.648902255639\\
44	363.085697744361\\
48	395.522493233083\\
52	427.959288721805\\
56	460.396084210526\\
60	492.832879699248\\
64	525.26967518797\\
68	557.706470676692\\
72	590.143266165414\\
76	622.580061654135\\
80	655.016857142857\\
};
\addplot [color=mycolor3, draw=none, mark=oplus, mark options={solid, mycolor3}]
  table[row sep=crcr]{%
4	54.292\\
8	86.868\\
12	114.638\\
16	150.564\\
20	190.504\\
24	235.774\\
28	272.906\\
32	313.62\\
36	354.998\\
40	399.124\\
44	439.362\\
48	472.926\\
52	522.698\\
56	560.93\\
60	600.496\\
64	640.426\\
68	656.658\\
72	703.816\\
76	729.462\\
80	772.986\\
};
\addlegendentry{$\text{n = 12; T}_{\text{est}}\text{ =  9.7m +  5.6}$}

\addplot [color=mycolor3, dashed, forget plot]
  table[row sep=crcr]{%
4	44.4852857142859\\
8	83.3449819548874\\
12	122.204678195489\\
16	161.06437443609\\
20	199.924070676692\\
24	238.783766917293\\
28	277.643463157895\\
32	316.503159398496\\
36	355.362855639098\\
40	394.222551879699\\
44	433.082248120301\\
48	471.941944360902\\
52	510.801640601504\\
56	549.661336842105\\
60	588.521033082707\\
64	627.380729323308\\
68	666.24042556391\\
72	705.100121804511\\
76	743.959818045113\\
80	782.819514285714\\
};
\addplot [color=mycolor4, draw=none, mark=diamond*, mark options={solid, mycolor4}]
  table[row sep=crcr]{%
4	73.816\\
8	111.708\\
12	137.546\\
16	167.872\\
20	203.85\\
24	256.054\\
28	301.752\\
32	341.282\\
36	387.202\\
40	434.808\\
44	474.526\\
48	523.204\\
52	563.33\\
56	606.3\\
60	638.842\\
64	690.854\\
68	729.82\\
72	774.096\\
76	809.514\\
80	853.214\\
};
\addlegendentry{$\text{n = 16; T}_{\text{est}}\text{ = 10.5m + 11.7}$}

\addplot [color=mycolor4, dashed, forget plot]
  table[row sep=crcr]{%
4	53.8596857142857\\
8	95.9775609022556\\
12	138.095436090226\\
16	180.213311278195\\
20	222.331186466165\\
24	264.449061654135\\
28	306.566936842105\\
32	348.684812030075\\
36	390.802687218045\\
40	432.920562406015\\
44	475.038437593985\\
48	517.156312781955\\
52	559.274187969925\\
56	601.392063157895\\
60	643.509938345865\\
64	685.627813533835\\
68	727.745688721805\\
72	769.863563909774\\
76	811.981439097744\\
80	854.099314285714\\
};
\end{axis}
\end{tikzpicture}%
		}
		\caption{An average completion time vs. $m$ (columns) on $n \times m$ grid with a constant ratio $\rho = 0.6$ of 
			empty spaces.}
		\label{fig:FreeRatio.M.12}
	\end{figure}
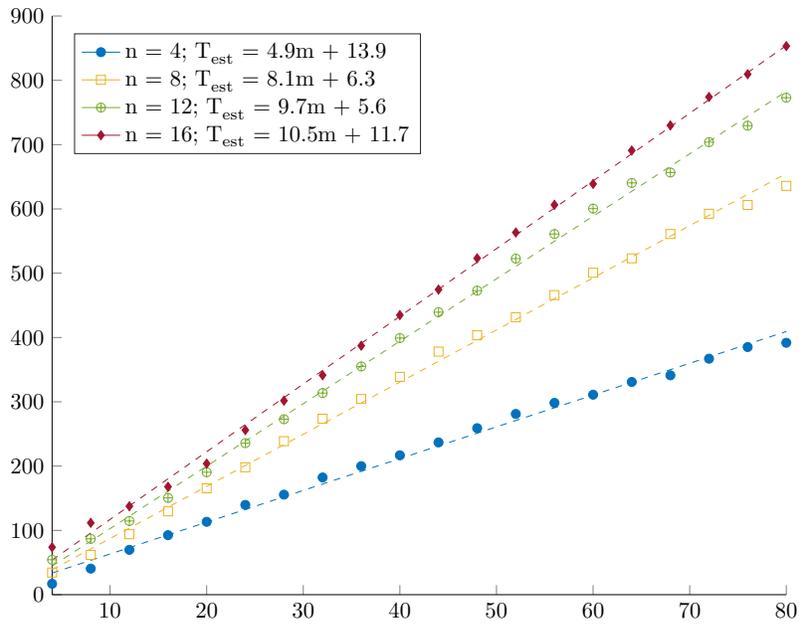
	
	\section{Sorting a two-column grid}\label{section:two-column sorting}
	
		
		In this section we consider a special case: a freeway road with two lanes ($n=2$). It could be shown
		that an Algorithm $\mathcal{A}$, presented in \cref{sec:Multi-lane Algorithm A}, does not sort all
		the possible bi-lane grid configurations. In particular, in some initial configurations exiting agents
		in the first column are	\enquote{starved} in the following sense: after some time $t$ an agent never observes
		an empty space, and, hence, is unable to move into the exiting lane right next to it.
		
		Below we present a different $3$-bit visibility $1$ algorithm ($\mathcal{A}_2$, 
		that solves ER(Exit the Road) problem on the road with two lanes. Like previously, 
		an agent is not required to communicate its intentions in any way, nor does it require 
		centralized orders to execute in-frame movement. The actual cellular automata state machines
		that makes the solution possible could be found in \cref{sec:Two.Lane-Algorithm.Exit} and 
		\cref{sec:Two.Lane-Algorithm.Continue}.
		
		The proposed algorithm is simple and is based on few principles:
		\begin{itemize}
			\item
				Agents move \textbf{only} into empty slots.
			\item
				All agents, that move at time $t$, move in the \textbf{non-conflicting} 
				directions at time $t$. This rule and the one above ensure no collision could 
				happen at time $t$.
			\item
				Exiting agents move horizontally inside a frame \textbf{only} from the non-exiting
				lane to the exiting one.
			\item
				Continue agents, on the contrary, move horizontally \textbf{only} from
				the exiting lane to the non-exiting one. The rule and the previous sibling rule
				ensure, that agents sort themselves to the target lane (TC) according to own type.
			\item
				Agent that moved in the certain direction does not return back in less than one
				cycle of timer (we further denote by CC(clock cycle); spans $4$ time ticks). The rule ensures 
				that an agent is not starved while
				waiting for an empty slot neighbor, and, eventually, the rule will serve us to show
				that an empty space \enquote{visits} every cell that \textit{should} be visited.
		\end{itemize}
		
		An agent that executes $\mathcal{A}_2$ is exactly the same agent as in the multi-lane ER Problem. 
		With exactly the same capabilities: one unit $L_1$ visibility and $3$ bits of memory 
		employed to track the timer and the last movement direction.
		
		Memory bits are split into a $1-2$ tuple. A couple of bits is used to track CC
		(\enquote{modulo $4$ timer}). These bits are initialized to $0$, so all agents are initially
		synchronized. An implementation should merely add $1$ to the current timer value, so an 
		agent \enquote{proceeds} to the next time tick. Since all the agents implement the 
		same logic, they remain time-synchronized during the algorithm execution. We arbitrary
		treat timer as the lower bits of the state id encoding (see \cref{fig:fg_3191_1} and 
		following table representation of $\mathcal{A}_2$). The remaining bit,
		aka a Movement Direction bit, is utilized to store the last memorized movement direction.
		This bit allows agents to continue unobstructed movement in the same direction.
		However, an agent will eventually change the direction to the opposite if the 
		movement is blocked	(by another agent or by the edge of the frame). To prevent sporadic
		changes in direction, we require a whole cycle to pass between the last movement tick and 
		a Direction Bit flip tick.
		
		\begin{theorem}\label{thm:Two.Lane-the Main Theorem}[The Algorithm $\mathcal{A}_2$ Soundness Theorem]
			Consider an initial configuration $\mathcal{C}$ of an $n \times 2$ grid $G$ that 
			contains at least a single empty slot. Let the number of exiting vehicles be $N_1$.
			Then Algorithm $\mathcal{A}_2$ solves the ER Problem and transitions $G$ to a 
			target configuration. Where target configuration is defined in the following sense:
			in case $N_1 \leq n$, then no exiting vehicles remain in the first column,
			otherwise no continue vehicles remain in the second column.
		\end{theorem}
		
		\subsection{Time complexity}
			In this section we analyze the time complexity of the two-lane Algorithm and provide
			time-to-solution evaluation based on computer simulations. 
			
			We are interested to estimate the time span of the Algorithm for some initial configuration
			of an $n \times 2$ grid with $N_0$ empty cells and $N_1$ exiting agents. Note, that if both
			lanes are exiting, then, after the vehicle rearrangement, all exiting vehicles will be able
			to exit the road. Hence, there is no limitation on the $N_1$, except, that $N_0 + N_1 \leq 2n$ 
			and that $N_0 > 0$, i.e. it is possible to rearrange vehicles at all. However, in case
			there a single exiting lane is present, the obvious limitation is $N_1 \leq n$.
			
\begin{figure}
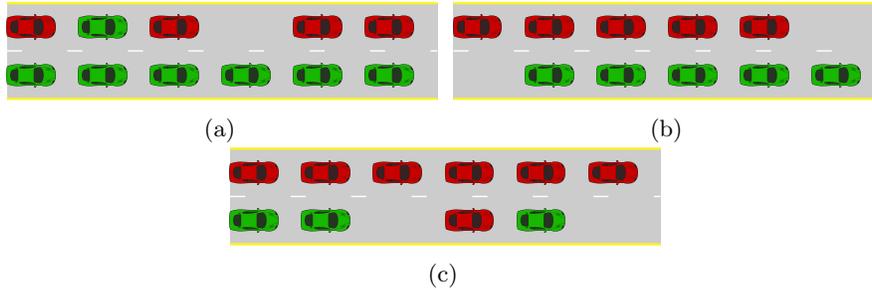

				\centering
				\begin{subfigure}{.48\textwidth}
					\resizebox{\textwidth}{!}{\subimport{./images}{TwoLanes.Sorted.Case1}}
					\caption{}
					\label{sub:two.lane-case 1}
				\end{subfigure}
				\begin{subfigure}{.48\textwidth}
					\resizebox{\textwidth}{!}{\subimport{./images}{TwoLanes.Sorted.Case2}}
					\caption{}
					\label{sub:two.lane-case 2}
				\end{subfigure}
				\begin{subfigure}{.48\textwidth}
					\resizebox{\textwidth}{!}{\subimport{./images}{TwoLanes.Sorted.Case3}}
					\caption{}
					\label{sub:two.lane-case 3}
				\end{subfigure}
				\caption{Two lane road. A sorted configuration (\subref{sub:two.lane-case 1}) right lane
				is filled by exiting vehicles,  (\subref{sub:two.lane-case 2}) vehicles are split according to
				type, (\subref{sub:two.lane-case 3}) left lane is fully occupied by continue vehicles. }
				\label{fig:two.lane.sorted.configs}
			\end{figure}

			We first consider a simple case: an initial configuration has a single empty slot. $N_0 = 1$.
			An empty slot \enquote{travels} at the speed of one slot per timer cycle ($4$ time ticks).
			Exiting vehicles are expected to move into the right (exiting) lane, unless $N_1 > n$. Then
			we expect exactly $n$ exiting vehicles to move on the right, and the rest to stay in the left
			lane. In the opposite case, where $N_1 < n$ we expect all the exiting vehicles to move on the
			right. See \cref{fig:two.lane.sorted.configs} for different possible sorted configurations.
			
			All sorted configurations have a common property: no further progress of
			time will witness a lane change by an agent. Vehicles will continue to jiggle in their respective 
			column, however horizontal moves have ceased.
			
			In the worst case scenario an empty slot will need to traverse a whole grid column at most twice 
			before it could \enquote{find} a next candidate for a lane change, i.e. a single vehicle could
			be moved in at most $2n$ cycles. We need to move no more than \textit{all} the agents
			on the grid ($2n$), therefore the Algorithm running time is upper bounded by 
			\begin{equation*}
				T \leq 16n^2
			\end{equation*}
			time ticks.
			
			We assume that empty slots are \enquote{placed} and \enquote{moving} \textit{independently}. And,
			as such, each additional empty slot should \enquote{increase} the overall agent movement 
			count in a linear manner. Until the point, when empty slots \enquote{start} to interfere with each
			other. For example, a line of empty slots can not \enquote{process} an agent faster than a one
			cell at a cycle, no matter how many empty slots are queued in line. Therefore, we should observe
			some phase transition effect at the critical empty slot saturation point.
			
			Moreover, there is no need to move \textit{all} the agents. It is enough to move at 
			most $\min\{N_1, n\}$ exiting agents to the exiting lane. However, slots should be freed in the 
			exiting lane, i.e. the same number of continue agents should be moved to the left. We
			expect roughly half of exiting vehicles to reside initially in the right lane. Moreover, agent 
			spread in the column should be expected to be almost uniform on the average. Which, in turn,
			requires empty slots to \enquote{run} for only half a distance on the average. Hence,
			
			\begin{equation*}
				\mathbb{E}(T) \leq 4 \cdot \frac{1}{N_0} \cdot n \cdot N_1.
			\end{equation*}
			
		\subsubsection{Simulation Results}
			We further ran simulations to check our hypothesis. The simulation framework used was the same C++ 
			implementation that we have utilized in the preceding sections to assess Algorithm $\mathcal{A}$ 
			runtime complexity.
			
			\begin{figure}
				\centering
				\begin{subfigure}{.48\textwidth}
					\resizebox{\textwidth}{!}{
%
%
\definecolor{mycolor1}{rgb}{0.00000,0.44700,0.74100}%
\definecolor{mycolor2}{rgb}{0.85000,0.32500,0.09800}%
\definecolor{mycolor3}{rgb}{0.92900,0.69400,0.12500}%
\definecolor{mycolor4}{rgb}{0.49400,0.18400,0.55600}%
\definecolor{mycolor5}{rgb}{0.46600,0.67400,0.18800}%
\definecolor{mycolor6}{rgb}{0.30100,0.74500,0.93300}%
\definecolor{mycolor7}{rgb}{0.63500,0.07800,0.18400}%
\begin{tikzpicture}

\begin{axis}[%
width=4.521in,
height=3.527in,
at={(0.758in,0.519in)},
scale only axis,
xmin=1,
xmax=21,
xlabel style={font=\color{white!15!black}},
xlabel={$\text{N}_\text{0}$},
xtick=\empty,
ymin=0,
ymax=25000,
ylabel style={font=\color{white!15!black}},
ylabel={Time (seconds)},
ytick=\empty,
axis background/.style={fill=white},
title style={font=\bfseries},
axis x line*=bottom,
axis y line*=left,
legend style={legend cell align=left, align=left, draw=white!15!black}
]
\addplot [color=mycolor1, mark=*]
  table[row sep=crcr]{%
1	4333.392\\
2	2294.608\\
3	1576.096\\
4	1224.472\\
5	1043.018\\
6	901.006\\
7	821.968\\
8	751.276\\
9	705.92\\
10	656.588\\
11	625.322\\
12	589.65\\
13	578.096\\
14	550.922\\
15	530.672\\
16	521.22\\
17	523.172\\
18	505.634\\
19	499.25\\
20	491.564\\
21	474.954\\
};
\addlegendentry{n =  80}


\addplot [color=mycolor3, mark=oplus]
  table[row sep=crcr]{%
1	8128.12\\
2	4178.488\\
3	2911.736\\
4	2270.56\\
5	1857.508\\
6	1641.508\\
7	1466.73\\
8	1335.114\\
9	1227.944\\
10	1138.976\\
11	1076.894\\
12	1027.966\\
13	975.206\\
14	940.296\\
15	925.25\\
16	895.25\\
17	880.078\\
18	852.688\\
19	824.288\\
20	808.266\\
21	794.34\\
};
\addlegendentry{n = 120}


\addplot [color=mycolor5, mark=halfsquare left*]
  table[row sep=crcr]{%
1	12303.264\\
2	6506.328\\
3	4417.4\\
4	3477.72\\
5	2771.784\\
6	2461.212\\
7	2209.572\\
8	1983.318\\
9	1792.818\\
10	1692.08\\
11	1593.532\\
12	1505.284\\
13	1446.924\\
14	1388.786\\
15	1341.33\\
16	1306.998\\
17	1267.382\\
18	1222.476\\
19	1198.686\\
20	1159.674\\
21	1147.262\\
};
\addlegendentry{n = 160}


\addplot [color=mycolor7, mark=asterisk]
  table[row sep=crcr]{%
1	17268.432\\
2	9133.896\\
3	6229.92\\
4	4710.88\\
5	3993.208\\
6	3399.532\\
7	3022.178\\
8	2709.404\\
9	2473.754\\
10	2334.064\\
11	2138.576\\
12	2038.636\\
13	1932.94\\
14	1858.642\\
15	1768.886\\
16	1744.398\\
17	1712.068\\
18	1609.97\\
19	1585.252\\
20	1554.892\\
21	1515.892\\
};
\addlegendentry{n = 200}


\addplot [color=mycolor2, mark=triangle*]
  table[row sep=crcr]{%
1	22752.616\\
2	11970.208\\
3	8240.032\\
4	6274.536\\
5	5129.82\\
6	4482.89\\
7	3938.016\\
8	3499.014\\
9	3132.564\\
10	2998.85\\
11	2761.11\\
12	2628.102\\
13	2475.764\\
14	2352.118\\
15	2293.522\\
16	2206.024\\
17	2185.668\\
18	2072.072\\
19	2029.434\\
20	1964.528\\
21	1961.136\\
};
\addlegendentry{n = 240}


\end{axis}
\end{tikzpicture}%
}
					\caption{}
					\label{sub:two-lanes.ks}
				\end{subfigure}
				\begin{subfigure}{.48\textwidth}
					\resizebox{\textwidth}{!}{\subimport{\ProcessedDataRelativeToSortingTwoLaneDirectory}{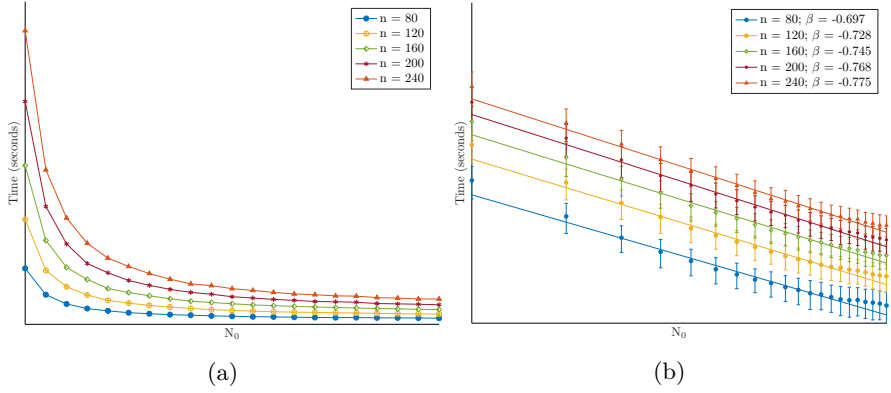}}
					\caption{}
					\label{sub:two-lanes.ks.power}
				\end{subfigure}
				\caption{$\mathcal{A}_2$ makespan as a function of $N_0$ (a number of empty slots on the $n \times 2$
				grid), in a (\subref{sub:two-lanes.ks}) linear and a (\subref{sub:two-lanes.ks.power}) log-log scale
				representation. A phase-transition is evident on the (\subref{sub:two-lanes.ks.power}), where $N_0 \to n$. 
				The $\beta$ is calculated according to $\beta = \frac{\log T - c}{\log N_0}$, and is
				the best-fit line slop, i.e. the power of $N_0$ in the $\mathbb{E}(T)$. ($N_1$ = $n$).}
				\label{fig:two-lanes.ks}
			\end{figure}
		
			\begin{figure}
				\centering
				\begin{subfigure}{.48\textwidth}
					\resizebox{\textwidth}{!}{
%
%
\definecolor{mycolor1}{rgb}{0.00000,0.44700,0.74100}%
\definecolor{mycolor2}{rgb}{0.92900,0.69400,0.12500}%
\definecolor{mycolor3}{rgb}{0.46600,0.67400,0.18800}%
\definecolor{mycolor4}{rgb}{0.63500,0.07800,0.18400}%
\definecolor{mycolor5}{rgb}{0.85000,0.32500,0.09800}%
\definecolor{mycolor6}{rgb}{0.49400,0.18400,0.55600}%
\definecolor{mycolor7}{rgb}{0.30100,0.74500,0.93300}%
\begin{tikzpicture}

\begin{axis}[%
width=4.521in,
height=3.566in,
at={(0.758in,0.481in)},
scale only axis,
xmin=10,
xmax=100,
xlabel style={font=\color{white!15!black}},
xlabel={n (row number)},
xtick=\empty,
ymin=0,
ymax=7000,
ylabel style={font=\color{white!15!black}},
ylabel={Time (seconds)},
ytick=\empty,
axis background/.style={fill=white},
title style={font=\bfseries},
axis x line*=bottom,
axis y line*=left,
legend style={at={(0.03,0.97)}, anchor=north west, legend cell align=left, align=left, draw=white!15!black}
]
\addplot [color=mycolor1, draw=none, mark=*, mark options={solid, mycolor1}]
  table[row sep=crcr]{%
10	160.326\\
20	503.114\\
30	946.892\\
40	1466.286\\
50	2083.02\\
60	2760.352\\
70	3512.494\\
80	4314.912\\
90	5155.892\\
100	6064.352\\
};
\addlegendentry{$N_0 =  1; T_{\text{est}} = 0.344 \cdot \tfrac{N_1}{N_0} \cdot n + O(1)$}

\addplot [color=mycolor1, dashed, forget plot]
  table[row sep=crcr]{%
10	127.555199999995\\
20	514.764266666664\\
30	970.870666666666\\
40	1495.8744\\
50	2089.77546666667\\
60	2752.57386666667\\
70	3484.2696\\
80	4284.86266666667\\
90	5154.35306666667\\
100	6092.7408\\
};
\addplot [color=mycolor2, draw=none, mark=square, mark options={solid, mycolor2}]
  table[row sep=crcr]{%
10	64.1065\\
20	194.4025\\
30	363.2265\\
40	564.829\\
50	782.4795\\
60	1030.6925\\
70	1289.725\\
80	1594.408\\
90	1894.056\\
100	2194.0745\\
};
\addlegendentry{$N_0 =  3; T_{\text{est}} = 0.335 \cdot \tfrac{N_1}{N_0} \cdot n + O(1)$}

\addplot [color=mycolor2, dashed, forget plot]
  table[row sep=crcr]{%
10	50.1255227272705\\
20	200.949780303029\\
30	374.137814393939\\
40	569.689625\\
50	787.605212121213\\
60	1027.88457575758\\
70	1290.52771590909\\
80	1575.53463257576\\
90	1882.90532575758\\
100	2212.63979545454\\
};
\addplot [color=mycolor3, draw=none, mark=oplus, mark options={solid, mycolor3}]
  table[row sep=crcr]{%
10	47.3535\\
20	135.697\\
30	249.2595\\
40	378.9655\\
50	531.132\\
60	688.229\\
70	862.9845\\
80	1045.8665\\
90	1240.6795\\
100	1440.5755\\
};
\addlegendentry{$N_0 = 5; T_{\text{est}} = 0.337 \cdot \tfrac{N_1}{N_0} \cdot n + O(1)$}

\addplot [color=mycolor3, dashed, forget plot]
  table[row sep=crcr]{%
10	37.6318999999981\\
20	140.398745454545\\
30	256.664886363636\\
40	386.430322727273\\
50	529.695054545455\\
60	686.459081818183\\
70	856.722404545455\\
80	1040.48502272727\\
90	1237.74693636364\\
100	1448.50814545454\\
};
%
\addplot [color=mycolor5, draw=none, mark=halfsquare left*, mark options={solid, mycolor5}]
  table[row sep=crcr]{%
10	39.553\\
20	102.2865\\
30	179.5785\\
40	267.1855\\
50	365\\
60	471.074\\
70	586.1915\\
80	695.177\\
90	819.7655\\
100	950.679\\
};
\addlegendentry{$N_0 =  9; T_{\text{est}} = 0.343 \cdot \tfrac{N_1}{N_0} \cdot n + O(1)$}

\addplot [color=mycolor5, dashed, forget plot]
  table[row sep=crcr]{%
10	33.4749863636373\\
20	105.184101515152\\
30	184.516803787879\\
40	271.473093181818\\
50	366.052969696969\\
60	468.256433333333\\
70	578.083484090909\\
80	695.534121969697\\
90	820.608346969697\\
100	953.30615909091\\
};
\addplot [color=mycolor2, draw=none, mark=triangle*, mark options={solid, mycolor2}]
  table[row sep=crcr]{%
10	36.398\\
20	92.079\\
30	149.8225\\
40	213.9785\\
50	280.833\\
60	355.863\\
70	432.0935\\
80	514.7235\\
90	595.1695\\
100	687.849\\
};
\addlegendentry{$N_0 = 17; T_{\text{est}} = 0.367 \cdot \tfrac{N_1}{N_0} \cdot n + O(1)$}

\addplot [color=mycolor2, dashed, forget plot]
  table[row sep=crcr]{%
10	36.2122909090908\\
20	91.2814272727271\\
30	150.672025757576\\
40	214.384086363636\\
50	282.417609090909\\
60	354.772593939394\\
70	431.449040909091\\
80	512.44695\\
90	597.766321212121\\
100	687.407154545455\\
};
%
\end{axis}
\end{tikzpicture}%
}
					\caption{}
					\label{sub:two-lanes.ns}
				\end{subfigure}
				\begin{subfigure}{.48\textwidth}
					\resizebox{\textwidth}{!}{
%
%
\definecolor{mycolor1}{rgb}{0.00000,0.44700,0.74100}%
\definecolor{mycolor2}{rgb}{0.92900,0.69400,0.12500}%
\definecolor{mycolor3}{rgb}{0.46600,0.67400,0.18800}%
\definecolor{mycolor4}{rgb}{0.63500,0.07800,0.18400}%
\definecolor{mycolor5}{rgb}{0.85000,0.32500,0.09800}%
\begin{tikzpicture}

\begin{axis}[%
width=4.521in,
height=3.527in,
at={(0.758in,0.519in)},
scale only axis,
xmin=5,
xmax=100,
xlabel style={font=\color{white!15!black}},
xlabel={$\text{N}_\text{1}\text{ (exiting vehicle number)}$},
xtick=\empty,
ymin=0,
ymax=800,
ylabel style={font=\color{white!15!black}},
ylabel={Time (seconds)},
ytick=\empty,
axis background/.style={fill=white},
title style={font=\bfseries},
axis x line*=bottom,
axis y line*=left,
legend style={at={(0.03, 0.97)}, anchor=north west, legend cell align=left, align=left, draw=white!15!black}
]
\addplot [color=mycolor1, draw=none, mark=*, mark options={solid, mycolor1}]
  table[row sep=crcr]{%
5	70.084\\
10	100.224\\
15	122.712\\
20	143.6\\
25	166.208\\
30	177.834\\
35	203.656\\
40	228.458\\
45	243.29\\
50	259.042\\
55	268.08\\
60	289.446\\
65	296.928\\
70	327.7\\
75	342.69\\
80	378.79\\
85	416.884\\
};
\addlegendentry{$n = 100;\; T_{\text{est}} = 0.390 \cdot \frac{n}{N_0} \cdot N_1 + O(1)$}

\addplot [color=mycolor1, dashed, forget plot]
  table[row sep=crcr]{%
5	80.8823529411765\\
10	100.445779411765\\
15	120.009205882353\\
20	139.572632352941\\
25	159.136058823529\\
30	178.699485294118\\
35	198.262911764706\\
40	217.826338235294\\
45	237.389764705882\\
50	256.953191176471\\
55	276.516617647059\\
60	296.080044117647\\
65	315.643470588235\\
70	335.206897058824\\
75	354.770323529412\\
80	374.33375\\
85	393.897176470588\\
};
\addplot [color=mycolor2, draw=none, mark=square, mark options={solid, mycolor2}]
  table[row sep=crcr]{%
5	81.928\\
10	122.396\\
15	139.992\\
20	167.264\\
25	197.648\\
30	232.704\\
35	238.48\\
40	259.98\\
45	290.544\\
50	293.098\\
55	334.104\\
60	336.896\\
65	356.872\\
70	381.73\\
75	395.228\\
80	394.984\\
85	415.36\\
90	447.21\\
95	468.034\\
100	499.962\\
};
\addlegendentry{$n = 120;\; T_{\text{est}} = 0.341 \cdot \frac{n}{N_0} \cdot N_1 + O(1)$}

\addplot [color=mycolor2, dashed, forget plot]
  table[row sep=crcr]{%
5	110.201457142857\\
10	130.466640601504\\
15	150.73182406015\\
20	170.997007518797\\
25	191.262190977444\\
30	211.52737443609\\
35	231.792557894737\\
40	252.057741353383\\
45	272.32292481203\\
50	292.588108270677\\
55	312.853291729323\\
60	333.11847518797\\
65	353.383658646617\\
70	373.648842105263\\
75	393.91402556391\\
80	414.179209022556\\
85	434.444392481203\\
90	454.70957593985\\
95	474.974759398496\\
100	495.239942857143\\
};
\addplot [color=mycolor3, draw=none, mark=oplus, mark options={solid, mycolor3}]
  table[row sep=crcr]{%
5	93.5\\
10	138.608\\
15	162.176\\
20	205.6\\
25	232.224\\
30	268.074\\
35	300.992\\
40	326.856\\
45	343.024\\
50	367.88\\
55	370.84\\
60	384.344\\
65	423.428\\
70	442.696\\
75	444.656\\
80	468.082\\
85	463.924\\
90	488.294\\
95	509.102\\
100	533.314\\
};
\addlegendentry{$n = 140;\; T_{\text{est}} = 0.307 \cdot \frac{n}{N_0} \cdot N_1 + O(1)$}

\addplot [color=mycolor3, dashed, forget plot]
  table[row sep=crcr]{%
5	143.450485714285\\
10	165.022087218045\\
15	186.593688721804\\
20	208.165290225564\\
25	229.736891729323\\
30	251.308493233083\\
35	272.880094736842\\
40	294.451696240601\\
45	316.023297744361\\
50	337.59489924812\\
55	359.16650075188\\
60	380.738102255639\\
65	402.309703759398\\
70	423.881305263158\\
75	445.452906766917\\
80	467.024508270677\\
85	488.596109774436\\
90	510.167711278196\\
95	531.739312781955\\
100	553.310914285714\\
};
\addplot [color=mycolor4, draw=none, mark=diamond*, mark options={solid, mycolor4}]
  table[row sep=crcr]{%
5	109.994\\
10	156.808\\
15	190.512\\
20	227.928\\
25	279.752\\
30	293.968\\
35	333.096\\
40	371.68\\
45	394.92\\
50	417.904\\
55	444.792\\
60	468.784\\
65	495.026\\
70	498.33\\
75	522.106\\
80	524.53\\
85	520.548\\
90	567.384\\
95	584.772\\
100	574.884\\
};
\addlegendentry{$n = 160;\; T_{\text{est}} = 0.306 \cdot \frac{n}{N_0} \cdot N_1 + O(1)$}

\addplot [color=mycolor4, dashed, forget plot]
  table[row sep=crcr]{%
5	167.935285714286\\
10	192.245876691729\\
15	216.556467669173\\
20	240.867058646616\\
25	265.17764962406\\
30	289.488240601504\\
35	313.798831578947\\
40	338.109422556391\\
45	362.420013533835\\
50	386.730604511278\\
55	411.041195488722\\
60	435.351786466165\\
65	459.662377443609\\
70	483.972968421053\\
75	508.283559398496\\
80	532.59415037594\\
85	556.904741353383\\
90	581.215332330827\\
95	605.525923308271\\
100	629.836514285714\\
};
\addplot [color=mycolor5, draw=none, mark=halfsquare left*, mark options={solid, mycolor5}]
  table[row sep=crcr]{%
5	115.766\\
10	187.22\\
15	221.352\\
20	256.416\\
25	302.912\\
30	343.44\\
35	355.848\\
40	408.392\\
45	449.552\\
50	496.112\\
55	477.362\\
60	516.776\\
65	553.392\\
70	540.17\\
75	608.792\\
80	585.272\\
85	615.058\\
90	637.41\\
95	638.434\\
100	663.444\\
};
\addlegendentry{$n = 180;\; T_{\text{est}} = 0.305 \cdot \frac{n}{N_0} \cdot N_1 + O(1)$}

\addplot [color=mycolor5, dashed, forget plot]
  table[row sep=crcr]{%
5	186.712857142857\\
10	214.285819548872\\
15	241.858781954887\\
20	269.431744360902\\
25	297.004706766917\\
30	324.577669172932\\
35	352.150631578947\\
40	379.723593984962\\
45	407.296556390977\\
50	434.869518796992\\
55	462.442481203007\\
60	490.015443609022\\
65	517.588406015037\\
70	545.161368421053\\
75	572.734330827068\\
80	600.307293233083\\
85	627.880255639098\\
90	655.453218045113\\
95	683.026180451128\\
100	710.599142857143\\
};
\end{axis}

\end{tikzpicture}%
}
					\caption{}
					\label{sub:two-lanes.n1ones}
				\end{subfigure}
				\caption{$\mathcal{A}_2$ makespan as a function of (\subref{sub:two-lanes.ns}) row number $n$ (exiting vehicle number $N_1$ is 
				equal to $n$) and  (\subref{sub:two-lanes.n1ones}) exiting vehicle number $N_1$ (for a constant $N_0 = 10$). }
				\label{fig:two-lanes.ns}
			\end{figure}
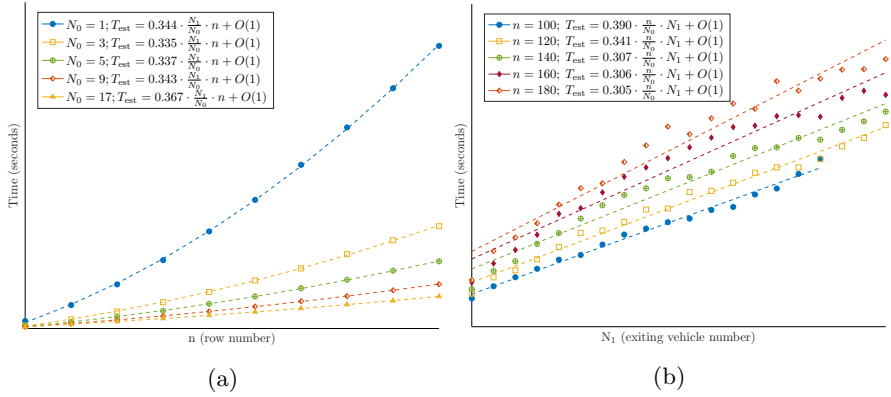

			We have studied a functional dependency between $\mathbb{E}(T)$ - an average time until $\mathcal{A}_2$ sorts a $n \times 2$ grid of 
			a random initial vehicle configuration. \cref{fig:two-lanes.ks} presents a time as a function of $N_0$ - the number of empty slots 
			on the grid. A regular linear representation (on the left) depicts an inverse dependency rate, i.e. the time required to sort a random
			initial vehicle configuration significantly drops once enough empty slots are \enquote{floating} around. The log-log representation
			(on the right) allows us to observe the predicted phase	transition effect, when $N_0 \to n$. Recall,
			\begin{equation*}
				\mathbb{E}(T) \leq 4 \cdot \frac{1}{N_0} \cdot n \cdot N_1.
			\end{equation*}
			Therefore, 
			\begin{equation*}
				\log \mathbb{E}(T) \propto -\log N_0,
			\end{equation*}
			where the proportionality coefficient $\beta \to -1$, and could be expected to be a limiting behavior for a higher values of $n$.
			
			Two other parameters: the number of rows in the frame $n$ and the number of exiting vehicles $N_1$ do produce a linear best fit line 
			dependency as predicted (see \cref{fig:two-lanes.ns}).
			Note, that on \cref{sub:two-lanes.ns} the number of exiting vehicles up-scales with the number of rows, hence the quadratic outlook
			of the image. Overall, the findings are in line with our theoretical reasoning.

	\section{Discussion}\label{section: MAGS. Discussion}
		We studied the problem of sorting vehicles on a multi-lane road, 
		sending vehicles with destination at the nearest exit to the 
		departing lane and leaving vehicles, that should continue down
		the road on the other lanes. We have shown that the problem is
		solvable in the distributed settings without communication. 
		Moreover, the hardware implementation requirements are minimal, 
		while the time to
		solution produced by the proposed algorithms is convenient.		
		
		We proved the correctness of the algorithms, thus ensuring
		that any feasible initial configuration gets solved under 
		physical constraints of the real world. We note, that sorting,
		constrained by physical reality, become more and more relevant
		with the advent of automation in such fields as warehouse
		management, traffic management, massive deployment of aerial 
		shipping solutions etc. Our work unequivocally shows
		that distributed approach is viable in the traffic context.
		Moreover, the technological requirements are few orders of 
		magnitude less demanding than previously thought.
		
		However, we note that our approach to the problem is not
		without a limitation. Our model relies on road infrastructure
		to support the so-called moving frames, where algorithms 
		meant to be applied. Therefore, it would be interesting to
		search for solutions that do not rely on infrastructure and
		inter-vehicle communication altogether.

\ifnum\mydoclevel=0
	\clearpage 

	\bookmarksetupnext{level=-1}
	\appendix
	\begin{appendices}
\else
	\begin{subappendices}
\fi
		\section{Local sensing algorithm}
			
	The algorithm that we here present is a set of rules. A rule
	in this set is an action of the following kind: \enquote{What
	an agent of type $T$ at state $S$ \textbf{should do} if it observes a specific neighborhood
	$N$}. Where state $S$ is the current value of an agent internal memory and
	type $T$ can either describe an exiting or a continue agent. 
	The rule additionally specifies how an agent should adjust its internal 
	memory state \textit{after}	an action got executed.
	
	From the mathematical perspective Algorithm $\mathcal{A}$ is a 
	function from triplets \textit{type-state-neighborhood} to pairs
	\textit{state-action}. Where \textit{type} is set arbitrary to $1$ for exiting vehicles,
	and to $-1$ for continuing vehicles, \textit{state} is one of the $2^b$ possible 
	memory values between $0$ and $2^b-1$, 
	and \textit{action} is either \textit{do nothing}, i.e., remain at place, or
	\textit{move} in one of the four primary directions \textit{North, 
	East, South} or \textit{West}.
	
	Further we shall explain what is a neighborhood, and how does an agent use it as an 
	input to the algorithm. 
	Per definition, an agent neighborhood
	at time $t$ is a collection of all the cells that an agent senses at time $t$. 
	We here develop an algorithm assuming an $L_1$ visibility
	of radius $1$. An example of one such $L_1$ neighborhood of radius $1$ 
	is presented on \cref{sub:L1.neighborhood}.
	
	\begin{figure}[h]
		\centering
		\begin{subfigure}{.32\textwidth}
			\centering
				\begin{tikzpicture}
		\pgfmathsetmacro{\offset}{0.2}
		\begin{scope}
			\clip (0-\offset, 0-\offset) --
				(0-\offset, 3+\offset) --
				(3+\offset, 3+\offset) --
				(3+\offset, 0) --
				(3, 0) -- (3, 3) -- (0, 3) -- (0, 0) -- (3, 0) --
				(3+\offset, 0) --
				(3+\offset, 0-\offset) --
				cycle;
			\draw[pattern=north west lines] (-1, -1) rectangle (4, 4);
		\end{scope}
		\draw (0, 0) grid (3, 3);
		\begin{scope}[shift={(0.5, 0.5)}]
			\foreach \coord[count=\i] in {(2,0), (0,0), (0, 2), (2,2), (1,0), (0, 1), (1, 2), (2, 1), (1, 1)}
			{
				\node at \coord{$\i$};
			}
		\end{scope}
	\end{tikzpicture}
			\caption{}
			\label{sub:grid.positions}
		\end{subfigure}
		\begin{subfigure}{.32\textwidth}
			\centering
				\begin{tikzpicture}
		\draw (0, 0) grid (1, 1);
		\draw[fill=black!60] (0, 1) rectangle (1, 2);
		\draw[fill=black!60] (1, 0) rectangle (2, 1);
		\draw[pattern=north west lines] (-1, 0) rectangle (0, 1);
		\draw[fill=black] (0.5, 0.5) circle (3pt);	
		\draw[ultra thick] (0,0) --
			++(0, -1) --
			++(1, 0) --
			++(0, 1) --
			++(1, 0) --
			++(0, 1) --
			++(-1, 0) --
			++(0, 1) --
			++(-1, 0) --
			++(0, -1) --
			++(-1, 0) --
			++(0, -1) --
			++(1, 0) --
			cycle;
	\end{tikzpicture}
			\caption{}
			\label{sub:L1.neighborhood}
		\end{subfigure}
		\begin{subfigure}{.32\textwidth}
			\centering
				\begin{tikzpicture}
		\begin{scope}[opacity=0.5]
			\draw (0, 0) rectangle (1, 1);
			\draw[pattern=north west lines] (-1, 0) rectangle (0, 1);
			\draw (1,0) rectangle (2, 1);
			\draw (0,-1) rectangle (1, 0);
			\node at ($(1, 0)+(0.5, 0.5)$) {X};		
			\node at ($(0, -1)+(0.5, 0.5)$) {X};		
		\end{scope}
		\draw[fill=black!60] (0, 1) rectangle (1, 2);
		\draw[fill=black] (0.5, 0.5) circle (3pt);	
		\draw[ultra thick] (0,0) rectangle (1, 2);
	\end{tikzpicture}
			\caption{}
			\label{sub:L1.neighborhood.example}
		\end{subfigure}
		\caption{(\subref{sub:grid.positions}) possible agent positions on the grid,
			(\subref{sub:L1.neighborhood}) a typical $L_1$ neighborhood 
			of radius $1$ (agent's sensor diagram), 
			(\subref{sub:L1.neighborhood.example}) a partial neighborhood that can be sensed by an 
			agent at position $6$. Essential for decision making neighbors are bold,
			non-essential neighbors are semi-transparent.
			Legend: 
			\protect\tikz{\protect\draw[fill=black] (0, 0) circle (3pt);} - the agent,
			\scalebox{0.3}{\insertimage{pattern=north west lines}} - an edge grid
			cell,
			\scalebox{0.3}{\insertimage{}} - an empty grid cell,
			\scalebox{0.3}{\insertimage{fill=black!60}} - a cell occupied by another agent,
			X - a \textit{do not care} reading (either an agent or an empty cell).
			}
		\label{img:legend}
	\end{figure}
	
	In our model, an agent, with the $L_1$ visibility of radius $1$, can distinguish 
	up to nine different types of grid positions (see \cref{sub:grid.positions}). 
	Position detection is done according to sensor readings of a simple $4$-neighborhood
	of an agent, where the presence of \textit{border} type neighbor cells is that, what
	determines the agent's grid position.
	E.g. on \cref{sub:L1.neighborhood.example} an agent detects only one
	border cell neighbor on its West, hence an agent self-identified grid
	position is $6$.
	
	Some visible neighboring cells could be irrelevant to the agent decision making process. 
	The relevance of a neighboring cell depends on its position relative to the agent location.
	For brevity, we replace irrelevant neighbors in the agent's sensor diagram with the
	X mark (see (\textit{do not care}), where X is set in place of East and South neighbors 
	on \cref{sub:L1.neighborhood.example}). However, in the actual algorithm implementation, 
	a rule corresponding to a neighborhood with an X mark should be replaced
	by a pair of rules. One rule, corresponding to a neighborhood, where the X mark is 
	replaced by an empty cell, and another rule, with an identical outcome, where X is 
	replaced by an agent occupied cell. Note, that a rule, corresponding to $k$ X marks at different
	neighborhood locations, should be replaced by $2^k$ rules, following the same
	substitution principles.
	
	For example, we present an agent sensor reading in \cref{sub:L1.neighborhood.example}, 
	which includes: an occupied cell on the North, 
	a border cell on the West and two do not care readings on the East and South. 
	Any rule corresponding to this neighborhood template is expanded into four possible
	rules in the \textbf{full} description of the algorithm. Where every possible combination
	of an empty cell and/or of an occupied cell instead of a do not care is listed. 
	The expanded rules share the same (original) output, i.e. a common
	\textit{state-direction} pair.
	
	The algorithm description minimization approach resembles a similar procedure in the 
	Karnaugh Map manipulation of incompletely specified switching functions, where the same output for
	two possible states of the same variable eliminates the variable from the input.
	
	Some rules in the short description of the algorithm are left undefined (see
	\cref{fig:UndefinedRules}). We later show that input triplets, corresponding to the 
	missing rules, are not attainable by the serial execution of the algorithm,
	therefore they could be arbitrary filled.
	
	\begin{figure}
		\centering
		\scalebox{0.8}{
			\subimport{../\StateMachineTablesDirectory/Ver.3101}{Table.Ver_3101.Position_3}
		}
		\caption{Rules for an exiting agent at grid position $3$. All memory states are 
		listed in binary, i.e. $011 \equiv 011_2 = 3$. Following agent actions are possible:
		$N$ - \textit{move} North, $E$ - \textit{move} East, $S$ - \textit{move} South,
		$W$ - \textit{move} West, $\varnothing$ - \textit{stay} in place. Undefined
		rules are left empty in the table.}
		\label{fig:UndefinedRules}
	\end{figure}

		\clearpage
		\section[Algorithm \texorpdfstring{$\mathcal{A}$}{} FSM]{Multi-column sorting algorithm state machine}
		\subsection{An exit agent}
			\label{sec:Sorting.Algorithm.Exit}
			\vspace{33mm}
			
	\begin{figure}[ht]
		\centering
		\begin{subfigure}[b]{0.27\textwidth}
			\centering
			\subimport{Ver.3101/}{Table.Ver_3101.Position_1}
			\caption{Position 1}
			\label{subfig:fg_3101_1}
		\end{subfigure}
		\hfill
		\begin{subfigure}[b]{0.69\textwidth}
			\centering
			\subimport{Ver.3101/}{Table.Ver_3101.Position_5}
			\caption{Position 5}
			\label{subfig:fg_3101_5}
		\end{subfigure}
		\caption{State machine(s) (\subref{subfig:fg_3101_1}) at position 1, (\subref{subfig:fg_3101_5}) at position 5.}
		\label{fig:fg_3101_1}
	\end{figure}
	\begin{figure}
		\centering
		\begin{subfigure}[b]{0.32\textwidth}
			\centering
			\subimport{Ver.3101/}{Table.Ver_3101.Position_2}
			\caption{Position 2}
			\label{subfig:fg_3101_2}
		\end{subfigure}
		\hfill
		\begin{subfigure}[b]{0.32\textwidth}
			\centering
			\subimport{Ver.3101/}{Table.Ver_3101.Position_3}
			\caption{Position 3}
			\label{subfig:fg_3101_3}
		\end{subfigure}
		\hfill
		\begin{subfigure}[b]{0.32\textwidth}
			\centering
			\subimport{Ver.3101/}{Table.Ver_3101.Position_4}
			\caption{Position 4}
			\label{subfig:fg_3101_4}
		\end{subfigure}
		\caption{State machine(s) (\subref{subfig:fg_3101_2}) at position 2, (\subref{subfig:fg_3101_3}) at position 3, (\subref{subfig:fg_3101_4}) at position 4.}
		\label{fig:fg_3101_2}
	\end{figure}
	\begin{figure}
		\centering
		\begin{subfigure}[b]{0.32\textwidth}
			\centering
			\subimport{Ver.3101/}{Table.Ver_3101.Position_6}
			\caption{Position 6}
			\label{subfig:fg_3101_6}
		\end{subfigure}
		\hfill
		\begin{subfigure}[b]{0.32\textwidth}
			\centering
			\subimport{Ver.3101/}{Table.Ver_3101.Position_7}
			\caption{Position 7}
			\label{subfig:fg_3101_7}
		\end{subfigure}
		\hfill
		\begin{subfigure}[b]{0.32\textwidth}
			\centering
			\subimport{Ver.3101/}{Table.Ver_3101.Position_8}
			\caption{Position 8}
			\label{subfig:fg_3101_8}
		\end{subfigure}
		\caption{State machine(s) (\subref{subfig:fg_3101_6}) at position 6, (\subref{subfig:fg_3101_7}) at position 7, (\subref{subfig:fg_3101_8}) at position 8.}
		\label{fig:fg_3101_3}
	\end{figure}
	\begin{figure}
		\centering
		\begin{subfigure}[b]{0.96\textwidth}
			\centering
	\pgfmathtruncatemacro{\memory}{3}%
	\pgfmathtruncatemacro{\visibility}{1}%
	\pgfmathtruncatemacro{\maxtablelength}{8}%

	\newcommand{\GetNeighborhoodSizes}[1]{%
		\xdef\toleft{0}
		\xdef\toright{0}
		\xdef\toup{0}
		\xdef\todown{0}

		\pgfmathtruncatemacro{\tableSize}{2 * \visibility + 1}
		\foreach \columnoffset/\rowoffset[count=\i] in #1 {%
			\ifnum\columnoffset<\toleft%
				\xdef\toleft{\columnoffset}%
			\fi%
			\ifnum\columnoffset>\toright%
				\xdef\toright{\columnoffset}%
			\fi%
			\ifnum\rowoffset<\todown%
				\xdef\todown{\rowoffset}%
			\fi%
			\ifnum\rowoffset>\toup%
				\xdef\toup{\rowoffset}%
			\fi%
			\coordinate (anchor_\i) at (\columnoffset, \rowoffset);
		}%

		\xdef\xbase{\toleft}
		\xdef\ybase{\todown}
		\pgfmathtruncatemacro{\pxsize}{\toright - \toleft + 1}
		\xdef\xsize{\pxsize}
		\pgfmathtruncatemacro{\pysize}{\toup - \todown + 1}
		\xdef\ysize{\pysize}

	}%

	\newcommand*{\ExtractCoordinate}[1]{\path[overlay] (#1); \pgfgetlastxy{\XCoord}{\YCoord}}%

	\newcommand{\DrawNeighborhood}[1]{%
		\begin{scope}[scale=\scalefactor]
			\begin{scope}[shift={(-\xbase, -\ybase)}]
				\draw[fill=black] (0.5, 0.5) circle (3pt);
				\draw (0, 0) rectangle (1,1);

				\foreach \value[count=\i] in #1 {%
					\ExtractCoordinate{anchor_\i}
					\begin{scope}[shift={(\XCoord, \YCoord)}]
						\ifnum\value=0%
							\draw (0, 0) rectangle (1, 1);
						\fi%
						\ifnum\value=1%
							\draw[fill=black!50] (0, 0) rectangle (1, 1);
						\fi%
						\ifnum\value=2%
							\begin{scope}[opacity=\opacityfactor, blend group=normal]
								\draw[pattern=north west lines, draw = none] (0, 0) rectangle (1, 1);
							\end{scope}
							\draw (0, 0) rectangle (1, 1);
						\fi%
						\ifnum\value=3%
							\begin{scope}[opacity=\opacityfactor]
								\clip (0, 0) rectangle (1, 1);
								\node[transform shape] at (0.5, 0.5) {\Large $\mathbf{X}$};
							\end{scope}
							\draw (0, 0) rectangle (1, 1);
						\fi%
					\end{scope}
				}%
			\end{scope}
		\end{scope}
	}%

	\newcommand{\DrawAtIndex}[1]{%
		\pgfmathtruncatemacro{\currentstate}{#1}%
		\pgfmathtruncatemacro{\tablenum}{ceil((\currentstate + 1) / \maxtablelength)}%
		\pgfmathtruncatemacro{\columnnum}{mod(\currentstate, \maxtablelength)}%

		\pgfmathsetmacro{\startoffset}{(1 - \xsize * \scalefactor) / 2}
		\begin{scope}[shift={($(left_upper_corner_\tablenum) + (\columnnum + \startoffset, 0)$)}]
			\DrawNeighborhood{\neighbors}
		\end{scope}
	}%

	\newcommand{\SetTableValues}[4]{%
		\pgfmathtruncatemacro{\currentstate}{#1}%
		\pgfmathtruncatemacro{\tablenum}{ceil((\currentstate + 1) / \maxtablelength)}%
		\pgfmathtruncatemacro{\columnnum}{mod(\currentstate, \maxtablelength)}%

		\xdef\pickedDirection{\text{,}}
		\ifnum#4=0%
			\xdef\pickedDirection{\pickedDirection\text{N}}%
		\fi%
		\ifnum#4=1%
			\xdef\pickedDirection{\pickedDirection\text{E}}%
		\fi%
		\ifnum#4=2%
			\xdef\pickedDirection{\pickedDirection\text{S}}%
		\fi%
		\ifnum#4=3%
			\xdef\pickedDirection{\pickedDirection\text{W}}%
		\fi%
		\ifnum#4=4%
			\xdef\pickedDirection{\pickedDirection\varnothing}%
		\fi%
		\begin{scope}[shift={($(left_upper_corner_\tablenum) + (\columnnum, -#2)$)}]
			\node at (\tiklabelxoffset, \tiklabelyoffset) {\small $\padzeroes[\memory]{\binarynum{#3}}\pickedDirection$};
		\end{scope}
	}%

	\pgfmathsetmacro{\scalefactor}{0.3}%
	\pgfmathsetmacro{\opacityfactor}{0.7}%
	\pgfmathsetmacro{\tiklabelxoffset}{0.5}%
	\pgfmathsetmacro{\tiklabelyoffset}{0.5}%
	\pgfmathsetmacro{\lengthlabelline}{0.8}%
	\pgfmathsetmacro{\labelxoffset}{0.5}%
	\pgfmathsetmacro{\labelyoffset}{0.1}%
	\pgfmathsetmacro{\mintablespace}{0.7}%
	\pgfmathsetmacro{\cellsize}{1}%

	\begin{tikzpicture}
		\pgfmathtruncatemacro{\numberofstates}{8}%
		\pgfmathtruncatemacro{\tableheight}{2^(\memory)}%
		\pgfmathtruncatemacro{\numberoftables}{ceil(\numberofstates / \maxtablelength)}%

		\newcommand{\neighborsCoordinates}{0/1, 0/-1, -1/0, 1/0}
		\newcommand{\neighbors}{}
		\GetNeighborhoodSizes{\neighborsCoordinates}

		\begin{scope}[scale=\cellsize]
			\foreach \tablenum in {1, ..., \numberoftables}{%
				\pgfmathtruncatemacro{\tablelength}{ifthenelse(\tablenum < \numberoftables, \maxtablelength, \numberofstates - (\numberoftables - 1) * \maxtablelength)}%
				\pgfmathsetmacro{\yoffset}{-(\tablenum - 1) * (\tableheight + \mintablespace + \scalefactor * \ysize)}%

				\begin{scope}[yshift=\yoffset cm]
					\coordinate (left_upper_corner_\tablenum) at (0, \tableheight);
					\draw (0, 0) grid (\tablelength, \tableheight);
					\begin{scope}[shift={(left_upper_corner_\tablenum)}]
						\draw (0, 0) -- ++(135:\lengthlabelline);
						\node[rotate=-45,scale=0.5] at ($(135:\labelxoffset)+(45:\labelyoffset)$) {input};
						\node[rotate=-45,scale=0.5] at ($(135:\labelxoffset)+(180+45:\labelyoffset)$) {memory};
					\end{scope}
					\foreach \state in { 1, ..., \tableheight}{%
						\pgfmathtruncatemacro{\memorystate}{\state-1}
						\node (memory_\tablenum_\state) at (-1 + \tiklabelxoffset, \tableheight - \state + \tiklabelyoffset) {$\padzeroes[\memory]{\binarynum{\memorystate}}$};
					}%
				\end{scope}
			}%

			\renewcommand{\neighbors}{0, 0, 0, 3}
			\DrawAtIndex{0}
			\renewcommand{\neighbors}{0, 0, 1, 3}
			\DrawAtIndex{1}
			\renewcommand{\neighbors}{0, 1, 0, 3}
			\DrawAtIndex{2}
			\renewcommand{\neighbors}{0, 1, 1, 3}
			\DrawAtIndex{3}
			\renewcommand{\neighbors}{1, 0, 0, 3}
			\DrawAtIndex{4}
			\renewcommand{\neighbors}{1, 0, 1, 3}
			\DrawAtIndex{5}
			\renewcommand{\neighbors}{1, 1, 0, 3}
			\DrawAtIndex{6}
			\renewcommand{\neighbors}{1, 1, 1, 3}
			\DrawAtIndex{7}

			\SetTableValues{0}{1}{1}{4}
			\SetTableValues{1}{1}{1}{4}
			\SetTableValues{2}{1}{1}{4}
			\SetTableValues{3}{1}{1}{4}
			\SetTableValues{4}{1}{1}{4}
			\SetTableValues{5}{1}{1}{4}
			\SetTableValues{6}{1}{1}{4}
			\SetTableValues{7}{1}{1}{4}
			\SetTableValues{0}{2}{2}{4}
			\SetTableValues{1}{2}{2}{4}
			\SetTableValues{2}{2}{2}{4}
			\SetTableValues{3}{2}{2}{4}
			\SetTableValues{4}{2}{2}{4}
			\SetTableValues{5}{2}{2}{4}
			\SetTableValues{6}{2}{2}{4}
			\SetTableValues{7}{2}{2}{4}
			\SetTableValues{0}{3}{3}{4}
			\SetTableValues{1}{3}{3}{2}
			\SetTableValues{2}{3}{7}{4}
			\SetTableValues{3}{3}{7}{4}
			\SetTableValues{4}{3}{3}{4}
			\SetTableValues{5}{3}{3}{2}
			\SetTableValues{6}{3}{3}{4}
			\SetTableValues{7}{3}{3}{4}
			\SetTableValues{0}{4}{4}{3}
			\SetTableValues{1}{4}{0}{4}
			\SetTableValues{2}{4}{4}{3}
			\SetTableValues{3}{4}{0}{4}
			\SetTableValues{4}{4}{4}{3}
			\SetTableValues{5}{4}{0}{4}
			\SetTableValues{6}{4}{4}{3}
			\SetTableValues{7}{4}{0}{4}
			\SetTableValues{0}{5}{5}{4}
			\SetTableValues{1}{5}{5}{4}
			\SetTableValues{2}{5}{5}{4}
			\SetTableValues{3}{5}{5}{4}
			\SetTableValues{4}{5}{5}{4}
			\SetTableValues{5}{5}{5}{4}
			\SetTableValues{6}{5}{5}{4}
			\SetTableValues{7}{5}{5}{4}
			\SetTableValues{0}{6}{6}{4}
			\SetTableValues{1}{6}{6}{0}
			\SetTableValues{2}{6}{6}{4}
			\SetTableValues{3}{6}{6}{0}
			\SetTableValues{4}{6}{2}{4}
			\SetTableValues{5}{6}{2}{4}
			\SetTableValues{6}{6}{6}{4}
			\SetTableValues{7}{6}{6}{4}
			\SetTableValues{0}{7}{7}{4}
			\SetTableValues{1}{7}{7}{4}
			\SetTableValues{2}{7}{7}{4}
			\SetTableValues{3}{7}{7}{4}
			\SetTableValues{4}{7}{7}{4}
			\SetTableValues{5}{7}{7}{4}
			\SetTableValues{6}{7}{7}{4}
			\SetTableValues{7}{7}{7}{4}
			\SetTableValues{0}{8}{4}{3}
			\SetTableValues{1}{8}{4}{4}
			\SetTableValues{2}{8}{4}{3}
			\SetTableValues{3}{8}{4}{4}
			\SetTableValues{4}{8}{4}{3}
			\SetTableValues{5}{8}{4}{4}
			\SetTableValues{6}{8}{4}{3}
			\SetTableValues{7}{8}{4}{4}

		\end{scope}
	\end{tikzpicture}
			\caption{Position 9}
			\label{subfig:fg_3101_9}
		\end{subfigure}
		\caption{State machine(s) (\subref{subfig:fg_3101_9}) at position 9.}
		\label{fig:fg_3101_4}
	\end{figure}

		\subsection{A continue agent}
			\label{sec:Sorting.Algorithm.Continue}
			\vspace{33mm}
			
	\begin{figure}[ht]
		\centering
		\begin{subfigure}[b]{0.36\textwidth}
			\centering
			\subimport{Ver.3102/}{Table.Ver_3102.Position_1}
			\caption{Position 1}
			\label{subfig:fg_3102_1}
		\end{subfigure}
		\hfill
		\begin{subfigure}[b]{0.60\textwidth}
			\centering
			\subimport{Ver.3102/}{Table.Ver_3102.Position_5}
			\caption{Position 5}
			\label{subfig:fg_3102_5}
		\end{subfigure}
		\caption{State machine(s) (\subref{subfig:fg_3102_1}) at position 1, (\subref{subfig:fg_3102_5}) at position 5.}
		\label{fig:fg_3102_1}
	\end{figure}
	\begin{figure}
		\centering
		\begin{subfigure}[b]{0.32\textwidth}
			\centering
			\subimport{Ver.3102/}{Table.Ver_3102.Position_2}
			\caption{Position 2}
			\label{subfig:fg_3102_2}
		\end{subfigure}
		\hfill
		\begin{subfigure}[b]{0.32\textwidth}
			\centering
			\subimport{Ver.3102/}{Table.Ver_3102.Position_3}
			\caption{Position 3}
			\label{subfig:fg_3102_3}
		\end{subfigure}
		\hfill
		\begin{subfigure}[b]{0.32\textwidth}
			\centering
			\subimport{Ver.3102/}{Table.Ver_3102.Position_4}
			\caption{Position 4}
			\label{subfig:fg_3102_4}
		\end{subfigure}
		\caption{State machine(s) (\subref{subfig:fg_3102_2}) at position 2, (\subref{subfig:fg_3102_3}) at position 3, (\subref{subfig:fg_3102_4}) at position 4.}
		\label{fig:fg_3102_2}
	\end{figure}
	\begin{figure}
		\centering
		\begin{subfigure}[b]{0.36\textwidth}
			\centering
			\subimport{Ver.3102/}{Table.Ver_3102.Position_6}
			\caption{Position 6}
			\label{subfig:fg_3102_6}
		\end{subfigure}
		\hfill
		\begin{subfigure}[b]{0.60\textwidth}
			\centering
			\subimport{Ver.3102/}{Table.Ver_3102.Position_7}
			\caption{Position 7}
			\label{subfig:fg_3102_7}
		\end{subfigure}
		\caption{State machine(s) (\subref{subfig:fg_3102_6}) at position 6, (\subref{subfig:fg_3102_7}) at position 7.}
		\label{fig:fg_3102_3}
	\end{figure}
	\begin{figure}
		\centering
		\begin{subfigure}[b]{0.48\textwidth}
			\centering
			\subimport{Ver.3102/}{Table.Ver_3102.Position_8}
			\caption{Position 8}
			\label{subfig:fg_3102_8}
		\end{subfigure}
		\hfill
		\begin{subfigure}[b]{0.48\textwidth}
			\centering
			\subimport{Ver.3102/}{Table.Ver_3102.Position_9}
			\caption{Position 9}
			\label{subfig:fg_3102_9}
		\end{subfigure}
		\caption{State machine(s) (\subref{subfig:fg_3102_8}) at position 8, (\subref{subfig:fg_3102_9}) at position 9.}
		\label{fig:fg_3102_4}
	\end{figure}

\ifnum\mydoclevel=0
	\section{\texorpdfstring{\cref{thm:The.A.Theorem}}{Algorithm A Soundness Theorem} proof}\label{sec:The.Proof}
\else
	\section{Algorithm A Soundness Theorem proof}\label{sec:The.Proof}
\fi
	\begin{reptheorem}{thm:The.A.Theorem}[Algorithm $\mathcal{A}$ Soundness Theorem]
		Consider an initial configuration $\mathcal{C}$ of an $n \times m$ grid $G$, where
		$m \geq 3, n \geq 2$. Suppose $\mathcal{C}$ meets \eqref{eq:C_1} constraint criteria. 
		Then Algorithm $\mathcal{A}$ solves the ER Problem and transitions $G$ to a 
		target configuration.
	\end{reptheorem}

	\subsection{Impossible inputs and collision avoidance}
		\begin{figure}[b!]
			\centering
				\begin{tikzpicture}
	
		\pgfmathsetmacro{\halfsqrootoften}{1.5815}
		\begin{scope}[xshift=-1.25 cm]
			\begin{scope}[scale = 0.5]
				\draw (0,0) grid (5, 3);
				\foreach \point in {(0,1),(1,1),(3,0),(3,1),(3,2)}
				{
					\begin{scope}
						\draw[fill=black!60] \point rectangle ++(1, 1);
					\end{scope}
				}
				\begin{scope}[shift={(4, 1)}]
					\draw[pattern=north west lines] (0, 0) rectangle (1, 1);
				\end{scope}
				
				\begin{scope}[opacity=0.7]
					\draw (1.5, 1.5) circle (\halfsqrootoften cm);
					\draw (3.5, 1.5) circle (\halfsqrootoften cm);
				\end{scope}
				
				\coordinate (A1) at ($(1.5, 1.5)+(270:\halfsqrootoften)$);
				\coordinate (A2) at ($(3.5, 1.5)+(270:\halfsqrootoften)$);
			\end{scope}
		\end{scope}
		
		\begin{scope}[shift={(-3, -3)}]
			\clip (0, 0) circle (2.1 cm);
			\draw (0, 0) circle (2 cm);
			\begin{scope}[shift={(-0.5, -0.5)}]
				\draw (-1, 0) grid +(3, 1);
				\draw (0, -1) grid +(1, 3);
				\begin{scope}[opacity=0.5]
					\node at ($(-1, 0) + (0.5, 0.5)$) {X};		
					\node at ($(0, -1) + (0.5, 0.5)$) {X};		
				\end{scope}
				\draw[fill=black] (0.5, 0.5) circle (3pt);	
			\end{scope}
			\draw[ultra thick, -latex] (0, 0) -- +(1, 0);
			\coordinate (B1) at ($(0,0)+(45:2)$);
		\end{scope}
		
		\begin{scope}[shift={(3, -3)}]
			\clip (0, 0) circle (2.1 cm);
			\draw (0, 0) circle (2 cm);
			\draw[fill=black] (0, 0) circle (3pt);	
			\begin{scope}[shift={(-0.5, -0.5)}]
				\draw (-1, 0) grid +(3, 1);
				\draw (0, -1) grid +(1, 3);
				\draw[pattern=north west lines] (1, 0) rectangle +(1, 1);
				\begin{scope}[opacity=0.5]
					\node at ($(0, 1) + (0.5, 0.5)$) {X};		
				\end{scope}
				\draw[fill=black!60] (0, -1) rectangle +(1, 1);
			\end{scope}
			\draw[ultra thick, -latex] (0, 0) -- +(-1, 0);
			\coordinate (B2) at ($(0,0)+(135:2)$);
		\end{scope}
		
		\draw (B1) -- (A1);
		\draw (B2) -- (A2);
	\end{tikzpicture}
			\caption{Possible collision under the $L_1$ visibility of radius $1$.
			Two agents decide to move into the same empty space. Observed neighborhoods
			and movement decisions.}
			\label{img:collision}
		\end{figure}
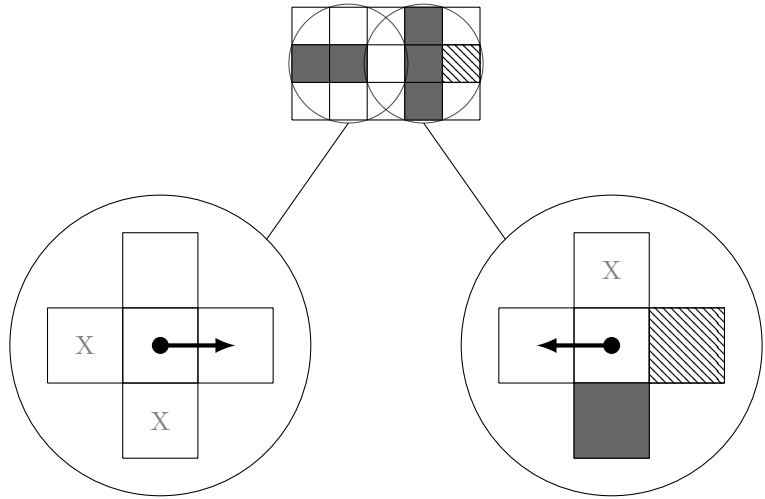
		
		Movement rules defined in short form in
		\cref{sec:Sorting.Algorithm.Exit} and \cref{sec:Sorting.Algorithm.Continue}
		result in a movement into an adjacent empty cells only. A fast enumeration ensures:
		no rule directs an agent to move into an occupied cell or into a neighboring border cell.
		However, a different type of collision could occur if two or more agents decide to move 
		into the same empty cell on the grid at the same time (see \cref{img:collision} for
		an illustration of a possible collision situation).
		Therefore, we shall thoroughly check all the potential moving agent pairs.
		Unsurprisingly, Algorithm $\mathcal{A}$ was designed with collision prevention in mind. 
		Since agent does not posses abilities to detect another agent \enquote{over the gap},
		agent movement should be synchronized in one way or another. We have implemented
		synchronization using a modulo $4$ timer, i.e. a timer counting $0, 1, 2, 3$, then
		count (cycle) repeats. Under the $4$ cycles clock regime, we allocate 
		specific times for the movement in the specific directions as follows:
		\begin{itemize}
			\item
				Cycles $0$ and $2$ will be assigned to southbound traffic.
			\item
				Cycle $1$ will be allocated to northbound traffic.
			\item
				The remaining cycle $3$ will be utilized for westward movement.
		\end{itemize}

	\begin{table}
		\centering
			{
			\rowcolors{2}{black!30}{black!5}
				\begin{tabular}{c ||c c c c c c c c c } 
					\hline
					Position & 1 & 2 & 3 & 4 & 5 & 6 & 7 & 8 & 9 \\ [0.5ex] 
					\hline\hline
					00 (0) & - & N & E & S & - &	N & E & S & - \\ 
					\hline
					01 (1) & - & N & - & S & N, W &	N & - & S & N \\
					\hline
					10 (2) & - & N & E & S & - &	N & E & S & S \\
					\hline
					11 (3) & - & - & E & S & W &	- & E & S & W \\ [1ex] 
					\hline
				\end{tabular}
			}
			\caption{Possible movement direction for exiting vehicle. Times on the left are given in the binary format.
				Columns enumeration is done according to agents' position on the grid (from $1$ to $9$). $N, E, S, W$
				are four possible directions (north, east, south, and west).}
			\label{tbl:MovementTable.Exit}
	\end{table}

	\begin{table}
		\centering
			{
			\rowcolors{2}{black!30}{black!5}
				\begin{tabular}{c ||c c c c c c c c c } 
					\hline
					Position & 1 & 2 & 3 & 4 & 5 & 6 & 7 & 8 & 9 \\ [0.5ex] 
					\hline\hline
					00 (0) & - & N & - & S & - &	N & - &		S &		S \\ 
					\hline
					01 (1) & W & - & - & S & N, W &	- & - &		S &		N \\
					\hline
					10 (2) & - & - & E & S & - &	- & E, S &	S &		- \\
					\hline
					11 (3) & - & - & E & S & - &	- & E &		S, W &	- \\ [1ex] 
					\hline
				\end{tabular}
			}
			\caption{Possible movement direction for continuing vehicle. Times on the left are given in the binary format.
				Columns enumeration is done according to agents' position on the grid (from $1$ to $9$). $N, E, S, W$
				are four possible directions (north, east, south, and west).}
			\label{tbl:MovementTable.Continue}
	\end{table}

		Nonetheless, we shall allow traffic in additional directions, if such traffic does not interfere
		with potential traffic in the designated directions at the same cycle. 
		For example, northbound traffic that enters the last $n$th row is impossible. Therefore
		a westward movement in the last row at tick $1$ does not pose a collision threat.
		Moreover, a southbound traffic could be allowed in the last column at the same tick.
		Since, the last column agents are unaffected by the two previous rules at tick $1$.
		
		Under the same principle, we can allow eastern movement in the first row at all times, except at
		cycle $1$. The already mentioned southern movement at the last column is possible around the clock. 
		Finally, we could speed up the movement in the first column. However, to make sure 
		that the Algorithm resolves the ER Problem, we prefer to implement speedup only for 
		the exiting agents.
		
		We list all possible movement directions in \cref{tbl:MovementTable.Exit} 
		and \cref{tbl:MovementTable.Continue}.
		Thus agents stay at place if the movement in the desired direction could lead to a possible
		collision with agents moving in the designated direction.
		We summarize the above ideas in the following Lemma.
		
		\begin{lemma}[No Collisions]\label{lemma: No Collisions}
			Let $\mathcal{C}$ be a legal initial configuration of an $n \times m$ grid $G$. 
			Then, simultaneous Algorithm $\mathcal{A}$ execution by the agents on the
			grid does not lead to collision at any future time $t > 0$.
		\end{lemma}	
		
		\cref{lemma: No Collisions} does not assume \eqref{eq:C_0} or \eqref{eq:C_1} requirements. 
		Thus it is a claim about Algorithm $\mathcal{A}$ and not about the geometry of the grid $G$.
		
		Algorithm $\mathcal{A}$ is only partially defined in \cref{sec:Sorting.Algorithm.Exit} and 
		\cref{sec:Sorting.Algorithm.Continue}. However, the missing input tuples are unattainable 
		from \textit{any} initial configuration $\mathcal{C}$ of the grid $G$. I.e. we further show
		that an \textit{agent type-state-neighborhood} tuple $I=(\uptau, S, N)$ with an undefined value 
		$\mathcal{A}(I)$ is not an attainable input tuple. Specifically, an agent of 
		\textit{agent type} $\uptau$, sensing a \textit{neighborhood} $N$ \textbf{can not} 
		possibly be in the \textit{state} $S$.
		Moreover, some other rules are in fact also unreachable, but are left in the tables for convenience. 

		\begin{lemma}[Algorithm is Well-Defined Lemma]
			Distributed parallel execution of Algorithm $\mathcal{A}$ does not lead to an undefined 
			behavior, due to the partial definition of the local rule of exchanges $\mathcal{A}$.
		\end{lemma}
		\begin{proof}
			Recall, that we conceptually treat an agent memory state, as the binary value resulting
			from concatenation of the Direction Bit ($d$) and the current value of a $4$-tick timer 
			held by an agent, where $d$ is treated as an MSB bit.
			Denote an agent memory state according to the MSB value of its binary representation as
			$0$-state or $1$-state respectively.
			
			We notice, that rules for both agent types are fully defined for all $0$-states (i.e. 
			for all states where Direction Bit ($d$) is set to $0$). Moreover, for a given pair 
			\textit{agent type - grid position} the following dichotomy holds:
			\begin{itemize}
				\item
					agent movement rules are defined for \textbf{all} possible neighborhoods
					and for \textbf{all} possible agent memory states, incl. $1$-states.
				\item
					only \textbf{zero} rules are defined for $1$-states in total.
			\end{itemize}
			
			Therefore, if an agent of type $\uptau$ moves in direction $D$ on the grid and its
			Direction Bit $d$ is set to $1$, we need to check that the first option holds.
			
			The full enumeration of rules is given in
			\cref{tbl:StateMachineHandling.Exit} and \cref{tbl:StateMachineHandling.Continue}.
			
	\begin{table}
		\centering
			{
			\rowcolors{2}{blue!30}{white}
				\begin{tabular}{c || c | c c | c c} 
					\hline \\ [1ex]
					Position & Direction & \MyHead{1.3 cm}{new \\ Position} & 
					\MyHead{1.3 cm}{handles \\ $d = 1$} & \MyHead{1.3 cm}{new \\ Position} & \MyHead{1.3 cm}{handles \\ $d = 1$} \\ [3ex] 
					\hline\hline
					5 & $\varnothing$ & 5 & \cmark & & \\ 
					\hline
					5 & 			N & 7 & \cmark & 9 & \cmark \\ 
					\hline
					5 & 			W & 2 & \cmark & 5 & \cmark \\
					\hline
					9 & $\varnothing$ & 9 & \cmark & & \\ 
					\hline
					9 & 			N & 7 & \cmark & 9 & \cmark \\ 
					\hline
					9 & 			W & 6 & \cmark & 9 & \cmark \\ 
					\hline
				\end{tabular}
			}
			\caption{Exit vehicle rules that set Direction Bit($d$) to $1$. An agent follows the
			prescribed motion and moves to, a possibly, different position on the grid (e.g. an agent moving 
			North from position $5$ could move into exactly	two grid positions: $7$ and $9$ respectively). 
			In case that the agent position's after the move rules support memory state(s) of the form $1\mathrm{xx}_2$,
			\space\cmark\space is set, otherwise \space\xmark\space is set. Possible agent actions[move to] are
			N - north, E - east, S - south, W - west and $\varnothing$ - stay.}
			\label{tbl:StateMachineHandling.Exit}
	\end{table}

	\begin{table}
		\centering
			{
			\rowcolors{2}{blue!30}{white}
				\begin{tabular}{c || c | c c | c c} 
					\hline \\ [1ex]
					Position & Direction & \MyHead{1.3 cm}{new \\ Position} & 
					\MyHead{1.3 cm}{handles \\ $d = 1$} & \MyHead{1.3 cm}{new \\ Position} & \MyHead{1.3 cm}{handles \\ $d = 1$} \\ [3ex] 
					\hline\hline
					1 & $\varnothing$ & 1 & \cmark & & \\ 
					\hline              
					1 & 			W & 5 & \cmark & & \\ 
					\hline              
					5 & $\varnothing$ & 5 & \cmark & & \\ 
					\hline              
					5 & 			N & 7 & \cmark & 9 & \cmark \\ 
					\hline              
					5 & 			W & 2 & \cmark & 5 & \cmark \\
					\hline              
					8 & 			W & 9 & \cmark & & \\ 
					\hline              
					9 & $\varnothing$ & 9 & \cmark & & \\ 
					\hline              
					9 & 			N & 7 & \cmark & 9 & \cmark \\ 
					\hline
				\end{tabular}
			}
			\caption{Continue agent rules that set Direction Bit($d$) to $1$. An agent follows the
			prescribed motion and moves to, a possibly, different position on the grid (e.g. an agent moving 
			North from position $5$ could move into exactly	two grid positions: $7$ and $9$ respectively). 
			In case that the agent position's after the move rules support memory state(s) of the form $1\mathrm{xx}_2$,
			\space\cmark\space is set, otherwise \space\xmark\space is set. Possible agent actions[move to] are
			N - north, E - east, S - south, W - west and $\varnothing$ - stay.}
			\label{tbl:StateMachineHandling.Continue}
	\end{table}

			In summary, every \textbf{possible} input tuple has an appropriate movement rule
			defined. Hence, no agent is stuck in the undefined state throughout the Algorithm execution.
		\end{proof}
	
	\subsection[Boundary empty spaces flow]{Empty spaces \enquote{turn-around} on a boundary and a hypothetical unsolvable ER problem Instance}
	The Algorithm is designed in a way to enforce a clock-wise agent movement on the
	edges of the grid. We elaborate this fact in the following straightforward observations.
	\begin{observation}
		Agents move East in the first (upper) row \textbf{only}.
	\end{observation}
	\begin{observation}
		Exiting agents do \textbf{not} leave the last (right-most) column.
	\end{observation}
	\begin{observation}
		Agents in the first (left-most) column move North \textbf{only}.
	\end{observation}
	
	However, sometimes agents move into the first row \enquote{unintentionally}, i.e. not 
	as a part of the routine clock-wise motion around the frame from the 
	left most column. In this case, exiting agents \textit{exploit} the opportunity to minimize the 
	traveling time to the target column	and remain in the first row. On the other hand, 
	continue agents have no immediate	benefit to reside in the first row and move back. 
	We summarize this insight below:
	
	\begin{lemma}[No Accidental Shortcuts to the Front Row Lemma]\label{lemma:Weird Move Lemma}
		Suppose continue agent $A$ occupies grid cell $a_{2, k}$ on a $n \times m$ grid,
		where $1 < k < m$.
		Assume $A$ moves North at time $t \equiv 01_2 \mod 4$. Then, $A$ moves back from
		$a_{1, k}$ to $a_{2, k}$ at time $t+1$.
	\end{lemma}
	\begin{proof}
		$a_{2,k}$ is a position $9$ grid cell, while $a_{1, k}$ - is a 
		position $7$ cell.
		A continue agent moving North from position $9$ sets the Direction bit to $1$. Therefore,
		an agent $A$ is in the state $110_2$ at $t+1$. According to \cref{subfig:fg_3102_7} a
		continue agent will move S(outh) at $(t+1) \equiv 10_2 \mod 4$, since two 
		possible neighborhood readings at time $t+1$ are: %
		\adjustbox{max width=1.5cm}{%

	\pgfmathtruncatemacro{\memory}{-1}%
	\pgfmathtruncatemacro{\visibility}{1}%
	\pgfmathtruncatemacro{\maxtablelength}{8}%

	\newcommand{\GetNeighborhoodSizes}[1]{%
		\xdef\toleft{0}
		\xdef\todown{0}

		\pgfmathtruncatemacro{\tableSize}{2 * \visibility + 1}
		\foreach \columnoffset/\rowoffset[count=\i] in #1 {%
			\ifnum\columnoffset<\toleft%
				\xdef\toleft{\columnoffset}%
			\fi%
			\ifnum\rowoffset<\todown%
				\xdef\todown{\rowoffset}%
			\fi%
			\coordinate (anchor_\i) at (\columnoffset, \rowoffset);
		}%

		\xdef\xbase{\toleft}
		\xdef\ybase{\todown}

	}%

	\newcommand*{\ExtractCoordinate}[1]{\path[overlay] (#1); \pgfgetlastxy{\XCoord}{\YCoord}}%

	\newcommand{\DrawNeighborhood}[1]{%
		\begin{scope}[scale=\scalefactor]
			\begin{scope}[shift={(-\xbase, -\ybase)}]
				\draw[fill=black] (0.5, 0.5) circle (3pt);
				\draw (0, 0) rectangle (1,1);

				\foreach \value[count=\i] in #1 {%
					\ExtractCoordinate{anchor_\i}
					\begin{scope}[shift={(\XCoord, \YCoord)}]
						\ifnum\value=1%
							\fill[black!50] (0, 0) rectangle (1, 1);
						\fi%
						\ifnum\value=2%
							\begin{scope}[opacity=\opacityfactor, blend group=normal]
								\draw[pattern=north west lines, draw = none] (0, 0) rectangle (1, 1);
							\end{scope}
						\fi%
						\ifnum\value=3%
							\begin{scope}[opacity=\opacityfactor]
								\node[transform shape] at (0.5, 0.5) {\Large $\mathbf{X}$};
							\end{scope}
						\fi%
						\draw (0, 0) rectangle (1, 1);
					\end{scope}
				}%
			\end{scope}
		\end{scope}
	}%

	\pgfmathsetmacro{\scalefactor}{0.3}%
	\pgfmathsetmacro{\opacityfactor}{0.7}%
	\pgfmathsetmacro{\tiklabelxoffset}{0.5}%
	\pgfmathsetmacro{\tiklabelyoffset}{0.5}%

	\begin{tikzpicture}
		\newcommand{\neighborsCoordinates}{1/0, 0/1, 0/-1, -1/0}
		\newcommand{\neighbors}{}
		\GetNeighborhoodSizes{\neighborsCoordinates}

		\renewcommand{\neighbors}{0, 2, 3, 3}
		\DrawNeighborhood{\neighbors}
	\end{tikzpicture}
		}
		or 
		\adjustbox{max width=1.5cm}{%

	\pgfmathtruncatemacro{\memory}{-1}%
	\pgfmathtruncatemacro{\visibility}{1}%
	\pgfmathtruncatemacro{\maxtablelength}{8}%

	\newcommand{\GetNeighborhoodSizes}[1]{%
		\xdef\toleft{0}
		\xdef\todown{0}

		\pgfmathtruncatemacro{\tableSize}{2 * \visibility + 1}
		\foreach \columnoffset/\rowoffset[count=\i] in #1 {%
			\ifnum\columnoffset<\toleft%
				\xdef\toleft{\columnoffset}%
			\fi%
			\ifnum\rowoffset<\todown%
				\xdef\todown{\rowoffset}%
			\fi%
			\coordinate (anchor_\i) at (\columnoffset, \rowoffset);
		}%

		\xdef\xbase{\toleft}
		\xdef\ybase{\todown}

	}%

	\newcommand*{\ExtractCoordinate}[1]{\path[overlay] (#1); \pgfgetlastxy{\XCoord}{\YCoord}}%

	\newcommand{\DrawNeighborhood}[1]{%
		\begin{scope}[scale=\scalefactor]
			\begin{scope}[shift={(-\xbase, -\ybase)}]
				\draw[fill=black] (0.5, 0.5) circle (3pt);
				\draw (0, 0) rectangle (1,1);

				\foreach \value[count=\i] in #1 {%
					\ExtractCoordinate{anchor_\i}
					\begin{scope}[shift={(\XCoord, \YCoord)}]
						\ifnum\value=1%
							\fill[black!50] (0, 0) rectangle (1, 1);
						\fi%
						\ifnum\value=2%
							\begin{scope}[opacity=\opacityfactor, blend group=normal]
								\draw[pattern=north west lines, draw = none] (0, 0) rectangle (1, 1);
							\end{scope}
						\fi%
						\ifnum\value=3%
							\begin{scope}[opacity=\opacityfactor]
								\node[transform shape] at (0.5, 0.5) {\Large $\mathbf{X}$};
							\end{scope}
						\fi%
						\draw (0, 0) rectangle (1, 1);
					\end{scope}
				}%
			\end{scope}
		\end{scope}
	}%

	\pgfmathsetmacro{\scalefactor}{0.3}%
	\pgfmathsetmacro{\opacityfactor}{0.7}%
	\pgfmathsetmacro{\tiklabelxoffset}{0.5}%
	\pgfmathsetmacro{\tiklabelyoffset}{0.5}%

	\begin{tikzpicture}
		\newcommand{\neighborsCoordinates}{1/0, 0/-1, 0/1, -1/0}
		\newcommand{\neighbors}{}
		\GetNeighborhoodSizes{\neighborsCoordinates}

		\renewcommand{\neighbors}{1, 0, 2, 3}
		\DrawNeighborhood{\neighbors}
	\end{tikzpicture}
		}
		, as claimed.
	\end{proof}
	
	Let us consider $\mathcal{P}(n, m, N_1, N_0, \mathcal{C}(0))$ an instance of the ER problem, such
	that $\mathcal{C}(0)$ is an initial configuration of an $n \times m$ grid with $N_1$ 
	exiting agents and $N_0$ empty cells. We assume problem parameters 
	$n$, $m$, $N_1$ and $N_0$ are natural numbers satisfying \eqref{eq:C_1}. 
	
	Let $\mathcal{C}(t)$ be a grid configuration at time $t \in \mathbb{N}$.
	
	We shall follow the \textit{proof by contradiction} approach to show the correctness of
	\cref{thm:The.A.Theorem}.
	Assume the opposite, i.e. that there exists
	an instance of the problem $\mathcal{P}(n, m, N_1, N_0, \mathcal{C}(0))$, such that no 
	grid configuration $\mathcal{C}(0), \mathcal{C}(1), \mathcal{C}(2), \ldots$ 
	is a grid target configurations. Where configuration $\mathcal{C}(t+1)$ succeeds the
	grid configuration $\mathcal{C}(t)$ at time $t+1$, and, is in fact, a result of decentralized
	simultaneous execution of Algorithm $\mathcal{A}$ by grid agents at time $t \in \mathbb{N}$.
	
	We now proceed to characterize a macroscopic emergent agent behavior.
	Exiting agents do not leave the first row, except to the last column (South movement
	from position $4$ to position $8$). And never leave the last column. Moreover, the first row
	is the only row in which agents execute eastward motion. Therefore, once 
	an exiting agent reaches to the first row, it either gets \enquote{stuck} in
	the first row forever, or, proceeds in the first row eastward and gets \enquote{stuck} in the 
	last column.
	
	Hence, we can make the following observation:
	\begin{observation} \label{observation: west-east exiting movements}
		An exiting agent $A$ executes at most $2m$ East/West moves during Algorithm $\mathcal{A}$
		execution.
	\end{observation}
	
	The observation holds true according to the behavior characteristic provided above, i.e. an
	exiting agent ceases to move West after a single East step. Furthermore, 
	the number of moves in each direction is bounded by the number of columns.
	
	\InsertEmptyLine
	Let $T_0$ - be the time tick just after the last column change by any exiting agent.
	$T_0$ is well defined according to the above observation. 
	
	Since $\mathcal{C}(T_0)$ is not a target configuration by our assumption, all the exiting 
	agents on the grid are \enquote{stuck} in their current column, i.e. they move in the North-South 
	direction only, if at all. And at least one exiting agent is not located in the right-most column.
	
	We claim, that :
	\begin{lemma}[No \enquote{Parking} at $a_{1, m}$ Lemma]\label{lemma:right-upper-open-corner}
		Grid cell $a_{1, m}$ becomes empty infinitely often, i.e.
		$\left|\{t | t > T_0 \text{ and } a_{1, m}(t) = 0\}\right| = \infty$.
	\end{lemma}

		Suppose this is not the case, let $T_1 > T_0$ be the time such that the upper right corner cell
		of the grid ($a_{1, n}$) is occupied forever after $T_1$, i.e.
		\begin{equation*}
			\forall t \geq T_1 \quad a_{1, m}(t) \neq 0
		\end{equation*}

		Then none of the last column cells is empty after $T_1$ due to the fact that all the agents
		in the last column of the grid (position $4$ and $8$) move South once neighbored by an empty cell.
		However, continue agents could also leave westward. Nonetheless, the observed behavior is
		a \enquote{North propagation} of an empty slot at the pace of one cell at a single time tick.		
		Moreover, no rule allows eastward moves into the last column. Except in the 
		first row, where an assumed \enquote{stuck} agent is located ($a_{1, m}$). Therefore, an iterative 
		application of Algorithm $\mathcal{A}$ by agents in the last column will inevitably \enquote{bring} 
		an empty slot to neighbor an agent occupying $a_{1, m}$ from the South.
		
		Since we assumed that an agent at $a_{1, m}$ is \enquote{stuck}, we are forced to conclude
		the whole $m$th column is \enquote{stuck}, i.e. no movement to/from/in it is possible
		after $T_1$.
		
		\InsertEmptyLine
		We observe, that the \enquote{flow} of empty spaces counters the movement of agents and 
		could be described as a motion in the counter clock-wise direction. 
		Moreover, the only area where empty spaces could \enquote{move} West is the first row. Therefore,
		it is unavoidable that East-West \enquote{movement} will cease, i.e. there is no source or sink for
		empty spaces on the grid. Denote $T_2 > T_1$ - the time tick of the last 
		East/West move by any agent.
		
		Let $k_g$ - be the highest column index, such that for $t > T_2$ columns $k_g+1, k_g+2, \ldots, m$ 
		do not contain, but column $k_g$ \textbf{does}. $k_g$ is well-defined, since there is at 
		least one empty cell on the grid. And, according to the previous proposition $1 \leq k_g < m$.
		
		\begin{observation}
			For $1 < i < m$ an agent $A$ moving North from $a_{n, i}$ at time $t > T_2$,
			does not return back before $t+5$.
		\end{observation}
		\begin{proof}
			Any agent in a column, except the first one, leaving North from $a_{n, i}$ sets its state to $110_2$. 
			After the East-West movement cessation, agent state could according to the following sequence only:
			\begin{equation*}	
				110_2 \rightarrow 111_2 \rightarrow 100_2 \rightarrow 101_2,
			\end{equation*}
			(see \cref{subfig:fg_3101_9} and \cref{subfig:fg_3102_9}).
			And only then a agent at position $9$ could change its state to $010_2$. In case $a_{n ,i}$
			remains empty, it is possible for agent $A$ to move back. However only an
			exiting agent could move back South at $t+5$. A continuing agent could not move back before
			$t + 7$.
			
			The only exception, is the $n=2$ case. An agent leaving $a_{n, i}$ moves from position 
			$5$ on the grid to the position $7$ and then immediately back. Although, a first row agent in the
			highest index column $j < i$ will immediately move East in the next tick, breaking no East-West 
			movement assumption.
			In the absence of such $j$, we have that an agent in the lowest-indexed non-empty column either:
			\begin{enumerate}[label=(\alph*)]
				\item
					moves West to the \textit{empty} column on the left,
				\item
					moves North, if $j = 1$, then eventually East.
			\end{enumerate}
			In both cases, after $T_2$ the above behavior will contradict our no East-West 
			movement assumption.
		\end{proof}

		\begin{observation}
			The previous claim holds true also for agents moving from cell $a_{j, i}$ at Position $9$ to 
			cell $a_{j - 1, i}$ at Position $9$. In other words, it takes at least
			five time ticks for a North-bound leaving agent to return to its previous position.
		\end{observation}
		\begin{observation}
			A similar claim holds for agents moving South in those internal columns. From position $9$
			to either position $5$ or $9$. With roles exchanged, i.e. it takes at least $7$
			time ticks for an exiting agent to return to its previous position, and at least $5$
			time ticks for a continue agent.
		\end{observation}
		
		Let us denote a grid cell that remains occupied (forever) after some time $T$ as a \textit{SOFT cell}.
		We then continue to investigate the dynamics in the column $k_g$, after all the SOFT grid cells
		were occupied for the last time. Curiously, we shall show, that all the $k_g$ column cells
		are non-SOFT grid cells.
		
		\begin{lemma}[No Partially Stuck Columns Lemma]\label{lemma:k_g column}
			Let $k_g < n$ be the index of a column, that contains empty cells after $T_2$,
			East-West movement termination point, but all
			columns right to the column $k_g$ are SOFT cell columns. Then all cells in the $k_g$
			column are non-SOFT cells. Except (maybe) for the $a_{1, k_g}$ cell.
		\end{lemma}
		\begin{proof}
			Suppose $a_{i, k_g}$ is a SOFT cell, that is occupied forever after w.l.o.g. $T_2$, by an agent $A$, 
			and $a_{i+1, k_g}$ is a non-SOFT cell. Suppose at some time $t \equiv 00_2$
			$a_{i+1, k_g}$ become empty [it either become empty, since an agent occupying it 
			moved South, or it was empty for all $t \geq T_2$]. By the observations above, an agent that
			left $a_{i+1, k_g}$ could not return to $a_{i+1, k_g}$ before $t+5$. Moreover, since East-Wast
			movement ceased at $T_1$ it is not possible that the cell $a_{i+1, k_g}$ becomes occupied by an agent 
			moving west from the column $k_g + 1$.
			
			If $A$ is a continue agent, it will either move North at $t+1$, or will set the
			state to $010_2$ and move South at $t+4$. Recall, that $A$ occupies a SOFT cell, hence
			leaving $a_{i, k_g}$ is a contradiction to assumptions.
			
			If $A$ is an exiting agent, the analysis is a bit more involved. Readings
			\adjustbox{max width=1.5cm}{%

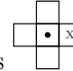
	\pgfmathtruncatemacro{\memory}{-1}%
	\pgfmathtruncatemacro{\visibility}{1}%
	\pgfmathtruncatemacro{\maxtablelength}{8}%

	\newcommand{\GetNeighborhoodSizes}[1]{%
		\xdef\toleft{0}
		\xdef\todown{0}

		\pgfmathtruncatemacro{\tableSize}{2 * \visibility + 1}
		\foreach \columnoffset/\rowoffset[count=\i] in #1 {%
			\ifnum\columnoffset<\toleft%
				\xdef\toleft{\columnoffset}%
			\fi%
			\ifnum\rowoffset<\todown%
				\xdef\todown{\rowoffset}%
			\fi%
			\coordinate (anchor_\i) at (\columnoffset, \rowoffset);
		}%

		\xdef\xbase{\toleft}
		\xdef\ybase{\todown}

	}%

	\newcommand*{\ExtractCoordinate}[1]{\path[overlay] (#1); \pgfgetlastxy{\XCoord}{\YCoord}}%

	\newcommand{\DrawNeighborhood}[1]{%
		\begin{scope}[scale=\scalefactor]
			\begin{scope}[shift={(-\xbase, -\ybase)}]
				\draw[fill=black] (0.5, 0.5) circle (3pt);
				\draw (0, 0) rectangle (1,1);

				\foreach \value[count=\i] in #1 {%
					\ExtractCoordinate{anchor_\i}
					\begin{scope}[shift={(\XCoord, \YCoord)}]
						\ifnum\value=1%
							\fill[black!50] (0, 0) rectangle (1, 1);
						\fi%
						\ifnum\value=2%
							\begin{scope}[opacity=\opacityfactor, blend group=normal]
								\draw[pattern=north west lines, draw = none] (0, 0) rectangle (1, 1);
							\end{scope}
						\fi%
						\ifnum\value=3%
							\begin{scope}[opacity=\opacityfactor]
								\node[transform shape] at (0.5, 0.5) {\Large $\mathbf{X}$};
							\end{scope}
						\fi%
						\draw (0, 0) rectangle (1, 1);
					\end{scope}
				}%
			\end{scope}
		\end{scope}
	}%

	\pgfmathsetmacro{\scalefactor}{0.3}%
	\pgfmathsetmacro{\opacityfactor}{0.7}%
	\pgfmathsetmacro{\tiklabelxoffset}{0.5}%
	\pgfmathsetmacro{\tiklabelyoffset}{0.5}%

	\begin{tikzpicture}
		\newcommand{\neighborsCoordinates}{0/1, 0/-1, -1/0, 1/0}
		\newcommand{\neighbors}{}
		\GetNeighborhoodSizes{\neighborsCoordinates}

		\renewcommand{\neighbors}{0, 0, 0, 3}
		\DrawNeighborhood{\neighbors}
	\end{tikzpicture}
			}
			and
			\adjustbox{max width=1.5cm}{%

	\pgfmathtruncatemacro{\memory}{-1}%
	\pgfmathtruncatemacro{\visibility}{1}%
	\pgfmathtruncatemacro{\maxtablelength}{8}%

	\newcommand{\GetNeighborhoodSizes}[1]{%
		\xdef\toleft{0}
		\xdef\todown{0}

		\pgfmathtruncatemacro{\tableSize}{2 * \visibility + 1}
		\foreach \columnoffset/\rowoffset[count=\i] in #1 {%
			\ifnum\columnoffset<\toleft%
				\xdef\toleft{\columnoffset}%
			\fi%
			\ifnum\rowoffset<\todown%
				\xdef\todown{\rowoffset}%
			\fi%
			\coordinate (anchor_\i) at (\columnoffset, \rowoffset);
		}%

		\xdef\xbase{\toleft}
		\xdef\ybase{\todown}

	}%

	\newcommand*{\ExtractCoordinate}[1]{\path[overlay] (#1); \pgfgetlastxy{\XCoord}{\YCoord}}%

	\newcommand{\DrawNeighborhood}[1]{%
		\begin{scope}[scale=\scalefactor]
			\begin{scope}[shift={(-\xbase, -\ybase)}]
				\draw[fill=black] (0.5, 0.5) circle (3pt);
				\draw (0, 0) rectangle (1,1);

				\foreach \value[count=\i] in #1 {%
					\ExtractCoordinate{anchor_\i}
					\begin{scope}[shift={(\XCoord, \YCoord)}]
						\ifnum\value=1%
							\fill[black!50] (0, 0) rectangle (1, 1);
						\fi%
						\ifnum\value=2%
							\begin{scope}[opacity=\opacityfactor, blend group=normal]
								\draw[pattern=north west lines, draw = none] (0, 0) rectangle (1, 1);
							\end{scope}
						\fi%
						\ifnum\value=3%
							\begin{scope}[opacity=\opacityfactor]
								\node[transform shape] at (0.5, 0.5) {\Large $\mathbf{X}$};
							\end{scope}
						\fi%
						\draw (0, 0) rectangle (1, 1);
					\end{scope}
				}%
			\end{scope}
		\end{scope}
	}%

	\pgfmathsetmacro{\scalefactor}{0.3}%
	\pgfmathsetmacro{\opacityfactor}{0.7}%
	\pgfmathsetmacro{\tiklabelxoffset}{0.5}%
	\pgfmathsetmacro{\tiklabelyoffset}{0.5}%

	
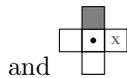
\begin{tikzpicture}
		\newcommand{\neighborsCoordinates}{0/1, 0/-1, -1/0, 1/0}
		\newcommand{\neighbors}{}
		\GetNeighborhoodSizes{\neighborsCoordinates}

		\renewcommand{\neighbors}{1, 0, 0, 3}
		\DrawNeighborhood{\neighbors}
	\end{tikzpicture}
			} at time tick $t+1$ could be sensed if one of the following is true 
			\begin{enumerate*}[label=\roman*)]
				\item
					an agent departing from cell $a_{i, k_g-1}$ South at time $t$,
				\item
					cell $a_{i, k_g-1}$ was empty at $t-1$.
			\end{enumerate*}
			However, in the former case $A$ will leave West, in contradiction to the definition of $T_1$. 
			And in the latter case, $A$ should have moved West at $t-1$ (or $t-3$, depends on the 
			$t \equiv 1,3 \mod 4$). Therefore, leading to a contradiction in one of our assumptions.
			
			Additionally, an exiting agent would behave exactly like a continue agent in two 
			remaining cases
			\adjustbox{max width=1.5cm}{%

	\pgfmathtruncatemacro{\memory}{-1}%
	\pgfmathtruncatemacro{\visibility}{1}%
	\pgfmathtruncatemacro{\maxtablelength}{8}%

	\newcommand{\GetNeighborhoodSizes}[1]{%
		\xdef\toleft{0}
		\xdef\todown{0}

		\pgfmathtruncatemacro{\tableSize}{2 * \visibility + 1}
		\foreach \columnoffset/\rowoffset[count=\i] in #1 {%
			\ifnum\columnoffset<\toleft%
				\xdef\toleft{\columnoffset}%
			\fi%
			\ifnum\rowoffset<\todown%
				\xdef\todown{\rowoffset}%
			\fi%
			\coordinate (anchor_\i) at (\columnoffset, \rowoffset);
		}%

		\xdef\xbase{\toleft}
		\xdef\ybase{\todown}

	}%

	\newcommand*{\ExtractCoordinate}[1]{\path[overlay] (#1); \pgfgetlastxy{\XCoord}{\YCoord}}%

	\newcommand{\DrawNeighborhood}[1]{%
		\begin{scope}[scale=\scalefactor]
			\begin{scope}[shift={(-\xbase, -\ybase)}]
				\draw[fill=black] (0.5, 0.5) circle (3pt);
				\draw (0, 0) rectangle (1,1);

				\foreach \value[count=\i] in #1 {%
					\ExtractCoordinate{anchor_\i}
					\begin{scope}[shift={(\XCoord, \YCoord)}]
						\ifnum\value=1%
							\fill[black!50] (0, 0) rectangle (1, 1);
						\fi%
						\ifnum\value=2%
							\begin{scope}[opacity=\opacityfactor, blend group=normal]
								\draw[pattern=north west lines, draw = none] (0, 0) rectangle (1, 1);
							\end{scope}
						\fi%
						\ifnum\value=3%
							\begin{scope}[opacity=\opacityfactor]
								\node[transform shape] at (0.5, 0.5) {\Large $\mathbf{X}$};
							\end{scope}
						\fi%
						\draw (0, 0) rectangle (1, 1);
					\end{scope}
				}%
			\end{scope}
		\end{scope}
	}%

	\pgfmathsetmacro{\scalefactor}{0.3}%
	\pgfmathsetmacro{\opacityfactor}{0.7}%
	\pgfmathsetmacro{\tiklabelxoffset}{0.5}%
	\pgfmathsetmacro{\tiklabelyoffset}{0.5}%

	\begin{tikzpicture}
		\newcommand{\neighborsCoordinates}{0/1, 0/-1, -1/0, 1/0}
		\newcommand{\neighbors}{}
		\GetNeighborhoodSizes{\neighborsCoordinates}

		\renewcommand{\neighbors}{0, 0, 1, 3}
		\DrawNeighborhood{\neighbors}
	\end{tikzpicture}
			}
			and
			\adjustbox{max width=1.5cm}{%

	\pgfmathtruncatemacro{\memory}{-1}%
	\pgfmathtruncatemacro{\visibility}{1}%
	\pgfmathtruncatemacro{\maxtablelength}{8}%

	\newcommand{\GetNeighborhoodSizes}[1]{%
		\xdef\toleft{0}
		\xdef\todown{0}

		\pgfmathtruncatemacro{\tableSize}{2 * \visibility + 1}
		\foreach \columnoffset/\rowoffset[count=\i] in #1 {%
			\ifnum\columnoffset<\toleft%
				\xdef\toleft{\columnoffset}%
			\fi%
			\ifnum\rowoffset<\todown%
				\xdef\todown{\rowoffset}%
			\fi%
			\coordinate (anchor_\i) at (\columnoffset, \rowoffset);
		}%

		\xdef\xbase{\toleft}
		\xdef\ybase{\todown}

	}%

	\newcommand*{\ExtractCoordinate}[1]{\path[overlay] (#1); \pgfgetlastxy{\XCoord}{\YCoord}}%

	\newcommand{\DrawNeighborhood}[1]{%
		\begin{scope}[scale=\scalefactor]
			\begin{scope}[shift={(-\xbase, -\ybase)}]
				\draw[fill=black] (0.5, 0.5) circle (3pt);
				\draw (0, 0) rectangle (1,1);

				\foreach \value[count=\i] in #1 {%
					\ExtractCoordinate{anchor_\i}
					\begin{scope}[shift={(\XCoord, \YCoord)}]
						\ifnum\value=1%
							\fill[black!50] (0, 0) rectangle (1, 1);
						\fi%
						\ifnum\value=2%
							\begin{scope}[opacity=\opacityfactor, blend group=normal]
								\draw[pattern=north west lines, draw = none] (0, 0) rectangle (1, 1);
							\end{scope}
						\fi%
						\ifnum\value=3%
							\begin{scope}[opacity=\opacityfactor]
								\node[transform shape] at (0.5, 0.5) {\Large $\mathbf{X}$};
							\end{scope}
						\fi%
						\draw (0, 0) rectangle (1, 1);
					\end{scope}
				}%
			\end{scope}
		\end{scope}
	}%

	\pgfmathsetmacro{\scalefactor}{0.3}%
	\pgfmathsetmacro{\opacityfactor}{0.7}%
	\pgfmathsetmacro{\tiklabelxoffset}{0.5}%
	\pgfmathsetmacro{\tiklabelyoffset}{0.5}%

	\begin{tikzpicture}
		\newcommand{\neighborsCoordinates}{0/1, 0/-1, -1/0, 1/0}
		\newcommand{\neighbors}{}
		\GetNeighborhoodSizes{\neighborsCoordinates}

		\renewcommand{\neighbors}{1, 0, 1, 3}
		\DrawNeighborhood{\neighbors}
	\end{tikzpicture}
			}, resulting in $A$ leaving $a_{i, k_g}$ no later than $t+4$. In contradiction
			to our assumption that $a_{i, k_g}$ is a SOFT cell.
			
			We still need to take care of a SOFT cell occupant with a non-SOFT Northern	neighbor cell. 
			W.l.o.g. we assume that the South neighbor is also a SOFT cell agent.
			
			Suppose $A$ at $a_{i, k_g}$ is a continue agent, and an agent $B$ that occupied $a_{i-1, k_g}$ 
			and just left it at $t \equiv 01_2 \mod 4$. Then, depending on the agent type, $B$ can
			move back to $a_{i-1,k_g}$ at $t+7$ if it is a continue agent, or at $t+5$ if $B$
			is an exiting agent. Although, $A$ will move North at $t+4$, contradicting our
			assumption, that $A$ resides at $a_{i, k_g}$ forever.
			
			However, a special case $i = n$ should be treated separately. We denote by
			$t \equiv 01_2 \mod 4$ a time tick when $B$, the North neighbor, moves further North.
			Then, the following three cases are possible.
			\begin{enumerate*} [label=\roman*)]
				\item
					the cell $a_{n, k_g - 1}$ is empty at $t$, and therefore was empty at $t-1$.
					Although, in this case $A$ observes
					\adjustbox{max width=1.5cm}{%

	\pgfmathtruncatemacro{\memory}{-1}%
	\pgfmathtruncatemacro{\visibility}{1}%
	\pgfmathtruncatemacro{\maxtablelength}{8}%

	\newcommand{\GetNeighborhoodSizes}[1]{%
		\xdef\toleft{0}
		\xdef\todown{0}

		\pgfmathtruncatemacro{\tableSize}{2 * \visibility + 1}
		\foreach \columnoffset/\rowoffset[count=\i] in #1 {%
			\ifnum\columnoffset<\toleft%
				\xdef\toleft{\columnoffset}%
			\fi%
			\ifnum\rowoffset<\todown%
				\xdef\todown{\rowoffset}%
			\fi%
			\coordinate (anchor_\i) at (\columnoffset, \rowoffset);
		}%

		\xdef\xbase{\toleft}
		\xdef\ybase{\todown}

	}%

	\newcommand*{\ExtractCoordinate}[1]{\path[overlay] (#1); \pgfgetlastxy{\XCoord}{\YCoord}}%

	\newcommand{\DrawNeighborhood}[1]{%
		\begin{scope}[scale=\scalefactor]
			\begin{scope}[shift={(-\xbase, -\ybase)}]
				\draw[fill=black] (0.5, 0.5) circle (3pt);
				\draw (0, 0) rectangle (1,1);

				\foreach \value[count=\i] in #1 {%
					\ExtractCoordinate{anchor_\i}
					\begin{scope}[shift={(\XCoord, \YCoord)}]
						\ifnum\value=1%
							\fill[black!50] (0, 0) rectangle (1, 1);
						\fi%
						\ifnum\value=2%
							\begin{scope}[opacity=\opacityfactor, blend group=normal]
								\draw[pattern=north west lines, draw = none] (0, 0) rectangle (1, 1);
							\end{scope}
						\fi%
						\ifnum\value=3%
							\begin{scope}[opacity=\opacityfactor]
								\node[transform shape] at (0.5, 0.5) {\Large $\mathbf{X}$};
							\end{scope}
						\fi%
						\draw (0, 0) rectangle (1, 1);
					\end{scope}
				}%
			\end{scope}
		\end{scope}
	}%

	\pgfmathsetmacro{\scalefactor}{0.3}%
	\pgfmathsetmacro{\opacityfactor}{0.7}%
	\pgfmathsetmacro{\tiklabelxoffset}{0.5}%
	\pgfmathsetmacro{\tiklabelyoffset}{0.5}%

	\begin{tikzpicture}
		\newcommand{\neighborsCoordinates}{-1/0, 0/1, 0/-1, 1/0}
		\newcommand{\neighbors}{}
		\GetNeighborhoodSizes{\neighborsCoordinates}

		\renewcommand{\neighbors}{0, 1, 2, 3}
		\DrawNeighborhood{\neighbors}
	\end{tikzpicture}
					}
					at $t-1$ and should have moved West unconditionally at time $t$.
				\item
					the cell $a_{n, k_g - 1}$ becomes empty at $t$, and an agent $C$, leaving it,
					could not return back until at least $t+5$. Therefore, $A$ will 
					unconditionally move either West or North at $t+4$ according to its 
					internal state.
				\item
					the cell $a_{n, k_g - 1}$ remains occupied, and $A$ observes
					\adjustbox{max width=1.5cm}{%

	\pgfmathtruncatemacro{\memory}{-1}%
	\pgfmathtruncatemacro{\visibility}{1}%
	\pgfmathtruncatemacro{\maxtablelength}{8}%

	\newcommand{\GetNeighborhoodSizes}[1]{%
		\xdef\toleft{0}
		\xdef\todown{0}

		\pgfmathtruncatemacro{\tableSize}{2 * \visibility + 1}
		\foreach \columnoffset/\rowoffset[count=\i] in #1 {%
			\ifnum\columnoffset<\toleft%
				\xdef\toleft{\columnoffset}%
			\fi%
			\ifnum\rowoffset<\todown%
				\xdef\todown{\rowoffset}%
			\fi%
			\coordinate (anchor_\i) at (\columnoffset, \rowoffset);
		}%

		\xdef\xbase{\toleft}
		\xdef\ybase{\todown}

	}%

	\newcommand*{\ExtractCoordinate}[1]{\path[overlay] (#1); \pgfgetlastxy{\XCoord}{\YCoord}}%

	\newcommand{\DrawNeighborhood}[1]{%
		\begin{scope}[scale=\scalefactor]
			\begin{scope}[shift={(-\xbase, -\ybase)}]
				\draw[fill=black] (0.5, 0.5) circle (3pt);
				\draw (0, 0) rectangle (1,1);

				\foreach \value[count=\i] in #1 {%
					\ExtractCoordinate{anchor_\i}
					\begin{scope}[shift={(\XCoord, \YCoord)}]
						\ifnum\value=1%
							\fill[black!50] (0, 0) rectangle (1, 1);
						\fi%
						\ifnum\value=2%
							\begin{scope}[opacity=\opacityfactor, blend group=normal]
								\draw[pattern=north west lines, draw = none] (0, 0) rectangle (1, 1);
							\end{scope}
						\fi%
						\ifnum\value=3%
							\begin{scope}[opacity=\opacityfactor]
								\node[transform shape] at (0.5, 0.5) {\Large $\mathbf{X}$};
							\end{scope}
						\fi%
						\draw (0, 0) rectangle (1, 1);
					\end{scope}
				}%
			\end{scope}
		\end{scope}
	}%

	\pgfmathsetmacro{\scalefactor}{0.3}%
	\pgfmathsetmacro{\opacityfactor}{0.7}%
	\pgfmathsetmacro{\tiklabelxoffset}{0.5}%
	\pgfmathsetmacro{\tiklabelyoffset}{0.5}%

	\begin{tikzpicture}
		\newcommand{\neighborsCoordinates}{-1/0, 0/1, 0/-1, 1/0}
		\newcommand{\neighbors}{}
		\GetNeighborhoodSizes{\neighborsCoordinates}

		\renewcommand{\neighbors}{1, 0, 2, 3}
		\DrawNeighborhood{\neighbors}
	\end{tikzpicture}
					} between $t+1$ and $t+4$. Consequently, $A$ moves North at $t+4$.
			\end{enumerate*}
			In all the above cases $A$ moves in contradiction to our assumption, that $A$ stays
			in $a_{n, k_g}$ forever.
			
			\InsertEmptyLine
			If $A$ is an exiting agent, we could w.l.o.g. suppose that $a_{n, k_g-1}$ is always 
			occupied (i.e. a SOFT cell), since otherwise $A$ will move West in at most two time ticks, 
			contradicting the definition of $T_0$. Moreover, an exiting agent in the observed neighborhood
			\adjustbox{max width=1.5cm}{%
			} 
			will leave North at $t+4$, in contradiction to our assumption.
			
			\InsertEmptyLine
			We have just shown that all cells in the column $k_g$ become empty eventually,
			and therefore will become empty infinitely often (i.e. non-SOFT cells).
		\end{proof}
		
		\InsertEmptyLine
		We have just established, that every cell in the $k_g$ column is empty infinitely often (
		i.e. is a non-SOFT cell, and except maybe to the cell $a_{1, k_g}$). In particular, a grid 
		cell $a_{n, k_g}$, hence an agent in $a_{n, k_g+1}$ at SOFT cell column $k_g + 1$ 
		observes a neighborhood
		\adjustbox{max width=1.5cm}{%
		} from time to time.
		Yet, after an occupying agent leaves the cell $a_{n, k_g}$ and moves North,
		an agent at cell $a_{n, k_g+1}$ (position $5$ on the grid) should move West in at most 
		$4$ time ticks. 
		The above conclusion contradicts the definition of $T_2$, except, whenever
		$k_g = m - 1$. In the latter case, an agent occupying $a_{n, m}$ could be an
		exiting agent, and will not leave the target column. Note, that in the edge case 
		whenever $n = 2$, we have that the current configuration is a target configuration and we 
		are done.
		
		However, by assumption, not all exiting agent moved
		into the last column at $T_0$. Therefore, there is at least one continue agent
		occupying one of the cells $a_{2, m}, \ldots, a_{n, m}$. Suppose, this is a
		continue agent $B$ at some cell $a_{i, m}$. 
		
		According to \cref{lemma:k_g column}, cell $a_{i, m-1}$ is a non-SOFT cell, hence
		it is empty infinitely often.
		Therefore, we have the following dichotomy: either no agent moves into cell 
		$a_{i, m-1}$ from North, then $B$ moves West at the next $11_2 \mod 4$ time tick, 
		upon $a_{i, m-1}$ is vacated. In contradiction to the definition of $T_2$. In the other
		case, agents periodically move into $a_{i, m-1}$ from North. However, it should
		be that at some time tick $t \equiv 01_2 \mod 4$ an agent occupying
		$a_{i, m-1}$ leaves North. According to our previous observations the same agent
		could not return to $a_{i, m-1}$ before $t+5$, nor could any other agent move into
		$a_{i, m-1}$ from $a_{i+1, m-1}$ before $t+4$. Hence, an agent $B$, that finds 
		itself at position $8$ on the grid, moves West at $t+2$.
		This contradicts the assumption, that an East-West movement ceased. Thus finally proving
		\cref{lemma:right-upper-open-corner}

	\subsection{Inner grid column}
	We have shown, that cell $a_{1, m}$ is empty infinitely often. Consequently, 
	every agent in the first row or in the first column will eventually reach $a_{1, m}$, 
	as all the agents in this \enquote{Great Outer Arc} move East and North respectively.	
	However, we are not yet done. We have established, that the first row or the first column 
	are free of exiting agents, but they (exiting agents) could be located in the 
	\enquote{inner} columns, i.e. columns with indices $1 < k < m$. 
	We, therefore, shall show, that these agents eventually move West.
	
	Recall, that cell $a_{1, m}$ is empty infinitely often, and that, by assumption, empty spaces 
	\enquote{move} East in the last row only. Hence, we have that $a_{n, 1}$, $a_{n, 2}$, $\ldots$, 
	$a_{n, m-1}$
	are also empty infinitely often. Therefore, as a corollary of \cref{lemma:k_g column}, all 
	cells in columns $2, 3, \ldots, m-2$ are empty infinitely often. Note, that the
	\cref{lemma:k_g column}'s assumptions do not hold for column $m-1$, since continue agents 
	\textbf{do} move West from column $m$ to column $m-1$ in any row besides the first.

	\InsertEmptyLine
	In the below analysis we shall show, that an exiting agent $A$, located in the column $1 < k < m$,
	is eventually \enquote{neighbored} by an empty cell from the West. Moreover, we claim that an exiting agent 
	at positions $5$ and $9$ that faces an empty cell from the West, necessarily moves into this
	empty cell. However, establishing the former is a much more involved task than the latter.
	
	We introduce a column spatio-temporal plane $D_k$ and show an existence of
	a \enquote{continuous} path along the column. That is we show that there exists a
	monotone function $h_e$, defined on some time interval 
	$TI \triangleq \{t_0, t_0 +1, \ldots, t_1\}$, that
	describes the row location of a \enquote{designated} empty space in the column. Later,
	we combine $h_e$ with a function $h_A$ - the row tracking function of an exiting
	agent $A$. It is a straightforward exercise in algebra to establish that function $g \triangleq h_e - h_A$ 
	has discrete changes bounded by $1$. Finally, \nameref{thm:A Discrete Intermediate Value theorem} is utilized 
	to demonstrate that $g$ takes a value of $0$ on $TI$. In other words, that $A$ 
	neighbors an empty cell	on its West.
	
	\begin{lemma}[No Missed Opportunities Lemma] \label{lemma:endgame}
		Let $A^*$ be a west-most exiting agent on the grid at time $T_0$. Suppose $A^*$
		occupies cell $a_{i, k+1}$, where $2 < i \leq n$ and $1 \leq k < m-1$. 
		Suppose $t \geq T_0$ is the earliest time when $A^*$ faces an empty cell on its West. 
		Then $A^*$ moves West before $t + 4$.
	\end{lemma}
	\begin{proof}
		Note, that all agents to the West of $A^*$ are continue agents.
		
		We analyze possible local changes around $A^*$ in the near past that \enquote{led}
		an empty space to neighbor $A^*$ from the West.
		
		\begin{enumerate}[label=(\alph*)]
		\item
			A continue agent $B$ left cell $a_{i, k}$ at time $t - 1$. 
			Then $a_{i, k}$ could be a cell at position $2, 5, 6$ or $9$ on the grid. However,
			a continue agent could leave $a_{i, k}$ at only two time ticks $00_2 \mod 4$ or 
			$01_2 \mod 4$. Since agents to the West of $A^*$ are all continue agents, none
			could re-occupy $a_{i, k}$ before $t+4$. On the other hand, $A^*$ is at position
			$5$ or $9$ and will move West no later than at the next $11_2 \mod 4$ time tick,
			which is at most $t+3$.
		
		\item
			Agent $A^*$ moved to $a_{i, k+1}$ at time $t-1$ and at 
			time $t$ faces the empty cell $a_{i, k}$. An exiting agent moves North at times
			$01_2$ and South at times $10_2$. Though, continue agents do not move at positions
			$2, 5, 6$ and $9$ at times $10_2$ and $11_2$. Therefore, $a_{i, k}$ could not
			become occupied before $A^*$'s move West at the next time tick $11_2$ (either $t+1$
			or $t+2$).
		\end{enumerate}
		
		In both cases, $A^*$ moves into an empty cell it faces on the West \textbf{before} any
		other agent have an opportunity to do so.
	\end{proof}
	
	The above \cref{lemma:endgame} holds true for an exiting agent at $a_{2, k+1}$,
	in case an adversary agent left $a_{2, k}$ to South. For $k =1$ this is always
	the case. Later we shall show, that an agent occupying $a_{2, k-1}$ eventually moves
	South also in the interior columns $2, 3, \ldots, m - 2$.
	
	From now on we focus on the column $k$ immediately West of $A^*$.
	
	We define a \textit{spatio-temporal block diagram} $D_k$ of column $k$ 
	on the $\{T, T +1, \ldots, \} \times \{1, 2, \ldots, n\}$ 
	in the following way:$D_k(t, i) = \mathrm{empty} \Leftrightarrow a_{i, k}(t) = 0$, the block 
	$D_k(t, i)$ is defined to be occupied otherwise (see \cref{img:tempo-spatial image}).
	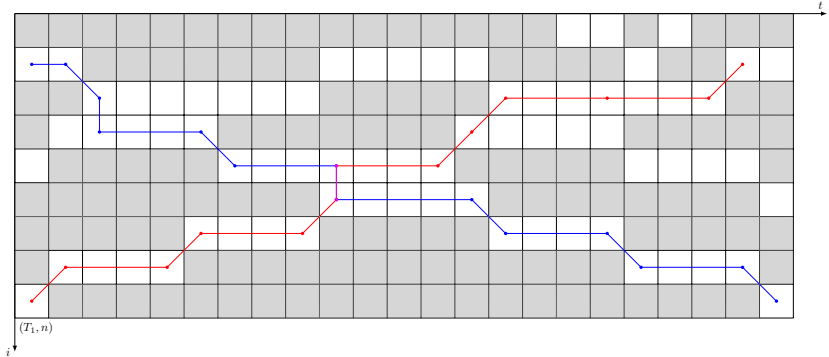
\begin{figure}
		\centering
		\scalebox{0.45}{	\begin{tikzpicture}[
		rcircle only/.style={circle,fill=red,inner sep=1pt,node contents={}},
		bcircle only/.style={circle,fill=blue,inner sep=1pt,node contents={}}
		]
		\draw (0, 0) grid (23, 9);
		\draw[thick, -latex] (0,9) -- (24,9) node[above left] {$t$};
		\draw[thick, -latex] (0,9) -- (0,-1) node[left] {$i$};
		\node[below right] at (0,0) {$(T_1, n)$};
		
		\begin{scope}[black!40, opacity = 0.4]
			\foreach \y in {1, 2, 3, 5, 6, 8}
			{
				\fill (0, \y) rectangle (0 + 1, \y + 1);
			}
			\foreach \y in {0, 2, 3, 4, 6, 8}
			{
				\fill (1, \y) rectangle (1 + 1, \y + 1);
			}
			\foreach \y in {0, 2, 3, 4, 7, 8}
			{
				\fill (2, \y) rectangle (2 + 3, \y + 1);
			}
			\foreach \y in {0, 1, 3, 4, 7, 8}
			{
				\fill (5, \y) rectangle (5 + 1, \y + 1);
			}
			\foreach \y in {0, 1, 3, 5, 7, 8}
			{
				\fill (6, \y) rectangle (6 + 3, \y + 1);
			}
			\foreach \y in {0, 1, 2, 5, 6, 8}
			{
				\fill (9, \y) rectangle (9 + 4, \y + 1);
			}
			\foreach \y in {0, 1, 2, 4, 6, 8}
			{
				\fill (13, \y) rectangle (13 + 1, \y + 1);
			}
			\foreach \y in {0, 1, 3, 4, 7, 8}
			{
				\fill (14, \y) rectangle (14 + 2, \y + 1);
			}
			\foreach \y in {0, 1, 3, 4, 7}
			{
				\fill (16, \y) rectangle (16 + 2, \y + 1);
			}
			\foreach \y in {0, 2, 3, 5, 8}
			{
				\fill (18, \y) rectangle (18 + 1, \y + 1);
			}
			\foreach \y in {0, 2, 3, 5, 7}
			{
				\fill (19, \y) rectangle (19 + 1, \y + 1);
			}
			\foreach \y in {0, 2, 3, 5, 7, 8}
			{
				\fill (20, \y) rectangle (20 + 1, \y + 1);
			}
			\foreach \y in {0, 2, 3, 5, 6, 8}
			{
				\fill (21, \y) rectangle (21 + 1, \y + 1);
			}
			\foreach \y in {1, 2, 4, 5, 6, 8}
			{
				\fill (22, \y) rectangle (22 + 1, \y + 1);
			}
		\end{scope}
		
		\begin{scope}[blend mode=lighten]
		\begin{scope}[red, thick, shift = {(0.5, 0.5)}]
			\draw 	(0, 0) node[rcircle only] --
					(1, 1) node[rcircle only] --
					(4, 1) node[rcircle only] --
					(5, 2) node[rcircle only] --
					(8, 2) node[rcircle only] --
					(9, 3) node[rcircle only] --
					(9, 4) node[rcircle only] --
					(12, 4) node[rcircle only] --
					(13, 5) node[rcircle only] --
					(14, 6) node[rcircle only] --
					(17, 6) node[rcircle only] --
					(20, 6) node[rcircle only] --
					(21, 7) node[rcircle only];
		\end{scope}
		\begin{scope}[blue, thick, shift = {(0.5, 0.5)}]
			\draw 	(0, 7) node[bcircle only] --
					(1, 7) node[bcircle only] --
					(2, 6) node[bcircle only] --
					(2, 5) node[bcircle only] --
					(5, 5) node[bcircle only] --
					(6, 4) node[bcircle only] --
					(9, 4) node[bcircle only] --
					(9, 3) node[bcircle only] --
					(13, 3) node[bcircle only] --
					(14, 2) node[bcircle only] --
					(17, 2) node[bcircle only] --
					(18, 1) node[bcircle only] --
					(21, 1) node[bcircle only] --
					(22, 0) node[bcircle only];
		\end{scope}
		\end{scope}
	\end{tikzpicture}}
		\caption{Spatio-temporal diagram of a grid column.}
		\label{img:tempo-spatial image}
	\end{figure}

	\begin{definition}
		Suppose $B = (t, i)$ is a singleton empty block in a spatio-temporal block diagram $D_k$. 
		A $B$'s \textit{following} block is another empty	block $C = (t', i')$ in $D_k$, 
		where $|i - i'| \leq 1$ and $t \leq t' \leq t+1$.
	\end{definition}
	In a layman terms, we would say: a following block is a block sharing at least one common 
	point with a given block, and not the block itself or blocks in a prior times.
	
	\begin{definition}
		A sequence of empty blocks in the spatio-temporal plane $D_k$ will be called a
		\textit{path}, if for each block $S$ in the sequence, except maybe for the last one, there 
		is another block $S'$ in the sequence, such that $S'$ is a following block of $S$.
	\end{definition}
	In other words, there are no isolated blocks on a \textit{path}, nor jumps. 
	See two piece-wise lines in \cref{img:tempo-spatial image} for path examples.
	
	We associate the time dimension of spatio-temporal plane $D_k$ with $x$-axis, extending from left to right.
	The row dimension of column $k$ is expressed on $y$-axis of $D_k$, stretching from top to bottom. 
	We define a \textit{right step} of the path to be a pair of two empty blocks $B_1$ and $B_2$ in $D_k$,
	such that $B_2$ is a following block of $B_1$, and both have the same $y$-coordinate, i.e. are on the same 
	row. Thus, a right step on the path is equivalent to a specific cell of the grid
	remaining empty for two consecutive time ticks.

	We define an \textit{up step} and a \textit{down step} in the similar manner.
	Except, that we shall assume $B_1$ and $B_2$ share the same $x$-coordinate. In grid terms, this 
	is equivalent to a pair of neighboring empty cells on the grid in the same column at the 
	same time tick.
	
	Finally, we define a \textit{diagonal up} and \textit{diagonal down} steps of the path
	to be a pair of empty blocks $B_1$ and $B_2$ that share a single common point in $D_k$.
	Both steps are equivalent to an agent moving either South or North respectively in 
	the $k$th column.
	
	\begin{definition}
		A path consisting entirely of \textit{up}, \textit{right} and \textit{diagonal up} steps
		is called an \textit{ascending} path.
	\end{definition}
	\begin{definition}
		A path consisting entirely of \textit{down}, \textit{right} and \textit{diagonal down} steps
		is called a \textit{descending} path.
	\end{definition}	
	
	We state and prove a following series of claims under the previous general assumption, that
	the East-West movement of exiting agents ceased, but at least one exiting agent is \enquote{stuck}
	in some internal column, i.e. all the claims are true for $t > T_0$.
	
	\begin{claim}[First Column Path Lemma]\label{lemma:firstrow}
		Suppose, that $a_{1, 1}$ is empty at time $t_0$. Then there exists a descending
		path in $D_1$ from cell $(t_0, 1)$ to cell $(t_1, n)$ for some $t_1 \geq t_0$.
	\end{claim}
	\begin{proof}
		We observe the first column of the grid. By the definition of $T_0$ 
		the first column could only contain continue agents. 
		Let us \textit{designate} an empty space that is 
		\enquote{located} at $a_{1,1}$ at time tick $t_0$. Note that the
		corresponding block in $D_1$ is $(t_0, 1)$.
		
		We interpret an agent move into $a_{1, 1}$, as though the designated empty
		space \enquote{moved} to $a_{2, 1}$, and so on.
		Such interpretation is equivalent to a sequence of right steps followed by
		a diagonal down step in the $D_1$ plane. In case the designated empty space
		reaches the $n$th row, we are done.		
		
		However, it could be possible that after $T_0$ only finitely many agents 
		will pass through $a_{1,1}$. In this case agents will stop entering the 
		first column (through $a_{n, 1}$, since by assumption only continue 
		agents are moving in the first column). 
		Then all cells beneath the designated empty space are already empty 
		(denote that time: $t_1$).
		Therefore, the path from that point $(t_1, i)$ in $D_1$ extends in a series 
		of down steps to $(t_1, n)$.		
	\end{proof}
		
	Unfortunately, the movement of agents in the column $k > 1$ is more complicated. 
	The agents could move North and South, moreover from time to
	time an agent at $a_{2, k}$ could \enquote{borrow} an empty space from the
	first row. The unexpectedly strange behavior is a direct consequence of the 
	agent characteristics in our model. The agents are quite myopic. 
	A visibility range of $1$ does not
	allow an agent to discriminate between second, third, or, for the matter,
	\textit{any} non-edge rows.
		
	Moreover, the paths from $(t_0, 2)$ could end before reaching $(t_1, n)$, where 
	$t_1 > t_0$, due to unfortunate conjunction of a grid configuration and an 
	agent internal state (see \cref{img:tempo-spatial image.bad} for a specific example).
	
	\begin{figure}
		\centering
		\scalebox{0.25}{	\begin{tikzpicture}[
		rcircle only/.style={circle,fill=red,inner sep=1pt,node contents={}},
		bcircle only/.style={circle,fill=blue,inner sep=1pt,node contents={}}
		]
		\draw (200, 0) grid (246, 13);
		\draw[thick, -latex] (200,13) -- (247,13) node[above left] {$t$};
		\draw[thick, -latex] (200,13) -- (200,-1) node[left] {$i$};
		\node[below right] at (200,0) {$(T_1, n)$};

		\begin{scope}[black!40, opacity = 0.4]
			\fill (200, 12) rectangle (246, 13);
			
			\foreach \y in {1,2,3,4,5,7,8,9,10}
			{
				\fill (200, \y) rectangle (201, \y + 1);
			}
			\foreach \y in {0,2,3,4,5,7,8,9,10}
			{
				\fill (201, \y) rectangle (202, \y + 1);
			}
			\foreach \y in {0,2,3,4,6,7,8,9,11}
			{
				\fill (202, \y) rectangle (203, \y + 1);
			}
			\foreach \y in {0,2,3,4,6,7,8,9,11}
			{
				\fill (203, \y) rectangle (204, \y + 1);
			}
			\foreach \y in {0,2,3,4,6,7,8,9,11}
			{
				\fill (204, \y) rectangle (205, \y + 1);
			}
			\foreach \y in {0,1,3,4,6,7,8,9,11}
			{
				\fill (205, \y) rectangle (206, \y + 1);
			}
			\foreach \y in {0,1,3,5,6,7,8,10,11}
			{
				\fill (206, \y) rectangle (207, \y + 1);
			}
			\foreach \y in {0,1,3,5,6,7,8,10,11}
			{
				\fill (207, \y) rectangle (208, \y + 1);
			}
			\foreach \y in {0,1,3,5,6,7,8,10,11}
			{
				\fill (208, \y) rectangle (209, \y + 1);
			}
			\foreach \y in {0,1,2,5,6,7,8,10,11}
			{
				\fill (209, \y) rectangle (210, \y + 1);
			}
			\foreach \y in {0,1,2,5,6,7,9,10,11}
			{
				\fill (210, \y) rectangle (211, \y + 1);
			}
			\foreach \y in {0,1,2,5,6,7,9,10,11}
			{
				\fill (211, \y) rectangle (212, \y + 1);
			}
			\foreach \y in {0,1,2,5,6,7,9,10,11}
			{
				\fill (212, \y) rectangle (213, \y + 1);
			}
			\foreach \y in {0,1,2,4,6,7,9,10,11}
			{
				\fill (213, \y) rectangle (214, \y + 1);
			}
			\foreach \y in {0,1,3,4,6,8,9,10,11}
			{
				\fill (214, \y) rectangle (215, \y + 1);
			}
			\foreach \y in {0,1,3,4,6,8,9,10,11}
			{
				\fill (215, \y) rectangle (216, \y + 1);
			}
			\foreach \y in {0,1,3,4,6,8,9,10,11}
			{
				\fill (216, \y) rectangle (217, \y + 1);
			}
			\foreach \y in {0,1,3,4,5,8,9,10,11}
			{
				\fill (217, \y) rectangle (218, \y + 1);
			}
			\foreach \y in {0,2,3,4,5,8,9,10,11}
			{
				\fill (218, \y) rectangle (219, \y + 1);
			}
			\foreach \y in {0,2,3,4,5,8,9,10,11}
			{
				\fill (219, \y) rectangle (220, \y + 1);
			}
			\foreach \y in {0,2,3,4,5,8,9,10,11}
			{
				\fill (220, \y) rectangle (221, \y + 1);
			}
			\foreach \y in {0,2,3,4,5,7,9,10,11}
			{
				\fill (221, \y) rectangle (222, \y + 1);
			}
			\foreach \y in {1,2,3,4,6,7,9,10,11}
			{
				\fill (222, \y) rectangle (223, \y + 1);
			}
			\foreach \y in {1,2,3,4,6,7,9,10,11}
			{
				\fill (223, \y) rectangle (224, \y + 1);
			}
			\foreach \y in {1,2,3,4,6,7,9,10,11}
			{
				\fill (224, \y) rectangle (225, \y + 1);
			}
			\foreach \y in {1,2,3,4,6,7,8,10,11}
			{
				\fill (225, \y) rectangle (226, \y + 1);
			}
			\foreach \y in {0,1,2,3,5,6,7,8,10,11}
			{
				\fill (226, \y) rectangle (227, \y + 1);
			}
			\foreach \y in {0,1,2,3,5,6,7,8,10,11}
			{
				\fill (227, \y) rectangle (228, \y + 1);
			}
			\foreach \y in {0,1,2,3,5,6,7,8,10,11}
			{
				\fill (228, \y) rectangle (229, \y + 1);
			}
			\foreach \y in {0,1,2,3,5,6,7,8,9,11}
			{
				\fill (229, \y) rectangle (230, \y + 1);
			}
			\foreach \y in {0,1,2,4,5,6,7,8,9,11}
			{
				\fill (230, \y) rectangle (231, \y + 1);
			}
			\foreach \y in {0,1,2,4,5,6,7,8,9,11}
			{
				\fill (231, \y) rectangle (232, \y + 1);
			}
			\foreach \y in {0,1,2,4,5,6,7,8,9,11}
			{
				\fill (232, \y) rectangle (233, \y + 1);
			}
			\foreach \y in {0,1,2,4,5,6,7,8,9,10}
			{
				\fill (233, \y) rectangle (234, \y + 1);
			}
			\foreach \y in {0,1,3,4,5,6,7,8,9,10}
			{
				\fill (234, \y) rectangle (235, \y + 1);
			}
			\foreach \y in {0,1,3,4,5,6,7,8,9,10}
			{
				\fill (235, \y) rectangle (236, \y + 1);
			}
			\foreach \y in {0,1,3,4,5,6,7,8,9,10}
			{
				\fill (236, \y) rectangle (237, \y + 1);
			}
			\foreach \y in {0,1,3,4,5,6,7,8,9,10}
			{
				\fill (237, \y) rectangle (238, \y + 1);
			}
			\foreach \y in {0,2,3,4,5,6,7,8,9,11}
			{
				\fill (238, \y) rectangle (239, \y + 1);
			}
			\foreach \y in {0,2,3,4,5,6,7,8,9,11}
			{
				\fill (239, \y) rectangle (240, \y + 1);
			}
			\foreach \y in {0,2,3,4,5,6,7,8,9,11}
			{
				\fill (240, \y) rectangle (241, \y + 1);
			}
			\foreach \y in {0,2,3,4,5,6,7,8,9,11}
			{
				\fill (241, \y) rectangle (242, \y + 1);
			}
			\foreach \y in {0,2,3,4,5,6,7,8,10,11}
			{
				\fill (242, \y) rectangle (243, \y + 1);
			}
			\foreach \y in {0,2,3,4,5,6,7,8,10,11}
			{
				\fill (243, \y) rectangle (244, \y + 1);
			}
			\foreach \y in {0,2,3,4,5,6,7,8,10,11}
			{
				\fill (244, \y) rectangle (245, \y + 1);
			}
			\foreach \y in {0,1,3,4,5,6,7,8,10,11}
			{
				\fill (245, \y) rectangle (246, \y + 1);
			}
		\end{scope}
		
		\begin{scope}[blue, thick, shift = {(200.5, 0.5)}]
			\draw 	(0, 11) node[bcircle only] --
					(1, 11) node[bcircle only] --
					(2, 10) node[bcircle only] --
					(5, 10) node[bcircle only] --
					(6, 9) node[bcircle only] --
					(9, 9) node[bcircle only] --
					(10, 8) node[bcircle only] --
					(13, 8) node[bcircle only] --
					(14, 7) node[bcircle only] --
					(17, 7) node[bcircle only] --
					(17, 6) node[bcircle only] --
					(21, 6) node[bcircle only] --
					(22, 5) node[bcircle only] --
					(25, 5) node[bcircle only] --
					(26, 4) node[bcircle only] --
					(29, 4) node[bcircle only] --
					(30, 3) node[bcircle only] --
					(33, 3) node[bcircle only] --
					(34, 2) node[bcircle only] --
					(37, 2) node[bcircle only] --
					(38, 1) node[bcircle only] --
					(44, 1) node[bcircle only];
			\node at (44, 1) {\Huge \xmark};
		\end{scope}

	\end{tikzpicture}}
		\caption{Tempo-spatial diagram of $k^{th}$ column. The path started from
		the second row, but not ending in the last row.}
		\label{img:tempo-spatial image.bad}
	\end{figure}
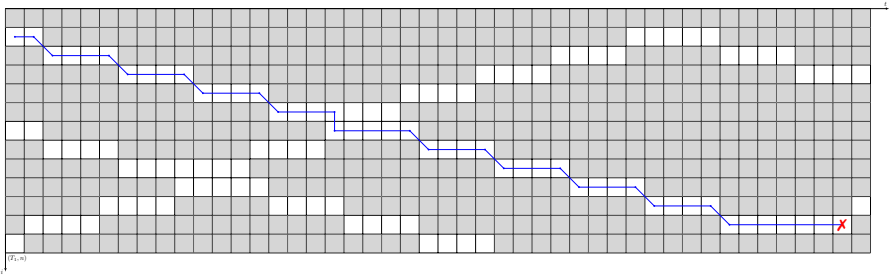
	
	Further we show, that for an internal column $k \in \{2, \ldots, m -1\}$ 
	there exists an ascending $D_k$ path that starts at 
	$(t_0, n-1)$ and ends at $(t_1, 2)$ for times $t_1 \geq t_0 \geq T_0$.
	We later demonstrate, that at least some paths from above could be pre-extended 
	by an empty block at $(t_0-1, n)$ or $(t_0, n)$, while remaining an
	ascending path, i.e. that there exists a path going up from $(t_0^*, n)$
	to $(t_1, 2)$. However, the latter will be useful only in the context of
	the columns containing continue agents only.

	\begin{claim}[Penetrable Column Lemma]\label{lemma:k column after T_0}
		Suppose, that an instance of the ER Problem is unsolvable by Algorithm
		$\mathcal{A}$, and the East-West movement of exiting agents ceased. 
		Then cell $a_{n - 1, k}$ is empty infinitely often, i.e. $a_{n - 1, k}$ 
		is not a SOFT-cell.
	\end{claim}
	\begin{proof}
		\cref{lemma:k column after T_0} is closely related to
		\cref{lemma:k_g column}, however the assumptions are slightly different. 
		We have already proved \cref{lemma:right-upper-open-corner}, i.e. that
		empty spaces \enquote{circulate} on the grid. Therefore, we have 
		established that $a_{n, k}$ is empty infinitely often.
		
		Suppose, that $a_{n-1, k}$ is a SOFT cell and is constantly occupied after 
		time $T$ by agent $A$, w.l.o.g. assume $T = T_0$. We consider the following
		scenario : there is an empty cell in the column $k$	at some time 
		$t \equiv 00_2 \mod 4$, i.e. an empty space that is not a \enquote{borrowed} 
		empty space. Then it could be shown, based on the ideas presented in the  proof
		of \cref{lemma:k_g column}, that all cells $a_{2, k}, \ldots,
		a_{n - 2, k}$ are empty infinitely often. Therefore, we list two possible 
		sequences of $A$'s states and decisions based on them
		(see \cref{tbl:k column stuck agent.one} and 
		\cref{tbl:k column stuck agent.two}). In all the possible scenarios $A$ moves 
		in at most $4$ time ticks, in contradiction to our assumption.

	\begin{table}
		\centering
			{
			\rowcolors{2}{blue!30}{blue!5}
				\begin{tabular}{c ||c c c c c c } 
					\hline
					Time tick & $01_2$ & $10_2$ & $11_2$ & $00_2$ & $01_2$ & $10_2$ \\ [0.5ex] 
					\hline\hline 
					Neighborhood & 
						\adjustbox{max width=1.5cm}{%

	\pgfmathtruncatemacro{\memory}{-1}%
	\pgfmathtruncatemacro{\visibility}{1}%
	\pgfmathtruncatemacro{\maxtablelength}{8}%

	\newcommand{\GetNeighborhoodSizes}[1]{%
		\xdef\toleft{0}
		\xdef\todown{0}

		\pgfmathtruncatemacro{\tableSize}{2 * \visibility + 1}
		\foreach \columnoffset/\rowoffset[count=\i] in #1 {%
			\ifnum\columnoffset<\toleft%
				\xdef\toleft{\columnoffset}%
			\fi%
			\ifnum\rowoffset<\todown%
				\xdef\todown{\rowoffset}%
			\fi%
			\coordinate (anchor_\i) at (\columnoffset, \rowoffset);
		}%

		\xdef\xbase{\toleft}
		\xdef\ybase{\todown}

	}%

	\newcommand*{\ExtractCoordinate}[1]{\path[overlay] (#1); \pgfgetlastxy{\XCoord}{\YCoord}}%

	\newcommand{\DrawNeighborhood}[1]{%
		\begin{scope}[scale=\scalefactor]
			\begin{scope}[shift={(-\xbase, -\ybase)}]
				\draw[fill=black] (0.5, 0.5) circle (3pt);
				\draw (0, 0) rectangle (1,1);

				\foreach \value[count=\i] in #1 {%
					\ExtractCoordinate{anchor_\i}
					\begin{scope}[shift={(\XCoord, \YCoord)}]
						\ifnum\value=1%
							\fill[black!50] (0, 0) rectangle (1, 1);
						\fi%
						\ifnum\value=2%
							\begin{scope}[opacity=\opacityfactor, blend group=normal]
								\draw[pattern=north west lines, draw = none] (0, 0) rectangle (1, 1);
							\end{scope}
						\fi%
						\ifnum\value=3%
							\begin{scope}[opacity=\opacityfactor]
								\node[transform shape] at (0.5, 0.5) {\Large $\mathbf{X}$};
							\end{scope}
						\fi%
						\draw (0, 0) rectangle (1, 1);
					\end{scope}
				}%
			\end{scope}
		\end{scope}
	}%

	\pgfmathsetmacro{\scalefactor}{0.3}%
	\pgfmathsetmacro{\opacityfactor}{0.7}%
	\pgfmathsetmacro{\tiklabelxoffset}{0.5}%
	\pgfmathsetmacro{\tiklabelyoffset}{0.5}%

	\begin{tikzpicture}
		\newcommand{\neighborsCoordinates}{0/1, 0/-1, -1/0, 1/0}
		\newcommand{\neighbors}{}
		\GetNeighborhoodSizes{\neighborsCoordinates}

		\renewcommand{\neighbors}{0, 1, 1, 3}
		\DrawNeighborhood{\neighbors}
	\end{tikzpicture}
						} &
						\adjustbox{max width=1.5cm}{%
						} &
						\adjustbox{max width=1.5cm}{%
						} &
						\adjustbox{max width=1.5cm}{%
						} &
						\adjustbox{max width=1.5cm}{%
						} &
					\\ 
					\hline\hline
					initial $d = 0$ & $001_2$ & $010_2$ & $011_2$ & $000_2$ & $101_2$ &	moved N \\
					\hline
					initial $d = 1$ & $101_2$ & $110_2$ & $111_2$ & $100_2$ & $101_2$ &	moved N \\ [1ex] 
					\hline
				\end{tabular}
			}
			\caption{Agent at $a_{n-1, k}$. North neighbor leaves at $t \equiv 01_2 \mod 4$.
				Agent's state and neighborhood readings, in case South neighbor does not leave.}
			\label{tbl:k column stuck agent.one}
	\end{table}

	\begin{table}
		\centering
			{
			\rowcolors{2}{blue!30}{blue!5}
				\begin{tabular}{c ||c c c c c c } 
					\hline
					Time tick & $01_2$ & $10_2$ & $11_2$ & $00_2$ & $01_2$ & $10_2$ \\ [0.5ex] 
					\hline\hline 
					Neighborhood & 
						\adjustbox{max width=1.5cm}{%
						} &
						\adjustbox{max width=1.5cm}{%
						} &
						\adjustbox{max width=1.5cm}{%
						} &
						\adjustbox{max width=1.5cm}{%
						} &
						\adjustbox{max width=1.5cm}{%
						} &
					\\ 
					\hline\hline
					initial $d = 0$ & $001_2$ & $010_2$ & $011_2$ & $000_2$ & moved S &	 \\
					\hline
					initial $d = 1$ & $101_2$ & $110_2$ & $111_2$ & $100_2$ & $101_2$ &	moved N \\ [1ex] 
					\hline
				\end{tabular}
			}
			\caption{Agent at $a_{n-1, k}$. Both North and South neighbors leave at $t \equiv 01_2 \mod 4$.
				Agent's state and neighborhood readings.}
			\label{tbl:k column stuck agent.two}
	\end{table}

		In the remaining case, there are no non-\enquote{borrowed} empty spaces in 
		column $k$. Therefore, $A$ only ever faces an empty space from South. However, after at
		most the second \enquote{encounter} $A$ will unconditionally move South before 
		an empty space could be occupied by an agent from cell $a_{n, k+1}$.
		
		However, two special cases still require our attention. The case of $n = 4$ and 
		the case of $n = 3$.
		
		In the former case, $n = 4$, the North neighbor of $A$ at cell $a_{2, k}$ could 
		\enquote{borrow} an empty cell \enquote{passing} in the first row and leave North.
		However, a continue agent that \enquote{erroneously} enters the first row
		immediately	returns back. Hence, $A$ is regularly neighbored by an empty space from 
		North at $10_2$ ticks. On the other hand, $A$'s Direction Bit could only change 
		at $00_2$ ticks. Therefore,	$A$ leaves South on at most the second encounter with 
		an empty cell from South.
		
		In the latter case, $n = 3$, we claim that there exists a time tick $T$, such that
		for any $t > T$ such that $t\equiv 00_2$ $A$'s Direction Bit is set to $0$. Indeed, if
		$A$ ever moved North, then $A$ immediately returned and set its Direction Bit to $0$.
		Moreover, $A$ resets the Direction Bit to $1$ only at times $t \equiv 00_2$. Hence,
		in case $a_{n, k}$ is empty at time $t = 00_2$, we have that $A$ will immediately 
		move South (see \cref{subfig:fg_3102_9}).
		
		Hence, $a_{n-1, k}$ eventually becomes empty in all the cases.
	\end{proof}
	
	We conclude from \cref{lemma:k_g column} and \cref{lemma:k column after T_0}
	that all cells in columns $1$ through $m-2$ are empty infinitely often.
	
	\begin{claim}[Start of the Path]\label{lemma:path in the desired direction}
		Consider grid column $k$.
		There exists $t \geq T_0$ such that at least one of the following is true:
		\begin{enumerate*}[label=\arabic*)]
			\item
				cell $a_{n-1, k}$ and cell $a_{n-2, k}$ are empty at time $t$
			\item
				cell $a_{n-1, k}$ is empty at time $t$, and cell $a_{n-2, k}$ 
				is empty at the next time tick $t+1$.
		\end{enumerate*}
	\end{claim}
	\begin{proof}
		Assume that former never happens, and that no agent moves from $a_{n-2,k}$
		into $a_{n-1, k}$ after $T_0$. Recall, that according to 
		\cref{lemma:k column after T_0} cell $a_{n-2,k}$ is empty infinitely often. 
		
		Consider a time tick $t \equiv 10_2 \mod 4$, a moment when $a_{n-2,k}$ became 
		empty due to the northbound move of the agent $B$ occupying $a_{n-2,k}$ at $t-1$. 
		In line with our assumption cell $a_{n-1, k}$ is occupied at time $t$. Let
		$A$ be an agent occupying cell $a_{n-1, k}$ at time $t$.
		
		In case $a_{n, k}$ is empty at $t$, we have that $A$ necessarily 
		executes one of the following moves. $A$ will either move South at $t+2$, 
		and since $B$ could not return before time tick $t+4$, we have a contradiction
		- both $a_{n-1, k}$ and $a_{n-2, k}$ are empty at $t+3$.
		
		Otherwise, $A$ could move North at $t + 3$. Since continue agents
		at position $9$	could not leave a column, the number of such northbound
		moves (unidirectional moves) is finite. Denote the last claim the \enquote{
		Finite Column Capacity Argument}. Ultimately, some agent ought to move 
		in the opposite direction. However, such a move from  cell $a_{n - 2, k}$ to 
		cell $a_{n - 1, k}$ proves the claim.
		
		In the complementary case, $a_{n, k}$ is occupied at time $t$, therefore $A$ moves
		North at $t+3$. See the \enquote{Finite Column Capacity Argument} above, 
		that leads to the desired conclusion.
	\end{proof}
	
	\cref{lemma:path in the desired direction} clears up the path 
	to the existence of an ascending $D_k$ path, starting at $(t_0, n-1)$ and all
	the way up to $(t_1, 2)$.
	
	\begin{definition}
		Suppose $a_{i, j}$ is empty at time $t$. We define the \textit{leading} agent
		of $a_{i, j}$ to be the closest northern agent located at the same column $j$ 
		at time $t$. 
	\end{definition}
	\begin{definition}
		A \textit{leading block} of an empty block $(t, i)$ in the spatio-temporal 
		plane $D_j$ is a block $(t, i')$ corresponding to the leading agent of an empty
		cell $a_{i, j}$.
	\end{definition}
	
	Note, that the leading agent should not necessarily exist. Though in that case,
	cells $a_{i, j}, a_{i - 1, j}, \ldots, a_{1, j}$ are simultaneously empty at time $t$.
	
	\begin{claim}[Path Extension Lemma]\label{lemma:Path Extension in D_k}
		Let $P$ be an ascending path in the tempo-spatial plane $D_k$,
		starting at $(t_0, n-1)$ and ending at $(t_1, n-2)$.
		Then the path extends to an ascending path ending at $(t_2, 2)$. Where
		$t_2 \geq t_1 \geq t_0 \geq T_0$.
	\end{claim}
	\begin{proof}
	
		Define a \textit{leading block} (and a corresponding \textit{leading
		agent}) of a path at time $t$ to be the leading block of one of the empty
		blocks on the path at time $t$.
		
		There are two possible ways a leading agent could move and \enquote{extend} the path.
		First, it could go south, then the path could be extended by a diagonal 
		step, followed by possibly one or more up steps, due to empty spaces configuration. 
		This move necessarily forces the leading agent change. Second, 
		a leading agent could move North, then the path extends by a diagonal step,
		though the leading agent remains the same.
		
		Otherwise the leading agent does not move, and the path could be extended by the
		right steps. However, we should take caution and show that another agent 
		does not move North at the same just under the leading agent, thus terminating 
		the ascending path. 

		A leading agent observes exactly one of the two possible neighborhoods:
		\adjustbox{max width=1.5cm}{%
		}
		or
		\adjustbox{max width=1.5cm}{%

	\pgfmathtruncatemacro{\memory}{-1}%
	\pgfmathtruncatemacro{\visibility}{1}%
	\pgfmathtruncatemacro{\maxtablelength}{8}%

	\newcommand{\GetNeighborhoodSizes}[1]{%
		\xdef\toleft{0}
		\xdef\todown{0}

		\pgfmathtruncatemacro{\tableSize}{2 * \visibility + 1}
		\foreach \columnoffset/\rowoffset[count=\i] in #1 {%
			\ifnum\columnoffset<\toleft%
				\xdef\toleft{\columnoffset}%
			\fi%
			\ifnum\rowoffset<\todown%
				\xdef\todown{\rowoffset}%
			\fi%
			\coordinate (anchor_\i) at (\columnoffset, \rowoffset);
		}%

		\xdef\xbase{\toleft}
		\xdef\ybase{\todown}

	}%

	\newcommand*{\ExtractCoordinate}[1]{\path[overlay] (#1); \pgfgetlastxy{\XCoord}{\YCoord}}%

	\newcommand{\DrawNeighborhood}[1]{%
		\begin{scope}[scale=\scalefactor]
			\begin{scope}[shift={(-\xbase, -\ybase)}]
				\draw[fill=black] (0.5, 0.5) circle (3pt);
				\draw (0, 0) rectangle (1,1);

				\foreach \value[count=\i] in #1 {%
					\ExtractCoordinate{anchor_\i}
					\begin{scope}[shift={(\XCoord, \YCoord)}]
						\ifnum\value=1%
							\fill[black!50] (0, 0) rectangle (1, 1);
						\fi%
						\ifnum\value=2%
							\begin{scope}[opacity=\opacityfactor, blend group=normal]
								\draw[pattern=north west lines, draw = none] (0, 0) rectangle (1, 1);
							\end{scope}
						\fi%
						\ifnum\value=3%
							\begin{scope}[opacity=\opacityfactor]
								\node[transform shape] at (0.5, 0.5) {\Large $\mathbf{X}$};
							\end{scope}
						\fi%
						\draw (0, 0) rectangle (1, 1);
					\end{scope}
				}%
			\end{scope}
		\end{scope}
	}%

	\pgfmathsetmacro{\scalefactor}{0.3}%
	\pgfmathsetmacro{\opacityfactor}{0.7}%
	\pgfmathsetmacro{\tiklabelxoffset}{0.5}%
	\pgfmathsetmacro{\tiklabelyoffset}{0.5}%

	\begin{tikzpicture}
		\newcommand{\neighborsCoordinates}{0/1, 0/-1, -1/0, 1/0}
		\newcommand{\neighbors}{}
		\GetNeighborhoodSizes{\neighborsCoordinates}

		\renewcommand{\neighbors}{0, 1, 0, 3}
		\DrawNeighborhood{\neighbors}
	\end{tikzpicture}
		}. This is due to our path extension analysis we have done above.
		
		Suppose a leading agent $L_1$ decides to move South at time $t \equiv 00_2$. Thus
		$L_2$ becomes a new leading agent, and additionally $L_1$ will not be able
		to move back until $t+5$. $L_2$ then either moves North at $t+1$ or otherwise 
		South at $t + 4$. In the former case, $L_2$ opens up a two-block distance
		at least from the next southern agent in the column. And the distance to 
		the nearest southern agent will be kept at value at least $2$, as long as $L_2$
		continues to move North. In case  $L_2$ stops North movement, it will change
		the movement direction and move South before $L_1$ (or the agent that could
		have replaced $L_1$ from the $k+1$ column) could occupy the southern neighbor 
		empty cell. However, if $L_2$ moves South we apply the same reasoning
		to the new leading agent.
		
		Note, that in the absence of the leading agent, we have that all cells 
		above the last leading agent are currently empty. And the path extends 
		vertically in $D_k$ to the first row.
		
		Notoriously, all the agents in the $k$th column become leading agents of the same
		ascending path.
		Until either no leading agent is defined, or the leading agent is the agent
		located in the first row.
		Thus, we have just showed that an ascending path extends to the second row. 
	\end{proof}
	
	The claim above establishes that any non-horizontal ascending path that starts at 
	the row $n-1$ extends to at least the second row. However, this path ends when a 
	leading agent at time $t_2$ \enquote{borrows} an empty space from the first row 
	of the grid, and then immediately returns back.
	
	Denote this path $\mathcal{P}_0$. We claim that there exists another path 
	$\mathcal{P}_1$ in $D_k$, coinciding with $\mathcal{P}_0$ from time $t_0$ to 
	$t_2 - 1$. $\mathcal{P}_1$ then could be extended to the second row at time
	$t_2 + 3$. Moreover, cell $a_{2, k}$ will remain empty throughout time interval
	$\{t_2+3, \ldots, t_2 + 8\}$.
	
	We assume that at time $t_2$ a leading agent $L_1$ moved North from 
	position $9$ to position $7$. According to Algorithm $\mathcal{A}$, $L_1$ 
	returns back immediately at $t_2+1$. In addition $L_1$ sets its 
	Direction bit to $0$ and will move South at	$t_2 + 3$, since the gap between
	the agent $L_1$ and the former leading agent $L_2$ stayed 
	of size at least $1$ (the gap was either of size at least two before the agent $L_1$ 
	decided to push North into the first row. The gap could have narrowed down at $t_2 + 1$
	in case $L_2$ also moved North at $t_2$. Or the gap was of size $1$ at $t_2$. However,
	in the latter case, it is the result of the lead change at time $t_2 - 1$. Consequently,
	$L_2$ is unable to return to the third row until at least $t_2 + 4$).
	
	Once the agent $L_1$ at the second row moves South at $t_2+3$ it could not return 
	back until at least $t_2+8$, nor any agent could move South into $a_{2, k}$ from 
	$a_{1, k}$.
	
	Therefore, we define a new path $\mathcal{P}_1$ in the following manner. The path
	$\mathcal{P}_1$ coincides exactly with $\mathcal{P}_0$ in $D_k$ from $(t_0, n-1)$ through
	$(t_1, n-2)$ until $(t_2, 3)$. Then $\mathcal{P}_1$ deviates horizontally right until 
	$(t_2+2, 3)$, followed by a diagonal step to $(t_2+3,2)$, and finally to $(t_2+8, 2)$
	where it ends. Note that by definition $\mathcal{P}_1$ is also an ascending path.
	
	\InsertEmptyLine
	We claim, that at least one such ascending path $\mathcal{P}_1$ in $D_k$ could be 
	pre-extended to a longer ascending path by a diagonal-up from $(t_0-1, n)$ or by 
	an up step from $(t_0, n)$. Thus, we claim that 
	there exists an ascending path in $D_k$ that spans from the row $n$ to the row $2$.
	
	We assume below, that column $k$ agents do not leave cell $a_{n-1,k}$ in
	the southern direction, i.e. that the column $k$ is kind of a \textit{plugged from 
	below} column.
	Additionally, we shall assume that $a_{n, k}$ and $a_{n-1, k}$ are not simultaneously 
	empty. Denote these assumptions above $\mathcal{C}_2$.
	
	\begin{lemma}[Limited Agent Throughput]\label{lemma:ClosedColumnPairing}
		Let $A$ be an agent that occupies $a_{n-1, k}$ during time interval $(t_0, t_1]$,
		i.e. $A$ moves to $a_{n-1,k}$ at time $t_0$ and leaves North at $t_1$. Then, under
		the $\mathcal{C}_2$ assumptions, at most two different agents occupy $a_{n, k}$
		during $(t_0, t_1]$.
	\end{lemma}
	\begin{proof}
		Suppose the opposite is true, and at least three different agents occupy cell
		$a_{n, k}$ during time interval $(t_0, t_1]$.

		If $A$ neighbors an empty cell on the North for the $4$ continuous ticks it 
		leaves the $a_{n-1,k}$ (North or South, depends on the actual neighborhood).
		Therefore, we conclude that $A$ senses only the following two neighborhoods
		during time interval $(t_0 + 4, t_1 - 4]$
		\adjustbox{max width=1.5cm}{%
		} and
		\adjustbox{max width=1.5cm}{%
		}.
		
		However, in that time interval at least two agents left $a_{n,k}$ to 
		the West, leaving $a_{n, k}$ empty for at least $4$ consecutive time
		ticks each time. In the first $4$ consecutive time ticks term $A$ should
		have changed its Direction bit to $0$. Subsequently, it should have
		left South during the second term, in contradiction to $\mathcal{C}_2$
		assumption.		
	\end{proof}
	
	\begin{lemma}[No Closed Columns Lemma]\label{lemma:ClosedColumn}
		Suppose that the western-most exiting agent is located at column $k+1$. Then there 
		exists $t \geq T_0$, such that cell $a_{n, k}$ is empty at time $t$, and 
		cell $a_{n - 1, k}$ is either empty at the same time $t$, or is empty at the 
		next time tick $t+1$. 
	\end{lemma}
	
	Note, that the latter condition implies that an agent leaves South from $a_{n-1, k}$ to
	$a_{n, k}$.
	
	\begin{proof}
		We note, that cells $a_{n, k}$ and $a_{n-1, k}$ could become empty simultaneously,
		if agents occupying them leave West and North respectively at the same time.
		On the other hand, the cells could be empty in two consecutive time ticks,
		if an agent occupying cell $a_{n-1, k}$ leaves South to cell $a_{n ,k}$.
		
		Assume $\mathcal{C}_2$ on the contrary, that such time tick does not exist, 
		i.e. no	agent leaves cell $a_{n-1, k}$ to the South after the time $T_0$, 
		nor both cells are empty at the same time.
		
		Note, that an agent $M$ moving from cell $a_{n, k + 1}$ to $a_{n, k}$ sets its
		own Direction bit to $1$. If cell $a_{n-1, k}$ is empty at $t + 1$, then we have 
		the following dichotomy: either $a_{n-1, k}$ was empty at time $t$ or it became
		empty following an agent departing North at time $t$. In the former case,
		at time $t$ $a_{n ,k}$ and $a_{n - 1, k}$ were simultaneously empty, in
		contradiction to $\mathcal{C}_2$. 
		
		In the latter case, $M$ will move North
		at $t+4$. However, in the periodic state of the grid, agents could
		not enter the column $k$ from $a_{n, k}$ in the North direction, without
		some agents eventually leaving in the opposite direction (according to 
		Algorithm $\mathcal{A}$). Hence, at some point in time an agent moves 
		South to the $a_{n, k}$, which is also in contradiction to the assumption.
		
		Therefore, we conclude that cell $a_{n - 1, k}$ is occupied at time $t + 1$ 
		(and was occupied at time $t$). Let $A$ be an agent occupying cell 
		$a_{n - 1, k}$ at time $t$.
		
		Once the agent $M$ moved to $a_{n ,k}$, it could only change its Direction
		Bit value to $0$ if faced by an empty cell on the West, while facing 
		an occupied cell on the North. Note, that the second option of going
		up the column and then back, is assumed to be non-existent (see the
		latter case above).
		
		We want to emphasis, that the opposite case when $M$ faces an empty
		cell on the North, and an occupied cell on the West, leads to an 
		imminent move of the agent $M$ to the North, and a 
		contradiction to assumptions.
		
		Moreover, suppose that $M$ faces an empty cell on its North and/or West 
		at time	$T \equiv 00_2 \mod 4$. Then, the only viable way $M$ does not leave 
		cell $a_{n,k}$ at $T + 1$ is in the case that all the following 
		conditions are satisfied:
		\begin{itemize}
			\item
				the $M$'s Direction bit is set to value $0$ at time $T$.
			\item
				at time $T$ a North-West (invisible-to-$M$) neighbor
				moves South and becomes a West neighbor of $M$ (visible
				only from time $T+1$). We further denote the-invisible-at-$T$
				agent as agent $B$.
		\end{itemize}
		
		Then, we conclude based on \cref{lemma:ClosedColumnPairing} that 
		the number of agents leaving $a_{n - 1, k}$ to the North is at least
		equal to the number of agents leaving $a_{n, k}$ to the West.
		Moreover, each such northbound excursion of the agent $A$ happens when
		an agent at $a_{n, k}$ already set its Direction bit value to $0$.
		
		However, $M$ sets the Direction bit to $0$ at time $t \equiv 00_2$ and moves
		to the West at the next time tick, unless $a_{n, k - 1}$ is already occupied
		by an agent $B$ that moved at $t$ from $a_{n - 1, k - 1}$. We note, that
		the southbound motion of an agent in the $k-1$ column is only possible
		following the westward departure of an agent that previously occupied
		$a_{n, k - 1}$ at $t - 3$, otherwise agent $B$ was the last agent that left
		$a_{n, k - 1}$ and it should have happen at time $t-7$. Though in this case $M$ 
		would have already moved West to occupy $a_{n, k-1}$.
		
		We summarize, $M$ observes at least two agents leaving westward (the fact 
		unknown to agent $M$) from column $k-1$ before finally leaving itself to the $k-1$ column. 
		Therefore, for every agent leaving column $k$, at least two agents leave column
		$k-1$, which is unsustainable in the long run, when the only source
		of those column $k-1$ \enquote{leaving} agents are the agents moving
		West from cell $a_{n,k}$.
		
		Since this is an obvious fallacy, we conclude, that $a_{n, k}$ become empty 
		at one of those times $T+1$. Consequently, agent $A$ would eventually
		move South, contradicting our assumptions.
		
		Therefore, our initial assumption turned out to not hold. Except if $n=3$ or
		$n=4$. However, In the former case, an agent at $a_{2, k}$ sets the value
		of the Direction bit to $0$ each time it \enquote{borrows} an empty space, and
		each time it neighbors an empty space from the South. Hence,
		inevitably, an agent occupying cell $a_{2, k}$ will move South to $a_{3, k}$ on the next tick 
		after cell $a_{3, k}$ becomes empty.
		
		In the latter case, an agent $A$ at $a_{3, k}$ either never observes an empty
		space at time tick $01_2$, hence will move eventually South, as in case $n=3$.
		Otherwise, an agent at $a_{3, k}$ is the agent moving back and forth
		from $a_{3, k}$ to $a_{2, k}$ (and, possibly, \enquote{borrowing} an empty space
		from $a_{1, k}$ from time to time). However, in this specific case, denote
		by $T$ - time when an occupying agent leaves $a_{4, k}$ to the West. If $A$ at  
		time $T$ is located in cell $a_{2, k}$ - we are done. Hence suppose $A$ occupied
		$a_{3, k}$ at time $T$. It could either move North at time $T$, or to move South 
		at $T+3$ - in both cases the claim is established.
	\end{proof}
	
	Recall our assumption that $\mathcal{A}$ does not lead this specific ER Problem 
	instance $P$ to target configuration. Therefore, at least one exiting agent
	remains in the same non-edge column forever. However, if at least one such
	\enquote{stuck} exiting agent moves to a new column, we deduce that our
	assumption was wrong, and $\mathcal{A}$ does lead $P$ to a target configuration.
	We then plan to show that a left-most exiting agent $V$ faces an empty
	cell as its West neighbor and subsequently leaves to the West (\cref{lemma:endgame}),
	thus breaking our assumption and proving, that $\mathcal{A}$ brings grid 
	of an arbitrary size ($m \geq 3, n \geq 2$) to a target configuration.
	
	Re-define the time (in $D_k$ plane), so that up steps on the path become
	diagonal steps, i.e. we add virtual time ticks to make this happen. Define a 
	function $H(\hat{t})$ - that designates the row index of an empty space on the path 
	at new adjusted time $\hat{t}$. Finally, define function $G$ - that tracks the
	row index of an exiting agent $V$ (in column $k+1$) at $\hat{t}$. Note that both 
	functions are discrete functions, such that $H$ is monotone, and both 
	have increments of $0$ and $1$. Though $G$ could also decrease by $1$.
	
	Let $T_s$ - be the time tick when cell $a_{n, k}$ is empty, and $a_{n-1, k}$ is
	empty either at $T_s$ or at $T_s+1$ (see \cref{lemma:ClosedColumn}).
	Note, that we define a virtual version of time $\hat{t}$ after the ascending 
	path $\mathcal{P}$ has been started, i.e. at either $T_s$ or $T_s+1$.
	Define, $f := H - G$. Let $T_e$ - be the time when the ascending path $\mathcal{P}$
	reaches the second row (see \cref{lemma:Path Extension in D_k} and the 
	conclusion thereafter).
	
	Note, that in case one of the following equalities $G(T_s) = n$ or $G(T_e) = 2$ hold, 
	we are done (\cref{lemma:endgame}), since an exiting agent $V$ neighbors an 
	empty space on its West.
	Therefore, we can assume w.l.o.g. that 
	\begin{equation*}
		\begin{array}{l c r}
			G(T_s) < n & & G(T_e) > 2.
		\end{array}
	\end{equation*}
	However, according to the definition of times $T_s$ and $T_e$ we have that
	\begin{equation*}
		\begin{array}{l c r}
			H(T_s) = n & & H(T_e) = 2.
		\end{array}
	\end{equation*}

	We also claim that $|f(\hat{t} + 1) - f(\hat{t})| \leq 1|$. 
	
	\begin{table}
		\centering
			{
			\rowcolors{2}{blue!30}{white}
				\begin{tabular}{c || c | l l | l } 
					\toprule \\ [1ex]
					Time tick & Virtual tick & \multicolumn{1}{>{\centering\arraybackslash}m{20mm}}{$\Delta G$} & 
					\multicolumn{1}{>{\centering\arraybackslash}m{20mm}|}{$\Delta H$} & 
					\multicolumn{1}{>{\centering\arraybackslash}m{20mm}}{$\Delta f$} \\ [3ex] 
					\midrule
					$0 \mod 4$ & no & 0 & $0, -1$ & $0, -1$ \\ 
					\hline
					$1 \mod 4$ & no & $0, -1$ & $0, -1$ & $0, 1, -1$ \\ 
					\hline
					$2 \mod 4$ & no & $0, 1$ & 0 & $0, -1$ \\
					\hline
					$3 \mod 4$ & no & 0 & 0 & 0 \\ 
					\hline
					$\hat{t}$ & yes & 0 & $-1$ & $-1$ \\ 
					\bottomrule
				\end{tabular}
			}
			\caption{Changes in functions $f, G, H$ split by time ticks. $G$ - tracks the $V$'s row;
			$H$ - tracks the \enquote{designated} empty space's row; $\Delta f = \Delta H - \Delta G$.}
			\label{tbl:Path catching exiting agent}
	\end{table}

	See \cref{tbl:Path catching exiting agent} for $f$ changes summary. The changes 
	are as follows: if $\hat{t}$ is	derived from the real time (e.g. on diagonal up step
	of an ascending path), than the change is a consequence of a leading agent move 
	(time ticks $0,1 \mod 4$), $V$'s move (time ticks
	$1, 2 \mod 4$), or both. In case the time tick is a virtual time tick then only 
	the function $H$ value changes, as it tracks a \enquote{designated} empty space.
	
	Therefore, we can apply \nameref{thm:A Discrete Intermediate Value theorem}
	to the function $f$.
	
	\begin{theorem}[The Discrete Intermediate Value Theorem] \label{thm:A Discrete Intermediate Value theorem}
		For integers $a < b$, let $f$ be a function from integers in $[a, b]$ to 
		$\mathbb{Z}$, that satisfies the property
		\begin{equation*}
			|f(i+1) - f(i)| \leq 1
		\end{equation*}
		for all $i$. If $f(a) < 0 < f(b)$, then there exists $c \in (a, b)$, such
		that $f(c) = 0$.
	\end{theorem}
	\begin{proof}
		Let $S = \{x \in \mathbb{Z} \cap [a, b]\;:\;f(x) < 0\}$ and let
		$c = \max S + 1$. We claim that $f(c) = 0$.
		
		Suppose otherwise $f(c) < 0$, then $c \in S$, a contradiction to the fact
		that $c - 1$ is an upper bound on $S$.
		If $f(c) > 0$, we have that $f(c-1) \geq 0$ (see the \enquote{continuity} property
		of $f$). But by definition of $c$, $c-1 \in S$. Contradiction.
	\end{proof}
	
	The application of the \nameref{thm:A Discrete Intermediate Value theorem} is as follows:
	we conclude that there exist a time $\hat{T}$ 
	when $H(\hat{T}) = G(\hat{T})$. In other words, $V$ an exiting agent is located
	East to an empty cell. If $\hat{T}$ happen to originate in real time, we conclude that an
	agent at column $k$ just moved North or South, and it will take at least one full cycle
	to another column $k$ agent to occupy the \enquote{designated} empty space.
	However, it will take $V$ at most $3$ time ticks to move West.
	
	Otherwise, the origin of time tick $\hat{T}$ is in the \enquote{virtualization} of the 
	time, i.e. a group of 
	empty cells was linked to the path under construction. According to the way we extended
	the ascending path, this event could happen after a continue agent moved South. The next possible
	move by continue agent in column $k$ could be North movement, but the just-linked 
	empty space is not the destination of such move. Hence, an empty space will \enquote{remain}
	near $V$, until the latter moves West in two time ticks.
	
	We have just proved, that $V$ moves West, in contradiction to the definition of $T_0$.
	
	Therefore, we conclude, that execution of Algorithm $\mathcal{A}$ leads to the solution
	of the ER Problem for any given grid ($m \geq 3$) initialized to any initial configuration
	under the following requirements: the number of exiting agents is at most $n-1$, and
	the number of empty cells is at least $1$.

		\clearpage
		
		\section[Algorithm \texorpdfstring{$\mathcal{A}_2$}{} FSM]{Two-column sorting algorithm state machine}
		\subsection{An exit agent}
			\label{sec:Two.Lane-Algorithm.Exit}
			\vspace{33mm}
			
	\begin{figure}[ht]
		\centering
		\begin{subfigure}[b]{0.26\textwidth}
			\centering
			\subimport{Ver.3191/}{Table.Ver_3191.Position_1}
			\caption{Position 1}
			\label{subfig:fg_3191_1}
		\end{subfigure}
		\hfill
		\begin{subfigure}[b]{0.44\textwidth}
			\centering
			\subimport{Ver.3191/}{Table.Ver_3191.Position_2}
			\caption{Position 2}
			\label{subfig:fg_3191_2}
		\end{subfigure}
		\hfill
		\begin{subfigure}[b]{0.26\textwidth}
			\centering
			\subimport{Ver.3191/}{Table.Ver_3191.Position_4}
			\caption{Position 4}
			\label{subfig:fg_3191_4}
		\end{subfigure}
		\caption{State machine(s) (\subref{subfig:fg_3191_1}) at position 1, (\subref{subfig:fg_3191_2}) at position 2, (\subref{subfig:fg_3191_4}) at position 4.}
		\label{fig:fg_3191_1}
	\end{figure}
	\begin{figure}
		\centering
		\begin{subfigure}[b]{0.48\textwidth}
			\centering
			\subimport{Ver.3191/}{Table.Ver_3191.Position_3}
			\caption{Position 3}
			\label{subfig:fg_3191_3}
		\end{subfigure}
		\hfill
		\begin{subfigure}[b]{0.48\textwidth}
			\centering
			\subimport{Ver.3191/}{Table.Ver_3191.Position_8}
			\caption{Position 8}
			\label{subfig:fg_3191_8}
		\end{subfigure}
		\caption{State machine(s) (\subref{subfig:fg_3191_3}) at position 3, (\subref{subfig:fg_3191_8}) at position 8.}
		\label{fig:fg_3191_2}
	\end{figure}
	\begin{figure}
		\centering
		\begin{subfigure}[b]{0.96\textwidth}
			\centering
	\pgfmathtruncatemacro{\memory}{3}%
	\pgfmathtruncatemacro{\visibility}{1}%
	\pgfmathtruncatemacro{\maxtablelength}{8}%

	\newcommand{\GetNeighborhoodSizes}[1]{%
		\xdef\toleft{0}
		\xdef\toright{0}
		\xdef\toup{0}
		\xdef\todown{0}

		\pgfmathtruncatemacro{\tableSize}{2 * \visibility + 1}
		\foreach \columnoffset/\rowoffset[count=\i] in #1 {%
			\ifnum\columnoffset<\toleft%
				\xdef\toleft{\columnoffset}%
			\fi%
			\ifnum\columnoffset>\toright%
				\xdef\toright{\columnoffset}%
			\fi%
			\ifnum\rowoffset<\todown%
				\xdef\todown{\rowoffset}%
			\fi%
			\ifnum\rowoffset>\toup%
				\xdef\toup{\rowoffset}%
			\fi%
			\coordinate (anchor_\i) at (\columnoffset, \rowoffset);
		}%

		\xdef\xbase{\toleft}
		\xdef\ybase{\todown}
		\pgfmathtruncatemacro{\pxsize}{\toright - \toleft + 1}
		\xdef\xsize{\pxsize}
		\pgfmathtruncatemacro{\pysize}{\toup - \todown + 1}
		\xdef\ysize{\pysize}

	}%

	\newcommand*{\ExtractCoordinate}[1]{\path[overlay] (#1); \pgfgetlastxy{\XCoord}{\YCoord}}%

	\newcommand{\DrawNeighborhood}[1]{%
		\begin{scope}[scale=\scalefactor]
			\begin{scope}[shift={(-\xbase, -\ybase)}]
				\draw[fill=black] (0.5, 0.5) circle (3pt);
				\draw (0, 0) rectangle (1,1);

				\foreach \value[count=\i] in #1 {%
					\ExtractCoordinate{anchor_\i}
					\begin{scope}[shift={(\XCoord, \YCoord)}]
						\ifnum\value=0%
							\draw (0, 0) rectangle (1, 1);
						\fi%
						\ifnum\value=1%
							\draw[fill=black!50] (0, 0) rectangle (1, 1);
						\fi%
						\ifnum\value=2%
							\begin{scope}[opacity=\opacityfactor, blend group=normal]
								\draw[pattern=north west lines, draw = none] (0, 0) rectangle (1, 1);
							\end{scope}
							\draw (0, 0) rectangle (1, 1);
						\fi%
						\ifnum\value=3%
							\begin{scope}[opacity=\opacityfactor]
								\clip (0, 0) rectangle (1, 1);
								\node[transform shape] at (0.5, 0.5) {\Large $\mathbf{X}$};
							\end{scope}
							\draw (0, 0) rectangle (1, 1);
						\fi%
					\end{scope}
				}%
			\end{scope}
		\end{scope}
	}%

	\newcommand{\DrawAtIndex}[1]{%
		\pgfmathtruncatemacro{\currentstate}{#1}%
		\pgfmathtruncatemacro{\tablenum}{ceil((\currentstate + 1) / \maxtablelength)}%
		\pgfmathtruncatemacro{\columnnum}{mod(\currentstate, \maxtablelength)}%

		\pgfmathsetmacro{\startoffset}{(1 - \xsize * \scalefactor) / 2}
		\begin{scope}[shift={($(left_upper_corner_\tablenum) + (\columnnum + \startoffset, 0)$)}]
			\DrawNeighborhood{\neighbors}
		\end{scope}
	}%

	\newcommand{\SetTableValues}[4]{%
		\pgfmathtruncatemacro{\currentstate}{#1}%
		\pgfmathtruncatemacro{\tablenum}{ceil((\currentstate + 1) / \maxtablelength)}%
		\pgfmathtruncatemacro{\columnnum}{mod(\currentstate, \maxtablelength)}%

		\xdef\pickedDirection{\text{,}}
		\ifnum#4=0%
			\xdef\pickedDirection{\pickedDirection\text{N}}%
		\fi%
		\ifnum#4=1%
			\xdef\pickedDirection{\pickedDirection\text{E}}%
		\fi%
		\ifnum#4=2%
			\xdef\pickedDirection{\pickedDirection\text{S}}%
		\fi%
		\ifnum#4=3%
			\xdef\pickedDirection{\pickedDirection\text{W}}%
		\fi%
		\ifnum#4=4%
			\xdef\pickedDirection{\pickedDirection\varnothing}%
		\fi%
		\begin{scope}[shift={($(left_upper_corner_\tablenum) + (\columnnum, -#2)$)}]
			\node at (\tiklabelxoffset, \tiklabelyoffset) {\small $\padzeroes[\memory]{\binarynum{#3}}\pickedDirection$};
		\end{scope}
	}%

	\pgfmathsetmacro{\scalefactor}{0.3}%
	\pgfmathsetmacro{\opacityfactor}{0.7}%
	\pgfmathsetmacro{\tiklabelxoffset}{0.5}%
	\pgfmathsetmacro{\tiklabelyoffset}{0.5}%
	\pgfmathsetmacro{\lengthlabelline}{0.8}%
	\pgfmathsetmacro{\labelxoffset}{0.5}%
	\pgfmathsetmacro{\labelyoffset}{0.1}%
	\pgfmathsetmacro{\mintablespace}{0.7}%
	\pgfmathsetmacro{\cellsize}{1}%

	\begin{tikzpicture}
		\pgfmathtruncatemacro{\numberofstates}{8}%
		\pgfmathtruncatemacro{\tableheight}{2^(\memory)}%
		\pgfmathtruncatemacro{\numberoftables}{ceil(\numberofstates / \maxtablelength)}%

		\newcommand{\neighborsCoordinates}{0/1, 1/0, 0/-1, -1/0}
		\newcommand{\neighbors}{}
		\GetNeighborhoodSizes{\neighborsCoordinates}

		\begin{scope}[scale=\cellsize]
			\foreach \tablenum in {1, ..., \numberoftables}{%
				\pgfmathtruncatemacro{\tablelength}{ifthenelse(\tablenum < \numberoftables, \maxtablelength, \numberofstates - (\numberoftables - 1) * \maxtablelength)}%
				\pgfmathsetmacro{\yoffset}{-(\tablenum - 1) * (\tableheight + \mintablespace + \scalefactor * \ysize)}%

				\begin{scope}[yshift=\yoffset cm]
					\coordinate (left_upper_corner_\tablenum) at (0, \tableheight);
					\draw (0, 0) grid (\tablelength, \tableheight);
					\begin{scope}[shift={(left_upper_corner_\tablenum)}]
						\draw (0, 0) -- ++(135:\lengthlabelline);
						\node[rotate=-45,scale=0.5] at ($(135:\labelxoffset)+(45:\labelyoffset)$) {input};
						\node[rotate=-45,scale=0.5] at ($(135:\labelxoffset)+(180+45:\labelyoffset)$) {memory};
					\end{scope}
					\foreach \state in { 1, ..., \tableheight}{%
						\pgfmathtruncatemacro{\memorystate}{\state-1}
						\node (memory_\tablenum_\state) at (-1 + \tiklabelxoffset, \tableheight - \state + \tiklabelyoffset) {$\padzeroes[\memory]{\binarynum{\memorystate}}$};
					}%
				\end{scope}
			}%

			\renewcommand{\neighbors}{0, 0, 0, 2}
			\DrawAtIndex{0}
			\renewcommand{\neighbors}{0, 0, 1, 2}
			\DrawAtIndex{1}
			\renewcommand{\neighbors}{0, 1, 0, 2}
			\DrawAtIndex{2}
			\renewcommand{\neighbors}{0, 1, 1, 2}
			\DrawAtIndex{3}
			\renewcommand{\neighbors}{1, 0, 0, 2}
			\DrawAtIndex{4}
			\renewcommand{\neighbors}{1, 0, 1, 2}
			\DrawAtIndex{5}
			\renewcommand{\neighbors}{1, 1, 0, 2}
			\DrawAtIndex{6}
			\renewcommand{\neighbors}{1, 1, 1, 2}
			\DrawAtIndex{7}

			\SetTableValues{0}{1}{1}{1}
			\SetTableValues{1}{1}{1}{1}
			\SetTableValues{2}{1}{1}{0}
			\SetTableValues{3}{1}{1}{0}
			\SetTableValues{4}{1}{1}{1}
			\SetTableValues{5}{1}{1}{1}
			\SetTableValues{6}{1}{5}{4}
			\SetTableValues{7}{1}{1}{4}
			\SetTableValues{0}{2}{2}{1}
			\SetTableValues{1}{2}{2}{1}
			\SetTableValues{2}{2}{2}{4}
			\SetTableValues{3}{2}{2}{4}
			\SetTableValues{4}{2}{2}{1}
			\SetTableValues{5}{2}{2}{1}
			\SetTableValues{6}{2}{2}{4}
			\SetTableValues{7}{2}{6}{4}
			\SetTableValues{0}{3}{3}{4}
			\SetTableValues{1}{3}{3}{4}
			\SetTableValues{2}{3}{3}{4}
			\SetTableValues{3}{3}{3}{4}
			\SetTableValues{4}{3}{3}{4}
			\SetTableValues{5}{3}{3}{4}
			\SetTableValues{6}{3}{3}{4}
			\SetTableValues{7}{3}{3}{4}
			\SetTableValues{0}{4}{0}{4}
			\SetTableValues{1}{4}{0}{4}
			\SetTableValues{2}{4}{0}{4}
			\SetTableValues{3}{4}{0}{4}
			\SetTableValues{4}{4}{0}{4}
			\SetTableValues{5}{4}{0}{4}
			\SetTableValues{6}{4}{0}{4}
			\SetTableValues{7}{4}{0}{4}
			\SetTableValues{0}{5}{5}{1}
			\SetTableValues{1}{5}{5}{1}
			\SetTableValues{2}{5}{5}{4}
			\SetTableValues{3}{5}{5}{4}
			\SetTableValues{4}{5}{5}{1}
			\SetTableValues{5}{5}{5}{1}
			\SetTableValues{6}{5}{5}{4}
			\SetTableValues{7}{5}{1}{4}
			\SetTableValues{0}{6}{6}{1}
			\SetTableValues{1}{6}{6}{1}
			\SetTableValues{2}{6}{6}{2}
			\SetTableValues{3}{6}{2}{4}
			\SetTableValues{4}{6}{6}{1}
			\SetTableValues{5}{6}{6}{1}
			\SetTableValues{6}{6}{6}{2}
			\SetTableValues{7}{6}{6}{4}
			\SetTableValues{0}{7}{7}{4}
			\SetTableValues{1}{7}{7}{4}
			\SetTableValues{2}{7}{7}{4}
			\SetTableValues{3}{7}{7}{4}
			\SetTableValues{4}{7}{7}{4}
			\SetTableValues{5}{7}{7}{4}
			\SetTableValues{6}{7}{7}{4}
			\SetTableValues{7}{7}{7}{4}
			\SetTableValues{0}{8}{4}{4}
			\SetTableValues{1}{8}{4}{4}
			\SetTableValues{2}{8}{4}{4}
			\SetTableValues{3}{8}{4}{4}
			\SetTableValues{4}{8}{4}{4}
			\SetTableValues{5}{8}{4}{4}
			\SetTableValues{6}{8}{4}{4}
			\SetTableValues{7}{8}{4}{4}

		\end{scope}
	\end{tikzpicture}
			\caption{Position 6}
			\label{subfig:fg_3191_6}
		\end{subfigure}
		\caption{State machine(s) (\subref{subfig:fg_3191_6}) at position 6.}
		\label{fig:fg_3191_3}
	\end{figure}

		\subsection{A continue agent}
			\label{sec:Two.Lane-Algorithm.Continue}
			\vspace{33mm}
			
	\begin{figure}[ht]
		\centering
		\begin{subfigure}[b]{0.44\textwidth}
			\centering
			\subimport{Ver.3192/}{Table.Ver_3192.Position_1}
			\caption{Position 1}
			\label{subfig:fg_3192_1}
		\end{subfigure}
		\hfill
		\begin{subfigure}[b]{0.26\textwidth}
			\centering
			\subimport{Ver.3192/}{Table.Ver_3192.Position_2}
			\caption{Position 2}
			\label{subfig:fg_3192_2}
		\end{subfigure}
		\hfill
		\begin{subfigure}[b]{0.26\textwidth}
			\centering
			\subimport{Ver.3192/}{Table.Ver_3192.Position_3}
			\caption{Position 3}
			\label{subfig:fg_3192_3}
		\end{subfigure}
		\caption{State machine(s) (\subref{subfig:fg_3192_1}) at position 1, (\subref{subfig:fg_3192_2}) at position 2, (\subref{subfig:fg_3192_3}) at position 3.}
		\label{fig:fg_3192_1}
	\end{figure}
	\begin{figure}
		\centering
		\begin{subfigure}[b]{0.48\textwidth}
			\centering
			\subimport{Ver.3192/}{Table.Ver_3192.Position_4}
			\caption{Position 4}
			\label{subfig:fg_3192_4}
		\end{subfigure}
		\hfill
		\begin{subfigure}[b]{0.48\textwidth}
			\centering
			\subimport{Ver.3192/}{Table.Ver_3192.Position_6}
			\caption{Position 6}
			\label{subfig:fg_3192_6}
		\end{subfigure}
		\caption{State machine(s) (\subref{subfig:fg_3192_4}) at position 4, (\subref{subfig:fg_3192_6}) at position 6.}
		\label{fig:fg_3192_2}
	\end{figure}
	\begin{figure}
		\centering
		\begin{subfigure}[b]{0.96\textwidth}
			\centering
	\pgfmathtruncatemacro{\memory}{3}%
	\pgfmathtruncatemacro{\visibility}{1}%
	\pgfmathtruncatemacro{\maxtablelength}{8}%

	\newcommand{\GetNeighborhoodSizes}[1]{%
		\xdef\toleft{0}
		\xdef\toright{0}
		\xdef\toup{0}
		\xdef\todown{0}

		\pgfmathtruncatemacro{\tableSize}{2 * \visibility + 1}
		\foreach \columnoffset/\rowoffset[count=\i] in #1 {%
			\ifnum\columnoffset<\toleft%
				\xdef\toleft{\columnoffset}%
			\fi%
			\ifnum\columnoffset>\toright%
				\xdef\toright{\columnoffset}%
			\fi%
			\ifnum\rowoffset<\todown%
				\xdef\todown{\rowoffset}%
			\fi%
			\ifnum\rowoffset>\toup%
				\xdef\toup{\rowoffset}%
			\fi%
			\coordinate (anchor_\i) at (\columnoffset, \rowoffset);
		}%

		\xdef\xbase{\toleft}
		\xdef\ybase{\todown}
		\pgfmathtruncatemacro{\pxsize}{\toright - \toleft + 1}
		\xdef\xsize{\pxsize}
		\pgfmathtruncatemacro{\pysize}{\toup - \todown + 1}
		\xdef\ysize{\pysize}

	}%

	\newcommand*{\ExtractCoordinate}[1]{\path[overlay] (#1); \pgfgetlastxy{\XCoord}{\YCoord}}%

	\newcommand{\DrawNeighborhood}[1]{%
		\begin{scope}[scale=\scalefactor]
			\begin{scope}[shift={(-\xbase, -\ybase)}]
				\draw[fill=black] (0.5, 0.5) circle (3pt);
				\draw (0, 0) rectangle (1,1);

				\foreach \value[count=\i] in #1 {%
					\ExtractCoordinate{anchor_\i}
					\begin{scope}[shift={(\XCoord, \YCoord)}]
						\ifnum\value=0%
							\draw (0, 0) rectangle (1, 1);
						\fi%
						\ifnum\value=1%
							\draw[fill=black!50] (0, 0) rectangle (1, 1);
						\fi%
						\ifnum\value=2%
							\begin{scope}[opacity=\opacityfactor, blend group=normal]
								\draw[pattern=north west lines, draw = none] (0, 0) rectangle (1, 1);
							\end{scope}
							\draw (0, 0) rectangle (1, 1);
						\fi%
						\ifnum\value=3%
							\begin{scope}[opacity=\opacityfactor]
								\clip (0, 0) rectangle (1, 1);
								\node[transform shape] at (0.5, 0.5) {\Large $\mathbf{X}$};
							\end{scope}
							\draw (0, 0) rectangle (1, 1);
						\fi%
					\end{scope}
				}%
			\end{scope}
		\end{scope}
	}%

	\newcommand{\DrawAtIndex}[1]{%
		\pgfmathtruncatemacro{\currentstate}{#1}%
		\pgfmathtruncatemacro{\tablenum}{ceil((\currentstate + 1) / \maxtablelength)}%
		\pgfmathtruncatemacro{\columnnum}{mod(\currentstate, \maxtablelength)}%

		\pgfmathsetmacro{\startoffset}{(1 - \xsize * \scalefactor) / 2}
		\begin{scope}[shift={($(left_upper_corner_\tablenum) + (\columnnum + \startoffset, 0)$)}]
			\DrawNeighborhood{\neighbors}
		\end{scope}
	}%

	\newcommand{\SetTableValues}[4]{%
		\pgfmathtruncatemacro{\currentstate}{#1}%
		\pgfmathtruncatemacro{\tablenum}{ceil((\currentstate + 1) / \maxtablelength)}%
		\pgfmathtruncatemacro{\columnnum}{mod(\currentstate, \maxtablelength)}%

		\xdef\pickedDirection{\text{,}}
		\ifnum#4=0%
			\xdef\pickedDirection{\pickedDirection\text{N}}%
		\fi%
		\ifnum#4=1%
			\xdef\pickedDirection{\pickedDirection\text{E}}%
		\fi%
		\ifnum#4=2%
			\xdef\pickedDirection{\pickedDirection\text{S}}%
		\fi%
		\ifnum#4=3%
			\xdef\pickedDirection{\pickedDirection\text{W}}%
		\fi%
		\ifnum#4=4%
			\xdef\pickedDirection{\pickedDirection\varnothing}%
		\fi%
		\begin{scope}[shift={($(left_upper_corner_\tablenum) + (\columnnum, -#2)$)}]
			\node at (\tiklabelxoffset, \tiklabelyoffset) {\small $\padzeroes[\memory]{\binarynum{#3}}\pickedDirection$};
		\end{scope}
	}%

	\pgfmathsetmacro{\scalefactor}{0.3}%
	\pgfmathsetmacro{\opacityfactor}{0.7}%
	\pgfmathsetmacro{\tiklabelxoffset}{0.5}%
	\pgfmathsetmacro{\tiklabelyoffset}{0.5}%
	\pgfmathsetmacro{\lengthlabelline}{0.8}%
	\pgfmathsetmacro{\labelxoffset}{0.5}%
	\pgfmathsetmacro{\labelyoffset}{0.1}%
	\pgfmathsetmacro{\mintablespace}{0.7}%
	\pgfmathsetmacro{\cellsize}{1}%

	\begin{tikzpicture}
		\pgfmathtruncatemacro{\numberofstates}{8}%
		\pgfmathtruncatemacro{\tableheight}{2^(\memory)}%
		\pgfmathtruncatemacro{\numberoftables}{ceil(\numberofstates / \maxtablelength)}%

		\newcommand{\neighborsCoordinates}{0/1, 0/-1, -1/0, 1/0}
		\newcommand{\neighbors}{}
		\GetNeighborhoodSizes{\neighborsCoordinates}

		\begin{scope}[scale=\cellsize]
			\foreach \tablenum in {1, ..., \numberoftables}{%
				\pgfmathtruncatemacro{\tablelength}{ifthenelse(\tablenum < \numberoftables, \maxtablelength, \numberofstates - (\numberoftables - 1) * \maxtablelength)}%
				\pgfmathsetmacro{\yoffset}{-(\tablenum - 1) * (\tableheight + \mintablespace + \scalefactor * \ysize)}%

				\begin{scope}[yshift=\yoffset cm]
					\coordinate (left_upper_corner_\tablenum) at (0, \tableheight);
					\draw (0, 0) grid (\tablelength, \tableheight);
					\begin{scope}[shift={(left_upper_corner_\tablenum)}]
						\draw (0, 0) -- ++(135:\lengthlabelline);
						\node[rotate=-45,scale=0.5] at ($(135:\labelxoffset)+(45:\labelyoffset)$) {input};
						\node[rotate=-45,scale=0.5] at ($(135:\labelxoffset)+(180+45:\labelyoffset)$) {memory};
					\end{scope}
					\foreach \state in { 1, ..., \tableheight}{%
						\pgfmathtruncatemacro{\memorystate}{\state-1}
						\node (memory_\tablenum_\state) at (-1 + \tiklabelxoffset, \tableheight - \state + \tiklabelyoffset) {$\padzeroes[\memory]{\binarynum{\memorystate}}$};
					}%
				\end{scope}
			}%

			\renewcommand{\neighbors}{0, 0, 0, 2}
			\DrawAtIndex{0}
			\renewcommand{\neighbors}{0, 0, 1, 2}
			\DrawAtIndex{1}
			\renewcommand{\neighbors}{0, 1, 0, 2}
			\DrawAtIndex{2}
			\renewcommand{\neighbors}{0, 1, 1, 2}
			\DrawAtIndex{3}
			\renewcommand{\neighbors}{1, 0, 0, 2}
			\DrawAtIndex{4}
			\renewcommand{\neighbors}{1, 0, 1, 2}
			\DrawAtIndex{5}
			\renewcommand{\neighbors}{1, 1, 0, 2}
			\DrawAtIndex{6}
			\renewcommand{\neighbors}{1, 1, 1, 2}
			\DrawAtIndex{7}

			\SetTableValues{0}{1}{1}{4}
			\SetTableValues{1}{1}{1}{4}
			\SetTableValues{2}{1}{1}{4}
			\SetTableValues{3}{1}{1}{4}
			\SetTableValues{4}{1}{1}{4}
			\SetTableValues{5}{1}{1}{4}
			\SetTableValues{6}{1}{1}{4}
			\SetTableValues{7}{1}{1}{4}
			\SetTableValues{0}{2}{2}{4}
			\SetTableValues{1}{2}{2}{4}
			\SetTableValues{2}{2}{2}{4}
			\SetTableValues{3}{2}{2}{4}
			\SetTableValues{4}{2}{2}{4}
			\SetTableValues{5}{2}{2}{4}
			\SetTableValues{6}{2}{2}{4}
			\SetTableValues{7}{2}{2}{4}
			\SetTableValues{0}{3}{3}{3}
			\SetTableValues{1}{3}{3}{4}
			\SetTableValues{2}{3}{3}{3}
			\SetTableValues{3}{3}{3}{4}
			\SetTableValues{4}{3}{3}{3}
			\SetTableValues{5}{3}{3}{4}
			\SetTableValues{6}{3}{3}{3}
			\SetTableValues{7}{3}{7}{4}
			\SetTableValues{0}{4}{0}{3}
			\SetTableValues{1}{4}{0}{0}
			\SetTableValues{2}{4}{0}{3}
			\SetTableValues{3}{4}{0}{0}
			\SetTableValues{4}{4}{0}{3}
			\SetTableValues{5}{4}{4}{4}
			\SetTableValues{6}{4}{0}{3}
			\SetTableValues{7}{4}{0}{4}
			\SetTableValues{0}{5}{5}{4}
			\SetTableValues{1}{5}{5}{4}
			\SetTableValues{2}{5}{5}{4}
			\SetTableValues{3}{5}{5}{4}
			\SetTableValues{4}{5}{5}{4}
			\SetTableValues{5}{5}{5}{4}
			\SetTableValues{6}{5}{5}{4}
			\SetTableValues{7}{5}{5}{4}
			\SetTableValues{0}{6}{6}{4}
			\SetTableValues{1}{6}{6}{4}
			\SetTableValues{2}{6}{6}{4}
			\SetTableValues{3}{6}{6}{4}
			\SetTableValues{4}{6}{6}{4}
			\SetTableValues{5}{6}{6}{4}
			\SetTableValues{6}{6}{6}{4}
			\SetTableValues{7}{6}{6}{4}
			\SetTableValues{0}{7}{7}{3}
			\SetTableValues{1}{7}{7}{2}
			\SetTableValues{2}{7}{7}{3}
			\SetTableValues{3}{7}{3}{4}
			\SetTableValues{4}{7}{7}{3}
			\SetTableValues{5}{7}{7}{2}
			\SetTableValues{6}{7}{7}{3}
			\SetTableValues{7}{7}{7}{4}
			\SetTableValues{0}{8}{4}{3}
			\SetTableValues{1}{8}{4}{4}
			\SetTableValues{2}{8}{4}{3}
			\SetTableValues{3}{8}{4}{4}
			\SetTableValues{4}{8}{4}{3}
			\SetTableValues{5}{8}{4}{4}
			\SetTableValues{6}{8}{4}{3}
			\SetTableValues{7}{8}{0}{4}

		\end{scope}
	\end{tikzpicture}
			\caption{Position 8}
			\label{subfig:fg_3192_8}
		\end{subfigure}
		\caption{State machine(s) (\subref{subfig:fg_3192_8}) at position 8.}
		\label{fig:fg_3192_3}
	\end{figure}

\ifnum\mydoclevel=0
	\section{\texorpdfstring{\cref{thm:Two.Lane-the Main Theorem}}{Algorithm A2 Soundness Theorem} proof}\label{sec:Two.Lane-The.Proof}
\else
	\section{Algorithm \texorpdfstring{$\mathcal{A}_2$}{A2} Soundness Theorem proof}\label{sec:Two.Lane-The.Proof}
\fi
		In this section we provide a proof for the following theorem:
		\begin{reptheorem}{thm:Two.Lane-the Main Theorem}[Algorithm $\mathcal{A}_2$ Soundness Theorem]
			Consider an initial configuration $\mathcal{IC}$ of an $n \times 2$ grid $G$ that 
			contains at least a single empty slot. Let the number of exiting vehicles be $N_1$.
			Then Algorithm $\mathcal{A}_2$ solves the ER Problem and transitions $G$ to a 
			target configuration. Where target configuration is defined in the following sense:
			in case $N_1 \leq n$, then no exiting vehicles remain in the first column,
			otherwise no continue vehicles remain in the second column.
		\end{reptheorem}

	\subsection{Synchronization and Collision Prevention}
		Below we show that a distributed execution of the Algorithm $\mathcal{A}_2$ does not lead to an agent 
		collision. Recall that the Algorithm $\mathcal{A}_2$ is based on the LOOK-COMPUTE-MOVE paradigm,
		and is executed by the agents at discrete time ticks. I.e., the Algorithm $\mathcal{A}_2$ is a synchronized
		distributed algorithm.
		The movement direction summary is presented in 
		\cref{tbl:Two.Lane-MovementTable.Exit} and \cref{tbl:Two.Lane-MovementTable.Continue}. A succinct 
		agent flow overview is as follows:
		movement in North-South and East-West directions is equally split on the CC (see
		\cref{fig:Two.Lane-clock splits}). Column movement is allowed according to the East-West
		clock split, i.e. an agent is allowed to leave column $C$ at tick $t$ if the movement
		is allowed in column $C$ at tick $t$.
		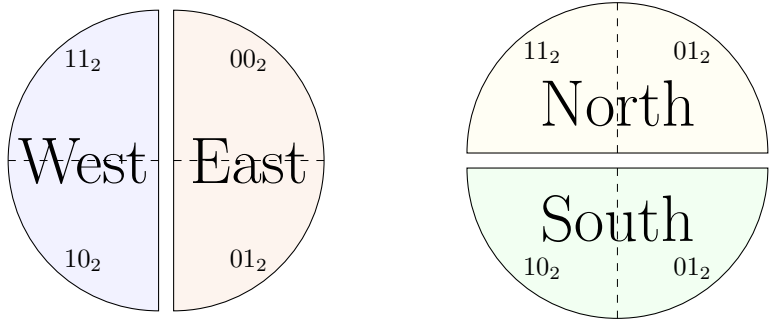
\begin{figure}
			\centering
				\begin{tikzpicture}
		\pgfmathsetmacro{\radius}{2}
		
		\begin{scope}[shift={(-0.1, 0)}]
			\draw[thin, fill = blue!5] (0, \radius) arc (90:270:\radius cm) -- cycle;
			\draw[ultra thin, dashed] (0,0) -- +(-\radius, 0);
			\node at (-\radius /2, 0) { \Huge West};
			\node at (-\radius / 2, \radius * 2 / 3) {$11_2$};
			\node at (-\radius / 2, -\radius * 2 / 3) {$10_2$};
		\end{scope}
		\begin{scope}[shift={(0.1, 0)}]
			\draw[thin, fill = yellow!5!red!5] (0, -\radius) arc (-90:90:\radius cm) -- cycle;
			\draw[ultra thin, dashed] (0,0) -- +(\radius, 0);
			\node at (\radius /2, 0) { \Huge East};
			\node at (\radius / 2, \radius * 2 / 3) {$00_2$};
			\node at (\radius / 2, -\radius * 2 / 3) {$01_2$};
		\end{scope}

		\begin{scope}[shift={(3 * \radius, 0)}]
			\begin{scope}[shift={(0, -0.1)}]
				\draw[thin, fill = green!5] (-\radius, 0) arc (180:360:\radius cm) -- cycle;
				\draw[ultra thin, dashed] (0,0) -- +(0, -\radius);
				\node at (0, -\radius / 3) { \Huge South};
				\node at (\radius / 2, -\radius * 2 / 3) {$01_2$};
				\node at (-\radius / 2, -\radius * 2 / 3) {$10_2$};
			\end{scope}
			\begin{scope}[shift={(0, 0.1)}]
				\draw[thin, fill = yellow!5] (\radius, 0) arc (0:180:\radius cm) -- cycle;
				\draw[ultra thin, dashed] (0,0) -- +(0, \radius);
				\node at (0, \radius / 3) { \Huge North};
				\node at (-\radius / 2, \radius * 2 / 3) {$11_2$};
				\node at (\radius / 2, \radius * 2 / 3) {$01_2$};
			\end{scope}
		\end{scope}
	\end{tikzpicture}
			\caption{(left) East-West clock split: East movement is allowed at times $0$ and $1$; 
			West movement is possible at times $2$ and $3$. (right) North-South clock split.
			Column movement is aligned with the East-West split.}
			\label{fig:Two.Lane-clock splits}
		\end{figure}

	\subsection{Non-starvation}
		The $2$-column case is much simpler than the general $m > 2$ one. We are essentially required to
		show that an agent, which should be in a target column $TC$, eventually leaves its initial non-target
		column $IC$. The motion on the East-West axes is a one-way only: East for exiting agents
		and West for continue agents. Hence, we conclude that every agent that \textit{has} changed 
		its initial column - is already located in the column it is expected to be in a target configuration. 
		
		Denote an agent located in its TC as a \textit{settled} agent, otherwise we denote
		an agent as an \textit{unsettled} agent.
		
		\begin{lemma}
			\label{lemma:Two.Lane-No.Fast.Return.Lemma}
			Let $j \in \{1, 2\}$ and $1 \leq i \leq n$. Suppose an agent $A$ left cell $a_{i, j}$ at time
			$t$ to either North or South. Then $a_{i, j}$ can not become occupied by an agent moving North
			or South until at least $t+4$.
		\end{lemma}
		\begin{proof}
			The proof is done by enumeration. We refer the reader to \cref{tbl:Two.Lane-Cell.Return.Table}.
		\end{proof}
		
	\begin{table}
		\centering
			{
				\begin{tabular}{c  c c c  c c  c c }
					\hline
					\multirow{3}{*}[3pt]{\tabhead{Category}} & 
					\multicolumn{3}{c}{\tabhead{Direction}} &
					\multicolumn{2}{c}{\tabhead{\textit{D} bit}} &
					\multicolumn{2}{c}{\tabhead{Earliest}} \\
						\cmidrule{2-4}
						\cmidrule(lr){5-6}
						\cmidrule(lr){7-8}
					& {from} & {to} & {$\rightarrow$} &
					{move} & {return} &
					{back} & {next} \\
					\hline\hline \\[-7pt]
						\multirow{4}{*}{
								\parbox[c]{1.2cm}{\centering first column}
						}
						& 2 & 3, 6 & N & 0 & 1 & t + 5& - \\
						& 3 & 2, 6 & S & 1 & 0 & t + 7 & - \\
						& 6 & 2, 6 & S & 1 & 0 & t + 7 & t + 4 \\
						& 6 & 3, 6 & N & 0 & 1 & t + 5 & t + 4 \\ [5pt]
						\multirow{4}{*}{
								\parbox[c]{1.2cm}{\centering second column}
						}
						& 1 & 4, 8 & N & 0 & 1 & t + 7 & - \\
						& 4 & 1, 8 & S & 1 & 0 & t + 5 & - \\
						& 8 & 1, 8 & S & 1 & 0 & t + 5 & t + 4 \\
						& 8 & 4, 8 & N & 0 & 1 & t + 7 & t + 4 \\
					\hline
				\end{tabular}
			}
			\caption{Cell $a_{i, j}$ vacancy/refill state. Cells in the first column are in position
			$2, 3, 6$, and in the second column $1, 4$ and $8$. An agent moves \textit{from} one position
			\textit{to} another; movement direction is the $\rightarrow$ column. Direction bit at
			departure and arrival are shown in the next two columns. Then briefly followed by the
			earliest possible return time, or the earliest arrival time of a following agent 
			that moves in the North/South direction.}
			\label{tbl:Two.Lane-Cell.Return.Table}
	\end{table}

		Let us name grid cell $a_{i, 3 - j}$ the \textit{side neighbor} cell of cell $a_{i,j}$.
		\begin{lemma}\label{lemma:Two.Lane-Unsettled-Patient-Agent}
			Consider an unsettled agent $A$ at cell $a_{i, j}$. Suppose that the side neighbor cell $a_{i, 3- j}$
			became empty at $t$, i.e. $A$ detects an empty cell on the East/West at the beginning of 
			time tick $t+1$. Then $A$ moves to its target column no later than at $t+2$.			
		\end{lemma}
		\begin{proof}
			Note, that the side neighbor cell became empty due to the North-South movement of an agent, that 
			occupied the $a_{i, 3 - j}$ cell at time $t$ and subsequently left. According to
			\cref{lemma:Two.Lane-No.Fast.Return.Lemma}, $a_{i, 3-j}$ will not be the target of the North-South
			agent movement at least until $t+4$. However, according to \cref{tbl:Two.Lane-MovementTable.Exit} 
			and \cref{tbl:Two.Lane-MovementTable.Continue}, an unsettled agent moves on the East-West axes
			in at most two time ticks after the side neighbor cell became empty. We observe, that agents in 
			columns move in \enquote{bursts}: first column agents move at ticks $00$ and $01$, followed
			by two ticks dedicated to the movement of the second column agents. Moreover, in the first tick
			out of two consecutive column-related ticks, an agent \textit{is} allowed to move to 
			its target column. Hence, $A$ will move to its target column no later than at $t+2$.
		\end{proof}
		
		Notice, that agents are active for two consecutive ticks during one CC ($4$ time ticks). Agents
		in the first column at ticks $00_2$ and $01_2$, while agents in the second column during the remaining 
		two ticks. Let denote those tick pairs as \textit{column activity interval}(s).
		
		An unsettled agent that neighbors an empty space at the start of its current column activity interval, 
		will unconditionally move to its target column immediately. 
		Furthermore, an unsettled agent that moved in the North-South direction at the start of its column
		activity interval, and consequently moved to neighbor an empty space, will also 
		unconditionally move to its target column in the closing tick of this activity interval. 
		
		We note, that an unsettled exiting agents move North in the first tick of the column activity interval, 
		while an unsettled continuing agents move South. Therefore, we have eliminated East-West movement options for
		positions $2$ and $4$ in the closing activity interval tick. This is due to the fact, that an 
		unsettled agent can not physically move to these two specific positions at the start of an
		activity interval.
		
		Our next goal is to show that at least one unsettled agent in the unsorted grid configuration
		eventually moves to its target column.
		
		Assume that an initial configuration $\mathcal{BC}$ can not be sorted by Algorithm
		$\mathcal{A}_2$ execution. Therefore, there exists a time $T$ such that for any $t > T$ no agent moves East or West.
		Moreover, grid configurations at all times $t$ are not sorted.

		Below we show, that empty cells are quite in demand despite the East-West movement termination. We prove
		that an agent near an empty cell does not ignore it for any substantial amount of time.
		\begin{lemma}
			\label{lemma:Two.Lane-No.Stalls}
			Suppose that an initial grid configuration is $\mathcal{BC}$.
			Consider an agent $A$ at cell $a_{i, j}$. Suppose, that a neighbor cell $a_{i-1, j}$ 
			(or $a_{i+1,j}$) becomes empty at time $t > T$, then $A$ leaves $a_{i,j}$ no later than at $t+4$.
		\end{lemma}
		The proof is by enumeration of all the possible states (see \cref{table:Two.Lane-for-lemma-no-stalls-E-148}-
		\cref{table:Two.Lane-for-lemma-no-stalls-C-236}).

	\begin{table}
		\pgfmathsetmacro{\NeighborhoodWidth}{1.2}
		\centering
\begin{tabular}{*{9}{c}}
			\toprule

			%
			%
			\multirow{3}{*}[3 pt]{\tabhead{Position}} &
			\multicolumn{3}{c}{\tabhead{Neighborhood}} &
			\multicolumn{5}{c}{\tabhead{Memory state}} \\

			%
			%
			\cmidrule(lr){2-4}
			\cmidrule(lr){5-9}
			 &
			{$t$} &
			{$t+1$} &
			{$t+2$} &
			{$t$} &
			{$t+1$} &
			{$t+2$} &
			{$t+3$} &
			{$t+4$} \\
			\midrule\midrule \\
\bottomrule
		\end{tabular}
		\caption{
		Enumeration of possible neighborhoods
		and memory state sequences of
		an exiting
		agent.
		After East-West movement cessation.
		Grid position; visible $L_1$ neighborhood and memory state in the binary format
		are provided for short periods of time starting at $t$.
		Where applicable agent movement direction is given instead of a memory state.
		Relevant agent positions: 
		1
		\!\!, 4
		 and 8.
		Impossible states are excluded from enumeration.
		}
		\label{table:Two.Lane-for-lemma-no-stalls-E-148}
	\end{table}

	\begin{table}
		\pgfmathsetmacro{\NeighborhoodWidth}{1.2}
		\centering
		\begin{tabular}{*{9}{c}}
			\toprule

			%
			%
			\multirow{3}{*}[3 pt]{\tabhead{Position}} &
			\multicolumn{3}{c}{\tabhead{Neighborhood}} &
			\multicolumn{5}{c}{\tabhead{Memory state}} \\

			%
			%
			\cmidrule(lr){2-4}
			\cmidrule(lr){5-9}
			 &
			{$t$} &
			{$t+1$} &
			{$t+2$} &
			{$t$} &
			{$t+1$} &
			{$t+2$} &
			{$t+3$} &
			{$t+4$} \\
			\midrule\midrule \\
\bottomrule
		\end{tabular}
		\caption{
		Enumeration of possible visible neighborhoods
		and memory state sequences of
		a continue
		agent.
		After East-West movement cessation.
		Grid position; visible $L_1$ neighborhood and memory state in the binary format
		are provided for short periods of time starting at $t$.
		Where applicable agent movement direction is given instead of a memory state.
		Relevant agent positions: 
		1
		\!\!, 4
		 and 8.
		Impossible states are excluded from enumeration.
		}
		\label{table:Two.Lane-for-lemma-no-stalls-C-148}
	\end{table}

	\begin{table}
		\pgfmathsetmacro{\NeighborhoodWidth}{1.2}
		\centering
\begin{tabular}{*{9}{c}}
			\toprule

			%
			%
			\multirow{3}{*}[3 pt]{\tabhead{Position}} &
			\multicolumn{3}{c}{\tabhead{Neighborhood}} &
			\multicolumn{5}{c}{\tabhead{Memory state}} \\

			%
			%
			\cmidrule(lr){2-4}
			\cmidrule(lr){5-9}
			 &
			{$t$} &
			{$t+1$} &
			{$t+2$} &
			{$t$} &
			{$t+1$} &
			{$t+2$} &
			{$t+3$} &
			{$t+4$} \\
			\midrule\midrule \\
\bottomrule
		\end{tabular}
		\caption{
		Enumeration of possible visible neighborhoods
		and memory state sequences of
		an exiting
		agent.
		After East-West movement cessation.
		Grid position; visible $L_1$ neighborhood and memory state in the binary format
		are provided for short periods of time starting at $t$.
		Where applicable agent movement direction is given instead of a memory state.
		Relevant agent positions: 
		2
		\!\!, 3
		 and 6.
		Impossible states are excluded from enumeration.
		}
		\label{table:Two.Lane-for-lemma-no-stalls-E-236}
	\end{table}

	\begin{table}
		\pgfmathsetmacro{\NeighborhoodWidth}{1.2}
		\centering
		\begin{tabular}{*{9}{c}}
			\toprule

			%
			%
			\multirow{3}{*}[3 pt]{\tabhead{Position}} &
			\multicolumn{3}{c}{\tabhead{Neighborhood}} &
			\multicolumn{5}{c}{\tabhead{Memory state}} \\

			%
			%
			\cmidrule(lr){2-4}
			\cmidrule(lr){5-9}
			 &
			{$t$} &
			{$t+1$} &
			{$t+2$} &
			{$t$} &
			{$t+1$} &
			{$t+2$} &
			{$t+3$} &
			{$t+4$} \\
			\midrule\midrule \\[-7 pt]

			%
			%
			\multirow{2}{*}[2 pt]{2} &
			\multirow{2}{*}[2 pt]{\adjustbox{max width=\NeighborhoodWidth cm}{%
				}} &
			\multirow{2}{*}[2 pt]{\adjustbox{max width=\NeighborhoodWidth cm}{%

	\pgfmathtruncatemacro{\memory}{-1}%
	\pgfmathtruncatemacro{\visibility}{1}%
	\pgfmathtruncatemacro{\maxtablelength}{8}%

	\newcommand{\GetNeighborhoodSizes}[1]{%
		\xdef\toleft{0}
		\xdef\todown{0}

		\pgfmathtruncatemacro{\tableSize}{2 * \visibility + 1}
		\foreach \columnoffset/\rowoffset[count=\i] in #1 {%
			\ifnum\columnoffset<\toleft%
				\xdef\toleft{\columnoffset}%
			\fi%
			\ifnum\rowoffset<\todown%
				\xdef\todown{\rowoffset}%
			\fi%
			\coordinate (anchor_\i) at (\columnoffset, \rowoffset);
		}%

		\xdef\xbase{\toleft}
		\xdef\ybase{\todown}

	}%

	\newcommand*{\ExtractCoordinate}[1]{\path[overlay] (#1); \pgfgetlastxy{\XCoord}{\YCoord}}%

	\newcommand{\DrawNeighborhood}[1]{%
		\begin{scope}[scale=\scalefactor]
			\begin{scope}[shift={(-\xbase, -\ybase)}]
				\draw[fill=black] (0.5, 0.5) circle (3pt);
				\draw (0, 0) rectangle (1,1);

				\foreach \value[count=\i] in #1 {%
					\ExtractCoordinate{anchor_\i}
					\begin{scope}[shift={(\XCoord, \YCoord)}]
						\ifnum\value=1%
							\fill[black!50] (0, 0) rectangle (1, 1);
						\fi%
						\ifnum\value=2%
							\begin{scope}[opacity=\opacityfactor, blend group=normal]
								\draw[pattern=north west lines, draw = none] (0, 0) rectangle (1, 1);
							\end{scope}
						\fi%
						\ifnum\value=3%
							\begin{scope}[opacity=\opacityfactor]
								\node[transform shape] at (0.5, 0.5) {\Large $\mathbf{X}$};
							\end{scope}
						\fi%
						\draw (0, 0) rectangle (1, 1);
					\end{scope}
				}%
			\end{scope}
		\end{scope}
	}%

	\pgfmathsetmacro{\scalefactor}{0.3}%
	\pgfmathsetmacro{\opacityfactor}{0.7}%
	\pgfmathsetmacro{\tiklabelxoffset}{0.5}%
	\pgfmathsetmacro{\tiklabelyoffset}{0.5}%

	\begin{tikzpicture}
		\newcommand{\neighborsCoordinates}{0/1, 0/-1, -1/0, 1/0}
		\newcommand{\neighbors}{}
		\GetNeighborhoodSizes{\neighborsCoordinates}

		\renewcommand{\neighbors}{0, 2, 2, 3}
		\DrawNeighborhood{\neighbors}
	\end{tikzpicture}
				}} &
			 &
			$000_2$ &
			\multirow{2}{*}[2 pt]{$001_2$} &
			\multirow{2}{*}[2 pt]{$010_2$} &
			\multirow{2}{*}[2 pt]{$011_2$} &
			\multirow{2}{*}[2 pt]{N} \\

			%
			%
			 &
			 &
			 &
			 &
			$100_2$ &
			 &
			 &
			 &
			 \\[10 pt]
			\midrule \\ [-5 pt]

			%
			%
			\multirow{2}{*}[2 pt]{3} &
			\multirow{2}{*}[2 pt]{\adjustbox{max width=\NeighborhoodWidth cm}{%

	\pgfmathtruncatemacro{\memory}{-1}%
	\pgfmathtruncatemacro{\visibility}{1}%
	\pgfmathtruncatemacro{\maxtablelength}{8}%

	\newcommand{\GetNeighborhoodSizes}[1]{%
		\xdef\toleft{0}
		\xdef\todown{0}

		\pgfmathtruncatemacro{\tableSize}{2 * \visibility + 1}
		\foreach \columnoffset/\rowoffset[count=\i] in #1 {%
			\ifnum\columnoffset<\toleft%
				\xdef\toleft{\columnoffset}%
			\fi%
			\ifnum\rowoffset<\todown%
				\xdef\todown{\rowoffset}%
			\fi%
			\coordinate (anchor_\i) at (\columnoffset, \rowoffset);
		}%

		\xdef\xbase{\toleft}
		\xdef\ybase{\todown}

	}%

	\newcommand*{\ExtractCoordinate}[1]{\path[overlay] (#1); \pgfgetlastxy{\XCoord}{\YCoord}}%

	\newcommand{\DrawNeighborhood}[1]{%
		\begin{scope}[scale=\scalefactor]
			\begin{scope}[shift={(-\xbase, -\ybase)}]
				\draw[fill=black] (0.5, 0.5) circle (3pt);
				\draw (0, 0) rectangle (1,1);

				\foreach \value[count=\i] in #1 {%
					\ExtractCoordinate{anchor_\i}
					\begin{scope}[shift={(\XCoord, \YCoord)}]
						\ifnum\value=1%
							\fill[black!50] (0, 0) rectangle (1, 1);
						\fi%
						\ifnum\value=2%
							\begin{scope}[opacity=\opacityfactor, blend group=normal]
								\draw[pattern=north west lines, draw = none] (0, 0) rectangle (1, 1);
							\end{scope}
						\fi%
						\ifnum\value=3%
							\begin{scope}[opacity=\opacityfactor]
								\node[transform shape] at (0.5, 0.5) {\Large $\mathbf{X}$};
							\end{scope}
						\fi%
						\draw (0, 0) rectangle (1, 1);
					\end{scope}
				}%
			\end{scope}
		\end{scope}
	}%

	\pgfmathsetmacro{\scalefactor}{0.3}%
	\pgfmathsetmacro{\opacityfactor}{0.7}%
	\pgfmathsetmacro{\tiklabelxoffset}{0.5}%
	\pgfmathsetmacro{\tiklabelyoffset}{0.5}%

	\begin{tikzpicture}
		\newcommand{\neighborsCoordinates}{1/0, 0/1, -1/0, 0/-1}
		\newcommand{\neighbors}{}
		\GetNeighborhoodSizes{\neighborsCoordinates}

		\renewcommand{\neighbors}{1, 2, 2, 3}
		\DrawNeighborhood{\neighbors}
	\end{tikzpicture}
				}} &
			\multirow{2}{*}[2 pt]{\adjustbox{max width=\NeighborhoodWidth cm}{%

	\pgfmathtruncatemacro{\memory}{-1}%
	\pgfmathtruncatemacro{\visibility}{1}%
	\pgfmathtruncatemacro{\maxtablelength}{8}%

	\newcommand{\GetNeighborhoodSizes}[1]{%
		\xdef\toleft{0}
		\xdef\todown{0}

		\pgfmathtruncatemacro{\tableSize}{2 * \visibility + 1}
		\foreach \columnoffset/\rowoffset[count=\i] in #1 {%
			\ifnum\columnoffset<\toleft%
				\xdef\toleft{\columnoffset}%
			\fi%
			\ifnum\rowoffset<\todown%
				\xdef\todown{\rowoffset}%
			\fi%
			\coordinate (anchor_\i) at (\columnoffset, \rowoffset);
		}%

		\xdef\xbase{\toleft}
		\xdef\ybase{\todown}

	}%

	\newcommand*{\ExtractCoordinate}[1]{\path[overlay] (#1); \pgfgetlastxy{\XCoord}{\YCoord}}%

	\newcommand{\DrawNeighborhood}[1]{%
		\begin{scope}[scale=\scalefactor]
			\begin{scope}[shift={(-\xbase, -\ybase)}]
				\draw[fill=black] (0.5, 0.5) circle (3pt);
				\draw (0, 0) rectangle (1,1);

				\foreach \value[count=\i] in #1 {%
					\ExtractCoordinate{anchor_\i}
					\begin{scope}[shift={(\XCoord, \YCoord)}]
						\ifnum\value=1%
							\fill[black!50] (0, 0) rectangle (1, 1);
						\fi%
						\ifnum\value=2%
							\begin{scope}[opacity=\opacityfactor, blend group=normal]
								\draw[pattern=north west lines, draw = none] (0, 0) rectangle (1, 1);
							\end{scope}
						\fi%
						\ifnum\value=3%
							\begin{scope}[opacity=\opacityfactor]
								\node[transform shape] at (0.5, 0.5) {\Large $\mathbf{X}$};
							\end{scope}
						\fi%
						\draw (0, 0) rectangle (1, 1);
					\end{scope}
				}%
			\end{scope}
		\end{scope}
	}%

	\pgfmathsetmacro{\scalefactor}{0.3}%
	\pgfmathsetmacro{\opacityfactor}{0.7}%
	\pgfmathsetmacro{\tiklabelxoffset}{0.5}%
	\pgfmathsetmacro{\tiklabelyoffset}{0.5}%

	\begin{tikzpicture}
		\newcommand{\neighborsCoordinates}{1/0, 0/1, -1/0, 0/-1}
		\newcommand{\neighbors}{}
		\GetNeighborhoodSizes{\neighborsCoordinates}

		\renewcommand{\neighbors}{0, 2, 2, 3}
		\DrawNeighborhood{\neighbors}
	\end{tikzpicture}
				}} &
			 &
			$001_2$ &
			\multirow{2}{*}[2 pt]{$110_2$} &
			\multirow{2}{*}[2 pt]{$111_2$} &
			\multirow{2}{*}[2 pt]{$100_2$} &
			\multirow{2}{*}[2 pt]{S} \\

			%
			%
			 &
			 &
			 &
			 &
			$101_2$ &
			 &
			 &
			 &
			 \\[10 pt]
			\midrule \\ [-5 pt]

			%
			%
			\multirow{10}{*}[-18 pt]{6} &
			\multirow{4}{*}[-3 pt]{\adjustbox{max width=\NeighborhoodWidth cm}{%
				}} &
			\multirow{4}{*}[-3 pt]{\adjustbox{max width=\NeighborhoodWidth cm}{%

	\pgfmathtruncatemacro{\memory}{-1}%
	\pgfmathtruncatemacro{\visibility}{1}%
	\pgfmathtruncatemacro{\maxtablelength}{8}%

	\newcommand{\GetNeighborhoodSizes}[1]{%
		\xdef\toleft{0}
		\xdef\todown{0}

		\pgfmathtruncatemacro{\tableSize}{2 * \visibility + 1}
		\foreach \columnoffset/\rowoffset[count=\i] in #1 {%
			\ifnum\columnoffset<\toleft%
				\xdef\toleft{\columnoffset}%
			\fi%
			\ifnum\rowoffset<\todown%
				\xdef\todown{\rowoffset}%
			\fi%
			\coordinate (anchor_\i) at (\columnoffset, \rowoffset);
		}%

		\xdef\xbase{\toleft}
		\xdef\ybase{\todown}

	}%

	\newcommand*{\ExtractCoordinate}[1]{\path[overlay] (#1); \pgfgetlastxy{\XCoord}{\YCoord}}%

	\newcommand{\DrawNeighborhood}[1]{%
		\begin{scope}[scale=\scalefactor]
			\begin{scope}[shift={(-\xbase, -\ybase)}]
				\draw[fill=black] (0.5, 0.5) circle (3pt);
				\draw (0, 0) rectangle (1,1);

				\foreach \value[count=\i] in #1 {%
					\ExtractCoordinate{anchor_\i}
					\begin{scope}[shift={(\XCoord, \YCoord)}]
						\ifnum\value=1%
							\fill[black!50] (0, 0) rectangle (1, 1);
						\fi%
						\ifnum\value=2%
							\begin{scope}[opacity=\opacityfactor, blend group=normal]
								\draw[pattern=north west lines, draw = none] (0, 0) rectangle (1, 1);
							\end{scope}
						\fi%
						\ifnum\value=3%
							\begin{scope}[opacity=\opacityfactor]
								\node[transform shape] at (0.5, 0.5) {\Large $\mathbf{X}$};
							\end{scope}
						\fi%
						\draw (0, 0) rectangle (1, 1);
					\end{scope}
				}%
			\end{scope}
		\end{scope}
	}%

	\pgfmathsetmacro{\scalefactor}{0.3}%
	\pgfmathsetmacro{\opacityfactor}{0.7}%
	\pgfmathsetmacro{\tiklabelxoffset}{0.5}%
	\pgfmathsetmacro{\tiklabelyoffset}{0.5}%

	\begin{tikzpicture}
		\newcommand{\neighborsCoordinates}{0/1, -1/0, 1/0, 0/-1}
		\newcommand{\neighbors}{}
		\GetNeighborhoodSizes{\neighborsCoordinates}

		\renewcommand{\neighbors}{1, 2, 3, 3}
		\DrawNeighborhood{\neighbors}
	\end{tikzpicture}
				}} &
			\multirow{2}{*}[2 pt]{\adjustbox{max width=\NeighborhoodWidth cm}{%
				}} &
			$000_2$ &
			\multirow{4}{*}[-3 pt]{$001_2$} &
			\multirow{4}{*}[-3 pt]{$010_2$} &
			\multirow{4}{*}[-3 pt]{$011_2$} &
			\multirow{4}{*}[-3 pt]{N} \\

			%
			%
			 &
			 &
			 &
			 &
			$100_2$ &
			 &
			 &
			 &
			 \\[5 pt]

			%
			%
			 &
			 &
			 &
			\multirow{2}{*}[2 pt]{\adjustbox{max width=\NeighborhoodWidth cm}{%

	\pgfmathtruncatemacro{\memory}{-1}%
	\pgfmathtruncatemacro{\visibility}{1}%
	\pgfmathtruncatemacro{\maxtablelength}{8}%

	\newcommand{\GetNeighborhoodSizes}[1]{%
		\xdef\toleft{0}
		\xdef\todown{0}

		\pgfmathtruncatemacro{\tableSize}{2 * \visibility + 1}
		\foreach \columnoffset/\rowoffset[count=\i] in #1 {%
			\ifnum\columnoffset<\toleft%
				\xdef\toleft{\columnoffset}%
			\fi%
			\ifnum\rowoffset<\todown%
				\xdef\todown{\rowoffset}%
			\fi%
			\coordinate (anchor_\i) at (\columnoffset, \rowoffset);
		}%

		\xdef\xbase{\toleft}
		\xdef\ybase{\todown}

	}%

	\newcommand*{\ExtractCoordinate}[1]{\path[overlay] (#1); \pgfgetlastxy{\XCoord}{\YCoord}}%

	\newcommand{\DrawNeighborhood}[1]{%
		\begin{scope}[scale=\scalefactor]
			\begin{scope}[shift={(-\xbase, -\ybase)}]
				\draw[fill=black] (0.5, 0.5) circle (3pt);
				\draw (0, 0) rectangle (1,1);

				\foreach \value[count=\i] in #1 {%
					\ExtractCoordinate{anchor_\i}
					\begin{scope}[shift={(\XCoord, \YCoord)}]
						\ifnum\value=1%
							\fill[black!50] (0, 0) rectangle (1, 1);
						\fi%
						\ifnum\value=2%
							\begin{scope}[opacity=\opacityfactor, blend group=normal]
								\draw[pattern=north west lines, draw = none] (0, 0) rectangle (1, 1);
							\end{scope}
						\fi%
						\ifnum\value=3%
							\begin{scope}[opacity=\opacityfactor]
								\node[transform shape] at (0.5, 0.5) {\Large $\mathbf{X}$};
							\end{scope}
						\fi%
						\draw (0, 0) rectangle (1, 1);
					\end{scope}
				}%
			\end{scope}
		\end{scope}
	}%

	\pgfmathsetmacro{\scalefactor}{0.3}%
	\pgfmathsetmacro{\opacityfactor}{0.7}%
	\pgfmathsetmacro{\tiklabelxoffset}{0.5}%
	\pgfmathsetmacro{\tiklabelyoffset}{0.5}%

	\begin{tikzpicture}
		\newcommand{\neighborsCoordinates}{0/1, -1/0, 1/0, 0/-1}
		\newcommand{\neighbors}{}
		\GetNeighborhoodSizes{\neighborsCoordinates}

		\renewcommand{\neighbors}{0, 2, 3, 3}
		\DrawNeighborhood{\neighbors}
	\end{tikzpicture}
				}} &
			$000_2$ &
			 &
			 &
			 &
			 \\

			%
			%
			 &
			 &
			 &
			 &
			$100_2$ &
			 &
			 &
			 &
			 \\[10 pt]

			%
			%
			 &
			\multirow{4}{*}[-3 pt]{\adjustbox{max width=\NeighborhoodWidth cm}{%
				}} &
			\multirow{2}{*}[2 pt]{\adjustbox{max width=\NeighborhoodWidth cm}{%
				}} &
			\multirow{2}{*}[2 pt]{} &
			$000_2$ &
			\multirow{2}{*}[2 pt]{$101_2$} &
			\multirow{2}{*}[2 pt]{$010_2$} &
			\multirow{2}{*}[2 pt]{$011_2$} &
			\multirow{2}{*}[2 pt]{N} \\

			%
			%
			 &
			 &
			 &
			 &
			$100_2$ &
			 &
			 &
			 &
			 \\[5 pt]

			%
			%
			 &
			 &
			\multirow{2}{*}[2 pt]{\adjustbox{max width=\NeighborhoodWidth cm}{%
				}} &
			\multirow{2}{*}[2 pt]{} &
			$000_2$ &
			\multirow{2}{*}[2 pt]{S} &
			 &
			 &
			 \\

			%
			%
			 &
			 &
			 &
			 &
			$100_2$ &
			 &
			 &
			 &
			 \\[10 pt]

			%
			%
			 &
			\multirow{2}{*}[2 pt]{\adjustbox{max width=\NeighborhoodWidth cm}{%
				}} &
			\multirow{2}{*}[2 pt]{\adjustbox{max width=\NeighborhoodWidth cm}{%
				}} &
			\multirow{2}{*}[2 pt]{} &
			$001_2$ &
			\multirow{2}{*}[2 pt]{$110_2$} &
			\multirow{2}{*}[2 pt]{$111_2$} &
			\multirow{2}{*}[2 pt]{$100_2$} &
			\multirow{2}{*}[2 pt]{S} \\

			%
			%
			 &
			 &
			 &
			 &
			$101_2$ &
			 &
			 &
			 &
			 \\[10 pt]

			\bottomrule
		\end{tabular}
		\caption{
		Enumeration of possible visible neighborhoods
		and memory state sequences of
		a continue
		agent.
		After East-West movement cessation.
		Grid position; visible $L_1$ neighborhood and memory state in the binary format
		are provided for short periods of time starting at $t$.
		Where applicable agent movement direction is given instead of a memory state.
		Relevant agent positions: 
		2
		\!\!, 3
		 and 6.
		Impossible states are excluded from enumeration.
		}
		\label{table:Two.Lane-for-lemma-no-stalls-C-236}
	\end{table}

		An immediate corollary of \cref{lemma:Two.Lane-No.Stalls} is the following:
		\begin{lemma}
			\label{lemma:Two.Lane-Infinitely.Often.Free.Cell}
			Suppose that an initial grid configuration is $\mathcal{BC}$.
			Let $i \in \{1, \ldots, n\}$, $j \in \{1, 2\}$. Suppose, there exists $i_0$, such that $a_{i_0, j}$
			is empty at time $t > T$, then grid cell $a_{i, j}$ is empty infinitely often for every $i$.
		\end{lemma}
		\begin{proof}
			Assume that $i$ is the lowest index such that $a_{i, j}$ is not empty i.o., but $a_{i - 1, j}$ is. 
			And in particular w.l.o.g. we can assume $a_{i, j}$ is occupied for all $t > T$ (denote an agent 
			at $a_{i, j}$ by $A$). Suppose agent $B$ leaves $a_{i-1, j}$ (North) at $t$, then by 
			\cref{lemma:Two.Lane-No.Fast.Return.Lemma} $B$ does not return back until at least $t + 5$. However, 
			according to \cref{lemma:Two.Lane-No.Stalls} $A$ will leave $a_{i,j}$ no later than at $t+4$. This is a
			contradiction to the assumptions.
			
			The case, when there are no cells that are constantly occupied with an i.o. empty North neighbor
			can be handled in a similar manner.
		\end{proof}
		
		We shall need to define a number of terms for simplicity:
		\begin{definition}
			A finite/infinite sequence of real numbers $(x_0, x_1, x_2, \ldots)$ 
			is called a \textit{bounded sequence}, if there exists a real number $M$, such that $|x_k| < M$ 
			for every $k$.
		\end{definition}
		\begin{definition}
			A finite/infinite sequence of real numbers $(x_0, x_1, x_2, \ldots)$ is called a \textit{bounded
			difference sequence}, if the sequence of differences $y_k := x_{k+1} - x_k$ is a bounded
			sequence.
		\end{definition}
		\begin{definition}
			A bounded sequence $\{x_k\}$ is called a \textit{special bounded sequence} if $x_k \leq 1$
			for all $k$.
		\end{definition}
		\begin{definition}
			A bounded difference sequence $\{x_k\}$ is called a \textit{special bounded difference sequence},
			if a sequence of differences is a special bounded sequence.
		\end{definition}
		
		\begin{definition}
			A cell sequence $a_{i_k, j}(t_k)_{k=1}^{N}$ will be called a \textit{path}, if the following
			holds:
			\begin{enumerate}
				\item $\{t_k\}_k$ is monotone non-decreasing special bounded difference time sequence.
				\item $\{i_k\}_k$ is a special bounded difference index sequence.
			\end{enumerate}
		\end{definition}
		
		\begin{definition}
			A path is called an \textit{empty path} if every cell in a sequence is empty.
		\end{definition}
		
		We shall prove \cref{thm:Two.Lane-the Main Theorem} in the following manner. First, we shall show
		the existence of an empty path that extends from one end of the grid column to the other (in columns 
		with at least one empty space). Then, we shall reveal, that an unsettled agent in the neighbor
		column necessarily observes an empty space near it in the target column. Finally, an unsettled agent
		will be demonstrated to move into its target column.
		
		\begin{proof}
			We assume that an initial grid configuration is $\mathcal{BC}$, and the column $j$ 
			contains at least one empty space at $t > T$ (i.e., the East-West agent movement ceased).
			
			Denote the \textit{leading agent} of the cell $a_{i, j}$ at time $t$ to be
			the agent at cell $a_{k, j}$ at time $t$, such that $k > i$ and there are no other agent at cell 
			$a_{k', j}$ at time $t$ where $i < k' < k$. It could be possible no cell leading agent at time $t$
			exists, i.e., all the cells with a higher row index are empty at time $t$.
			
			Our goal is to show the existence of an empty path that spans from $a_{1, j}$ at time $t_0$
			to $a_{n, j}$ at time $t_1$. By \cref{lemma:Two.Lane-Infinitely.Often.Free.Cell} every cell $a_{i, j}$ 
			is empty infinitely often. In particular, $a_{1, j}$. Fix time $t_0$, when $a_{1, j}$ is empty.
			We add $a_{1, j}(t_0)$ as the first cell in the sequence. In case $a_{2, j}, a_{3, j}, \ldots$
			are empty, add them to the sequence in that order.
			Inductively, we add cells to the sequence in the following manner. Denote $a_{i, j}(t-1)$ - the 
			last added cell to the sequence. If the leading agent of $a_{i, j}(t-1)$ moved South at time 
			$t-1$, we add $a_{i+1, j}(t)$ to the sequence and repeat the inductive step. If the leading 
			agent of $a_{i, j}(t-1)$ moved North, we add $a_{i+1, j}(t)$ and all the empty cells between 
			$a_{i+1, j}$ and its(\textit{new}) leading agent (or all the cells to the South, 
			if no leading agent exists). Otherwise, i.e., the leading agent did not move at all, we add
			$a_{i, j}(t)$ to the sequence.
			
			By the process description, we have just built an empty path. Note, that we still need to
			verify that a leading agent moves necessarily \textit{before} the previous leading agent could move
			back and ruin the construction process. However, see \cref{lemma:Two.Lane-No.Stalls} together with
			\cref{lemma:Two.Lane-No.Fast.Return.Lemma} that ensures this is not the case. For the first cell 
			in the sequence the fact is vacuously true.
			
			The path in the other direction could be build in the exact same way with appropriate changes.
			
			Now, suppose w.l.o.g. that an unsettled agent $A$ is an exiting agent in the first column, and that 
			an empty space is located at time $T$ in the second column. By the above claim, there exist an
			empty path in the second column (extending through space and time). Moreover, by definition,
			such path is a tempo-spatial sequence of empty cells, that either extends to spatially
			the south by at least one cell (in case a leading agent moves), or it extends in time 
			otherwise. If $A$ was in cell $a_{1,1}$ when the path started, it should have already left East. 
			Otherwise, $A$ is at row indexed at least $2$, i.e. $A$ is spatially below the path. At time tick 
			when the path reaches $a_{n, 2}$, $A$ is either at $a_{n ,1}$ and will leave East at the next time
			tick, or $A$ is spatially above the path.
			
			Is it possible for $A$ to somehow miss an empty side neighbor? We claim that this is impossible.
			Agents in the first column (such as $A$ above) move North at the start of the column activity 
			time interval, and South in the end. Hence, if $A$ is located below the path at the start of the 
			column activity interval, it($A$) can not find itself spatially above the 
			path at the end of the column activity interval. In case $A$ moves North at the start of
			such interval, it could only neighbor an empty cell belonging to the path, but it will move
			East then, in contradiction to our assumption.
			
			Therefore, we can safely assume that at the very start of the second column activity interval
			$A$ is neighbored by some agent $B$. The $B$ could be the leading agent of the path and move
			South at the start of some appropriate second column activity interval. In that case, according
			to \cref{lemma:Two.Lane-No.Fast.Return.Lemma}, the side neighbor of $A$ will remain empty for at
			least $4$ ticks, hence $A$ will not miss an empty neighbor. Otherwise, $B$ could move North
			in the end of its column activity interval, but then $A$ will also observe an empty neighbor
			during the following first column activity interval.
			
			In all cases, we have shown $A$ is neighbored by an empty cell, and, hence, will inevitably
			move East. This contradicts our assumption, that $\mathcal{BC}$ is one such \textit{unlucky}
			initial configuration, that could not be sorted by Algorithm $\mathcal{A}$.
		\end{proof}
			
		Moreover, the similar reasoning, but using the upward empty paths in the first column, shows, 
		that Algorithm $\mathcal{A}$ pushes continue vehicles out of the second column. Therefore, 
		Algorithm $\mathcal{A}$ moves the maximum number of unsettled agents to their target columns.

\ifnum\mydoclevel=0
	\end{appendices}
\else
	\end{subappendices}
\fi

\ifnum\mydoclevel=0
	\clearpage 
	\bibliographystyle{apalike}
	\bibliography{./bibliography/Biblio} 

\begin{thebibliography}{}

\bibitem[Amirgholy et~al., 2020]{amirgholy2020traffic}
Amirgholy, M., Shahabi, M., and Gao, H.~O. (2020).
\newblock Traffic automation and lane management for communicant, autonomous,
  and human-driven vehicles.
\newblock {\em Transportation research part C: emerging technologies},
  111:477--495.

\bibitem[Biham et~al., 1992]{biham1992self}
Biham, O., Middleton, A.~A., and Levine, D. (1992).
\newblock Self-organization and a dynamical transition in traffic-flow models.
\newblock {\em Physical Review A}, 46(10):R6124.

\bibitem[Buehler et~al., 2009]{buehler2009darpa}
Buehler, M., Iagnemma, K., and Singh, S. (2009).
\newblock {\em The DARPA urban challenge: autonomous vehicles in city traffic},
  volume~56.
\newblock springer.

\bibitem[Chen et~al., 2001]{chen2001causes}
Chen, C., Jia, Z., and Varaiya, P. (2001).
\newblock Causes and cures of highway congestion.
\newblock {\em IEEE Control Systems Magazine}, 21(6):26--32.

\bibitem[Chouhan et~al., 2020]{chouhan2020cooperative}
Chouhan, A.~P., Banda, G., and Jothibasu, K. (2020).
\newblock A cooperative algorithm for lane sorting of autonomous vehicles.
\newblock {\em IEEE Access}, 8:88759--88768.

\bibitem[Hedrick et~al., 1994]{hedrick1994control}
Hedrick, J.~K., Tomizuka, M., and Varaiya, P. (1994).
\newblock Control issues in automated highway systems.
\newblock {\em IEEE Control Systems Magazine}, 14(6):21--32.

\bibitem[Hsu et~al., 1993]{hsu1993protocol}
Hsu, A., Eskafi, F., Sachs, S., and Varaiya, P. (1993).
\newblock Protocol design for an automated highway system.
\newblock {\em Discrete Event Dynamic Systems}, 2(3):183--206.

\bibitem[Huang et~al., 2019]{huang2019determining}
Huang, Z., Zhang, Z., Li, H., Qin, L., and Rong, J. (2019).
\newblock Determining appropriate lane-changing spacing for off-ramp areas of
  urban expressways.
\newblock {\em Sustainability}, 11(7):2087.

\bibitem[Ji and Levinson, 2020]{ji2020review}
Ji, A. and Levinson, D. (2020).
\newblock A review of game theory models of lane changing.
\newblock {\em Transportmetrica A: transport science}, 16(3):1628--1647.

\bibitem[Jiang et~al., 2020]{jiang2020dynamic}
Jiang, X., Jin, P.~J., and Wang, Y. (2020).
\newblock A dynamic merge assistance method based on the concept of
  instantaneous virtual trajectory for vehicle-to-infrastructure connected
  vehicles.
\newblock {\em Journal of Intelligent Transportation Systems}, pages 1--20.

\bibitem[Jin et~al., 2017]{jin2017gap}
Jin, P.~J., Fang, J., Jiang, X., DeGaspari, M., and Walton, C.~M. (2017).
\newblock Gap metering for active traffic control at freeway merging sections.
\newblock {\em Journal of Intelligent Transportation Systems}, 21(1):1--11.

\bibitem[Johnson et~al., 1879]{johnson1879notes}
Johnson, W.~W., Story, W.~E., et~al. (1879).
\newblock Notes on the “15” puzzle.
\newblock {\em American Journal of Mathematics}, 2(4):397--404.

\bibitem[Kurzhanskiy and Varaiya, 2015]{kurzhanskiy2015traffic}
Kurzhanskiy, A.~A. and Varaiya, P. (2015).
\newblock Traffic management: An outlook.
\newblock {\em Economics of transportation}, 4(3):135--146.

\bibitem[Letter and Elefteriadou, 2017]{letter2017efficient}
Letter, C. and Elefteriadou, L. (2017).
\newblock Efficient control of fully automated connected vehicles at freeway
  merge segments.
\newblock {\em Transportation Research Part C: Emerging Technologies},
  80:190--205.

\bibitem[Lombard et~al., 2017]{lombard2017cooperative}
Lombard, A., Perronnet, F., Abbas-Turki, A., and El~Moudni, A. (2017).
\newblock On the cooperative automatic lane change: Speed synchronization and
  automatic “courtesy”.
\newblock In {\em Design, Automation \& Test in Europe Conference \& Exhibition
  (DATE), 2017}, pages 1655--1658. IEEE.

\bibitem[Nie et~al., 2016]{nie2016decentralized}
Nie, J., Zhang, J., Ding, W., Wan, X., Chen, X., and Ran, B. (2016).
\newblock Decentralized cooperative lane-changing decision-making for connected
  autonomous vehicles.
\newblock {\em IEEE Access}, 4:9413--9420.

\bibitem[Petig et~al., 2018]{petig2018changing}
Petig, T., Schiller, E.~M., and Suomela, J. (2018).
\newblock Changing lanes on a highway.
\newblock In {\em 18th Workshop on Algorithmic Approaches for Transportation
  Modelling, Optimization, and Systems (ATMOS 2018)}. Schloss
  Dagstuhl-Leibniz-Zentrum fuer Informatik.

\bibitem[Rabinovich et~al., 2021]{DBLP:journals/corr/abs-2111-06284}
Rabinovich, D., Amir, M., and Bruckstein, A.~M. (2021).
\newblock Optimal physical sorting of mobile vehicles.
\newblock {\em CoRR}, abs/2111.06284.

\bibitem[Ratner and Warmuth, 1990]{ratner1990n2}
Ratner, D. and Warmuth, M. (1990).
\newblock The (n2- 1)-puzzle and related relocation problems.
\newblock {\em Journal of Symbolic Computation}, 10(2):111--137.

\bibitem[Ratner and Warmuth, 1986]{ratner1986finding}
Ratner, D. and Warmuth, M.~K. (1986).
\newblock Finding a shortest solution for the n$\times$ n extension of the
  15-puzzle is intractable.
\newblock In {\em AAAI}, pages 168--172.

\bibitem[Shladover, 2009]{shladover2009cooperative}
Shladover, S.~E. (2009).
\newblock Cooperative (rather than autonomous) vehicle-highway automation
  systems.
\newblock {\em IEEE Intelligent Transportation Systems Magazine}, 1(1):10--19.

\bibitem[Toy et~al., 2002]{toy2002emergency}
Toy, C., Leung, K., Alvarez, L., and Horowitz, R. (2002).
\newblock Emergency vehicle maneuvers and control laws for automated highway
  systems.
\newblock {\em IEEE Transactions on intelligent transportation systems},
  3(2):109--119.

\bibitem[Varaiya, 1993]{varaiya1993smart}
Varaiya, P. (1993).
\newblock Smart cars on smart roads: problems of control.
\newblock {\em IEEE Transactions on automatic control}, 38(2):195--207.

\bibitem[Varaiya and Shladover, 1991]{varaiya1991sketch}
Varaiya, P. and Shladover, S.~E. (1991).
\newblock Sketch of an ivhs systems architecture.
\newblock In {\em Vehicle Navigation and Information Systems Conference, 1991},
  volume~2, pages 909--922. IEEE.

\bibitem[Wang et~al., 2019]{wang2019cooperative}
Wang, G., Hu, J., Li, Z., and Li, L. (2019).
\newblock Cooperative lane changing via deep reinforcement learning.
\newblock {\em arXiv preprint arXiv:1906.08662}.

\bibitem[Wang and Li, 2017]{wang2017dsolving}
Wang, G. and Li, R. (2017).
\newblock Dsolving: a novel and efficient intelligent algorithm for large-scale
  sliding puzzles.
\newblock {\em Journal of Experimental \& Theoretical Artificial Intelligence},
  29(4):809--822.

\bibitem[Zheng et~al., 2019]{zheng2019cooperative}
Zheng, Y., Ran, B., Qu, X., Zhang, J., and Lin, Y. (2019).
\newblock Cooperative lane changing strategies to improve traffic operation and
  safety nearby freeway off-ramps in a connected and automated vehicles
  environment.
\newblock {\em IEEE Transactions on Intelligent Transportation Systems},
  21(11):4605--4614.

\end{thebibliography}
\fi

\end{document}